\documentclass[12pt]{article}
\usepackage{amsmath,amssymb,amsfonts,color,graphicx,cite,color,feynarts}
\usepackage{axodraw,latexsym}
\usepackage{epsfig,psfrag,rotating,soul}
\usepackage{rotfloat}
\input paperdef

\evensidemargin \oddsidemargin
\allowdisplaybreaks

\begin{document}
\thispagestyle{empty}

\def\thefootnote{\fnsymbol{footnote}}

\begin{flushright}
IFT-UAM/CSIC-13-023\\
arXiv:1304.2783 [hep-ph]
\end{flushright}

\vspace{0.5cm}

\begin{center}

{\large\sc {\bf 
 New Constraints on General Slepton Flavor Mixing}} 

\vspace{1cm}

{\sc
M.~Arana-Catania$^{1}$%
\footnote{email: Miguel.Arana@uam.es}%
, S.~Heinemeyer$^{2}$%
\footnote{email: Sven.Heinemeyer@cern.ch}%
~and M.J.~Herrero$^{1}$%
\footnote{email: Maria.Herrero@uam.es}%
 
}

\vspace*{.7cm}

{\sl
$^1$Departamento de F\'isica Te\'orica and Instituto de F\'isica Te\'orica,
IFT-UAM/CSIC\\
Universidad Aut\'onoma de Madrid, Cantoblanco, Madrid, Spain

\vspace*{0.1cm}

$^2$Instituto de F\'isica de Cantabria (CSIC-UC), Santander, Spain


}

\end{center}

\vspace*{0.1cm}

\begin{abstract}
\noindent

We explore the phenomenological implications on charged lepton flavor
violating (LFV) processes from slepton flavor mixing within the Minimal
Supersymmetric Standard Model. We work under the model-independent
hypothesis of general flavor mixing in the slepton sector, being
parametrized by a complete set of dimensionless $\deABij$ ($A,B=L,R$;
$i,j=1,2,3$, $i\neq j$) parameters. The present upper bounds on
the most relevant 
LFV processes, together with the requirement of compatibility in the
choice of the MSSM parameters with the recent LHC and $(g-2)_\mu$ data,
lead to updated constraints on all slepton flavor mixing
parameters. A comparative discussion of the most effective LFV processes
to constrain the various generation mixings 
is included.
 
\end{abstract}

\def\thefootnote{\arabic{footnote}}
\setcounter{page}{0}
\setcounter{footnote}{0}

\newpage

\section{Introduction}

Lepton Flavor Violating (LFV) processes  provide 
one of the most 
challenging probes to physics beyond the Standard Model (SM) of particle
physics, and in particular to new physics involving non-vanishing flavor
mixing between the three generations. Within the SM, all interactions
preserve Lepton Flavor number and therefore the SM predicts zero rates
for all these LFV processes to all orders in perturbation theory. When
extending the SM to include neutrino masses and neutrino mixings in
agreement with the observed experimental values~\cite{pdg}, LFV
processes with external charged leptons of different generations can
then occur via one-loop diagrams with neutrinos in the internal
propagators, but the predicted rates are extremely tiny, far from being
ever reachable experimentally, due to the small masses of the
neutrinos. Therefore, a potential future
measurement of any of these 
(charged) LFV processes will be a clear signal of new physics and will
provide interesting information on the involved flavor mixing, as well
as on the underlying origin for this mixing (for a review see, for instance, \cite{Kuno:1999jp}). 

Within the Minimal Superymetric Standard Model
(MSSM)~\cite{mssm,Haber:1989xc}, there are clear candidates to produce
flavor mixings with important phenomenological implications on LFV
processes. The possible presence of soft Supersymmetry
(SUSY)-breaking parameters in the slepton sector, which are
off-diagonal in flavor space (mass parameters as well as trilinear
couplings) are
the most general way to introduce slepton flavor mixing within the
MSSM. The off-diagonality in the slepton mass matrix reflects the   
misalignment (in flavor space) between leptons and sleptons mass
matrices, that cannot be diagonalized 
simultaneously. This misalignment can be produced from various
origins. For instance, under the hypothesis of non-negligible
neutrino Yukawa 
couplings, as it happens in Seesaw models with three heavy  right
handed neutrinos and their SUSY partners, these off-diagonal slepton
mass matrix entries can be generated by Renormalization Group Equations
(RGE) running from the high energies, where the heavy right-handed
neutrinos are active, down to the low energies where the LFV processes
are explored~\cite{Hall:1985dx,Borzumati:1986qx}. 
The phenomenological implications of large neutrino Yukawa
couplings on LFV processes within the context of SUSY-Seesaw Models have
been studied exhaustively in the literature, and the absence of
experimental LFV signals sets stringent bounds on the parameters of
these models~\cite{Borzumati:1986qx,Hisano:1995nq,Hisano:1995cp,Hisano:1998fj,Illana:2000zy,Ellis:2002fe,Sher:2002ew,Brignole:2004ah,Chen:2006hp,Arganda:2005ji,Antusch:2006vw,Arganda:2007jw,Arganda:2008jj,Herrero:2009tm,Deppisch:2005rv,Abada:2010kj,Dinh:2012bp}.  

In this work we will not investigate the possible dynamical origin
of this slepton-lepton misalignment, nor the particular predictions for
the off-diagonal slepton soft SUSY-breaking mass terms in specific
SUSY models, but instead we parametrize the general
non-diagonal entries in the slepton mass matrices
in terms of generic soft SUSY-breaking terms, and we explore
here their phenomenological implications on LFV physics. In particular,
we explore the consequences of these general slepton mass matrices
that can produce, via radiative loop corrections, important
contributions to the rates of the LFV
processes~\cite{Borzumati:1986qx,Gabbiani:1996hi}. Specifically, we 
parametrize the non-diagonal slepton mass matrix entries in terms of a
complete set of generic dimensionless parameters, $\deABij$ ($A,B=L,R$;
$i,j=1,2,3$) where $L,R$ refer to the 
``left-'' and ``right-handed'' SUSY partners of the corresponding
leptonic degrees of freedom
and $i,j$ ($i \neq j$) are the involved generation
indexes. With this model-independent parametrization of general slepton
flavor mixing we  explore the sensitivity to the various $\deABij$'s
in different LFV processes and analyze comparatively which processes are
the most competitive ones. Previous studies of general slepton 
mixing within the MSSM have already set upper bounds for the
values of these $\deABij$'s that can be extracted from some selected
experimental LFV searches (for a review see, for instance,
\cite{Raidal:2008jk}). Some of these studies focus  on the LFV radiative
decays~\cite{Masina:2002mv}, $\mu \to e \gamma$, $\tau \to e \gamma$ and
$\tau \to \mu \gamma$, here denoted collectively as $l_j \to l_i
\gamma$, and others also take into account the leptonic LFV three body
decays, $\mu \to 3 e$, $\tau \to 3 e$ and $\tau \to 3 \mu$, referred
together here as  $l_j \to 3 l_i$, as well as the muon to electron
conversion in heavy nuclei\cite{Paradisi:2005fk}. There are also some 
studies that focus on the chirally-enhanced loop corrections that are 
induced in the MSSM in presence of general sources of lepton flavor 
violation~\cite{Crivellin:2010er,Crivellin:2011jt}.

One main aspect in this work is to update these studies of general
flavor mixing in the slepton sector of the MSSM, and to find new
constraints to the full set of $\deABij$'s mixing parameters in the
light of recent data, both on the most relevant LFV processes
\cite{Adam:2013mnn,Aubert:2009ag,Bertl:2006up,Bellgardt:1987du,Hayasaka:2010np,Hayasaka:2010et,Miyazaki:2008mw}
and also in view of the collected data at LHC\cite{LHCHiggs,LHCSusy,CMSpashig12050}, which has
provided very important information and constraints for the MSSM,
including the absence of SUSY particle experimental signals and
the discovery of a Higgs boson with a mass close to $125 - 126 \gev$.
We work consistently in MSSM scenarios that are compatible with
LHC data. In particular the analyzed scenarios have relatively
heavy SUSY spectra, which are naturally in
agreement with the present MSSM particle mass bounds (although
substantially lower masses, especially in the electroweak sector, are
allowed by LHC data). Furthermore the analyzed scenarios are chosen such
that the light $\cp$-even MSSM Higgs mass is around $125 - 126 \gev$ and
thus in agreement with the recent Higgs boson discovery~\cite{LHCHiggs}.
In addition we
require that our selected MSSM scenarios give a
prediction for the muon anomalous magnetic moment, $(g-2)_\mu$, in
agreement with current data~\cite{Bennett:2006fi}.
  
We present here a complete one-loop numerical analysis of the most
relevant LFV processes, including the three $l_j \to l_i \gamma$
radiative decays, the three $l_j \to 3 l_i$ leptonic decays, the muon to
electron conversion rates in heavy nuclei, and the two most promissing
semileptonic LFV tau decays, $\tau \to \mu \eta$ and $\tau \to e \eta$. 
Although the radiative decays are usually the most constraining
LFV processes, the leptonic and semileptonic decays are also of interest
because they can be mediated by the MSSM Higgs bosons, therefore  giving
access to the Higgs sector parameters and, presumably, with different
sensitivities to the various $\deABij$'s than those involved in the
radiative decays. From this complete one-loop analysis and the
requirement of compatibility with LFV searches, with LHC data and with
$(g-2)_\mu$ data, we  derive the general behavior of the
constraints on the $\deABij$'s.  

The paper is organized as follows: first we review the main features of
the MSSM with general slepton flavor mixing and set the relevant
notation for the $\deABij$'s in ~\refse{sec:scenarios}. The selection of
specific LFV processes and MSSM scenarios that we  work with here are
presented in  ~\refse{sec:scenariosLFV}. A summary on the present
experimental bounds on LFV, that will be used in our analysis are also
included in this section.  \refse{sec:results} contains the main results
of our numerical analysis and present the new constraints found on the
$\deABij$'s. Our conclusions are finally summarized in
\refse{sec:conclusions}.


\section{The MSSM with general slepton flavor mixing}
\label{sec:scenarios}

We work in SUSY scenarios with the same particle content as the
MSSM, but with 
general flavor mixing in the slepton sector. Within these
scenarios, besides the tiny lepton flavor violation induced by the PMNS matrix
of the neutrino sector and transmitted by the tiny neutrino Yukawa
couplings which we ignore here, this flavor mixing in the slepton sector is the main generator of LFV processes. The most general hypothesis for flavor
mixing in the slepton sector assumes a mass matrix that is not diagonal
in flavor space, both for charged sleptons and sneutrinos. 
In the charged slepton sector we have a $6 \times 6$ mass matrix,
since there are six electroweak interaction eigenstates, 
${\tilde l}_{L,R}$ with $l=e, \mu, \tau$. For the sneutrinos 
we have a $3 \times 3$ mass matrix, since within the MSSM we have
only three electroweak interaction eigenstates, ${\tilde \nu}_{L}$ with
$\nu=\nu_e, \nu_\mu, \nu_\tau$. 

The non-diagonal entries in this $6 \times 6$ general matrix for charged
sleptons can be described in a model-independent way in terms of a set
of 
dimensionless parameters $\deABij$ ($A,B=L,R$; $i,j=1,2,3$, 
$i \neq j$), where $L,R$ refer to the 
``left-'' and ``right-handed'' SUSY partners of the corresponding
leptonic degrees of freedom, and $i,j$
indexes run over the three generations. These scenarios with general
sfermion flavor mixing lead generally to larger LFV rates than in the
so-called Minimal Flavor Violation Scenarios, where the mixing is
induced exclusively by the Yukawa coupling of the corresponding fermion
sector. This is true for both squarks and sleptons but it is obviously of
special interest in the slepton case due to the extremely small size of
the lepton Yukawa couplings, suppressing LFV processes from this origin. 
Hence, in the present case of slepton mixing, we
assume that the $\deABij$'s  provide the unique origin
of LFV processes with potentially measurable rates.   

One usually starts  with the non-diagonal $6 \times 6$ slepton squared
mass matrix referred to the electroweak interaction basis,  
 that we order here as    $(\SelL, \SmuL, \StauL, \SelR, \SmuR,
 \StauR)$, and write this matrix in terms of left- and right-handed
 blocks $M^2_{\tilde l \, AB}$  
 ($A,B=L,R$), which are non-diagonal $3\times 3$ matrices,
\begin{equation}
{\mathcal M}_{\tilde l}^2 =\left( \begin{array}{cc}
M^2_{\tilde l \, LL} & M^2_{\tilde l \, LR} \\ 
M_{\tilde l \, LR}^{2 \, \dagger} & M^2_{\tilde l \,RR}
\end{array} \right),
\label{eq:slep-6x6}
\end{equation} 
 where:
 \begin{alignat}{5}
M_{\tilde l \, LL \, ij}^2 
  = &  m_{\tilde L \, ij}^2 + \left( m_{l_i}^2
     + (-\frac{1}{2}+ \sin^2 \theta_W ) M_Z^2 \cos 2\beta \right) \delta_{ij},  \notag\\
 M^2_{\tilde l \, RR \, ij}
  = &  m_{\tilde E \, ij}^2 + \left( m_{l_i}^2
     -\sin^2 \theta_W M_Z^2 \cos 2\beta \right) \delta_{ij} \notag, \\
  M^2_{\tilde l \, LR \, ij}
  = &  v_1 {\cal A}_{ij}^l- m_{l_{i}} \mu \tb \, \delta_{ij},
\label{eq:slep-matrix}
\end{alignat}
with flavor indexes $i,j=1,2,3$ corresponding to the first, second and
third generation respectively; $\theta_W$ is the weak angle; $M_Z$ is
the $Z$ gauge boson mass, and $(m_{l_1},m_{l_2},
m_{l_3})=(m_e,m_\mu,m_\tau)$ are the lepton masses; $\tb=v_2/v_1$
with  $v_1=\left< {\cal H}_1^0 \right>$ and $v_2=\left< {\cal H}_2^0
\right>$ being the two vacuum expectation values of the corresponding
neutral Higgs boson in the Higgs $SU(2)$ doublets, ${\cal H}_1= ({\cal
  H}^0_1\,\,\, {\cal H}^-_1)$ and ${\cal H}_2= ({\cal H}^+_2 \,\,\,{\cal
  H}^0_2)$; 
$\mu$ is the usual Higgsino mass term.
It should be noted that the non-diagonality in flavor comes
exclusively from the soft SUSY-breaking parameters, that could be
non-vanishing for $i \neq j$, namely: the masses $m_{\tilde L \, ij}$
for the slepton $SU(2)$ doublets, $(\tilde \nu_{Li}\,\,\, \tilde
l_{Li})$, the masses $m_{\tilde E \, ij}$ for the slepton $SU(2)$
singlets, $(\tilde l_{Ri})$, and the trilinear couplings, ${\cal
  A}_{ij}^l$.   

In the sneutrino sector there is, correspondingly, a one-block $3\times
3$ mass matrix, that is referred to the $(\tinu_{eL}, \tinu_{\mu L},
\tinu_{\tau L})$ electroweak interaction basis: 
\begin{equation}
{\mathcal M}_{\tilde \nu}^2 =\left( \begin{array}{c}
M^2_{\tilde \nu \, LL}   
\end{array} \right),
\label{eq:sneu-3x3}
\end{equation} 
 where:
\begin{equation} 
  M_{\tilde \nu \, LL \, ij}^2 
  =   m_{\tilde L \, ij}^2 + \left( 
      \frac{1}{2} M_Z^2 \cos 2\beta \right) \delta_{ij},   
\label{eq:sneu-matrix}
\end{equation} 
 
It should also be noted that, due to $SU(2)_L$ gauge invariance
the same soft masses $m_{\tilde L \, ij}$ enter in both the slepton and
sneutrino $LL$ mass matrices. 
If neutrino masses and neutrino flavor mixings (oscillations) were taken into account, the soft SUSY-breaking parameters for the
sneutrinos would differ from the corresponding ones for charged
sleptons by a rotation with the PMNS matrix. However, taking the neutrino masses and oscillations 
into account in the SM leads to LFV effects that are extremelly small. For instance, in $\mu \to e \gamma$  they are of \order{10^{-47}} in case
of Dirac neutrinos with mass around 1~eV and maximal mixing~\cite{Kuno:1999jp,DiracNu}, and of \order{10^{-40}} in case of
Majorana neutrinos~\cite{Kuno:1999jp,MajoranaNu}. Consequently we do not expect
large effects from the inclusion of neutrino mass effects here.
The general slepton flavor mixing is introduced via the
non-diagonal terms in the soft breaking slepton mass matrices and
trilinear coupling matrices, which are defined here as:

\noindent \begin{equation}  
m^2_{\tilde L}= \left(\begin{array}{ccc}
 m^2_{\tilde L_{1}} & \delta_{12}^{LL} m_{\tilde L_{1}}m_{\tilde L_{2}} & \delta_{13}^{LL} m_{\tilde L_{1}}m_{\tilde L_{3}} \\
 \delta_{21}^{LL} m_{\tilde L_{2}}m_{\tilde L_{1}} & m^2_{\tilde L_{2}}  & \delta_{23}^{LL} m_{\tilde L_{2}}m_{\tilde L_{3}}\\
\delta_{31}^{LL} m_{\tilde L_{3}}m_{\tilde L_{1}} & \delta_{32}^{LL} m_{\tilde L_{3}}m_{\tilde L_{2}}& m^2_{\tilde L_{3}} \end{array}\right)\end{equation}

\noindent \begin{equation}
v_1 {\cal A}^l  =\left(\begin{array}{ccc}
m_e A_e & \delta_{12}^{LR} m_{\tilde L_{1}}m_{\tilde E_{2}} & \delta_{13}^{LR} m_{\tilde L_{1}}m_{\tilde E_{3}}\\
\delta_{21}^{LR}  m_{\tilde L_{2}}m_{\tilde E_{1}} & m_\mu A_\mu & \delta_{23}^{LR} m_{\tilde L_{2}}m_{\tilde E_{3}}\\
\delta_{31}^{LR}  m_{\tilde L_{3}}m_{\tilde E_{1}} & \delta_{32}^{LR}  m_{\tilde L_{3}} m_{\tilde E_{2}}& m_{\tau}A_{\tau}\end{array}\right)\label{v1Al}\end{equation}

\noindent \begin{equation}  
m^2_{\tilde E}= \left(\begin{array}{ccc}
 m^2_{\tilde E_{1}} & \delta_{12}^{RR} m_{\tilde E_{1}}m_{\tilde E_{2}} & \delta_{13}^{RR} m_{\tilde E_{1}}m_{\tilde E_{3}}\\
 \delta_{21}^{RR} m_{\tilde E_{2}}m_{\tilde E_{1}} & m^2_{\tilde E_{2}}  & \delta_{23}^{RR} m_{\tilde E_{2}}m_{\tilde E_{3}}\\
\delta_{31}^{RR}  m_{\tilde E_{3}} m_{\tilde E_{1}}& \delta_{32}^{RR} m_{\tilde E_{3}}m_{\tilde E_{2}}& m^2_{\tilde E_{3}} \end{array}\right)\end{equation}

In all this work, for simplicity, we are assuming that all $\deABij$
parameters are real, therefore, hermiticity of 
${\mathcal M}_{\tilde l}^2$ and 
${\mathcal M}_{\tilde \nu}^2$ implies $\delta_{ij}^{AB}= \delta_{ji}^{BA}$.
Besides, in order to avoid extremely large off-diagonal matrix entries
we restrict ourselves to $|\deABij| \leq 1$. It is worth to have in mind for the rest of this work, that our parametrization of the off-diagonal in flavor space entries in the above mass matrices is purely phenomenological and does not rely on any specific assumption on the origin of the MSSM soft mass parameters. In particular, it should be noted that our parametrization for the LR and RL squared mass entries connecting different generations (i.e. for $i \neq j$) assumes a similar generic form as for the LL and RR entries. For instance, $M^2_{\tilde l \, LR \, 23}= \delta_{23}^{LR} m_{\tilde L_{2}}m_{\tilde E_{3}}$. This implies that our hypothesis for the trilinear off-diagonal couplings ${\cal A}^l_{ij}$ with $i \neq j$ (as derived from Eq.(\ref{v1Al})) is one among other possible definitions considered in the literature. In particular, it is related to the usual assumption $M^2_{\tilde l \, LR \, ij} \sim v_1 M_{\rm SUSY}$ by setting ${\cal A}^l_{ij} \sim {\cal O}(M_{\rm SUSY})$, where $v^2=v_1^2+v_2^2$ and $M_{\rm SUSY}$ is a typical SUSY mass scale, as it is done for instance in Ref.~\cite{Crivellin:2011jt}.  

The next step is to rotate the sleptons and sneutrinos from the electroweak interaction basis to the physical mass eigenstate basis, 
\BE
\VL  \til_{1} \\ \til_{2}  \\ \til_{3} \\
                                    \til_{4}   \\ \til_{5}  \\\til_{6}   \VR
  \; = \; R^{\til}  \VL \SelL \\ \SmuL \\\StauL \\ 
  \SelR \\ \SmuR \\ \StauR \VR ~,~~~~
\VL  \tinu_{1} \\ \tinu_{2}  \\  \tinu_{3}  \VR             \; = \; R^{\tinu}  \VL \tinu_{eL} \\ \tinu_{\mu L}  \\  \tinu_{\tau L}   \VR ~,
\label{newsquarks}
\end{equation} 
with $R^{\til}$ and $R^{\tinu}$ being the respective $6\times 6$ and
$3\times 3$ unitary rotating matrices that yield the diagonal
mass-squared matrices as follows, 
\BEA
{\rm diag}\{m_{\til_1}^2, m_{\til_2}^2, 
          m_{\til_3}^2, m_{\til_4}^2, m_{\til_5}^2, m_{\til_6}^2 
           \}  & = &
R^{\til}  \;  {\cal M}_{\til}^2   \; 
 R^{\til \dagger}    ~,\\
{\rm diag}\{m_{\tinu_1}^2, m_{\tinu_2}^2, 
          m_{\tinu_3}^2  
          \}  & = &
R^{\tinu}  \;   {\cal M}_{\tinu}^2   \; 
 R^{\tinu \dagger}    ~.
\EEA 
The physics must not depend on the ordering of the masses. However,
  in our numerical analysis we work with mass ordered states, 
$m_{\til_i} \le m_{\til_j}$ for $i < j$ and $m_{\tinu_k} \le  m_{\tinu_l}$ 
for $k < l$.


\section{Selection of LFV processes and MSSM parameters} 
\label{sec:scenariosLFV}

The general slepton flavor mixing introduced above produce interactions
among mass eigenstates of different generations, therefore changing
flavor. 
In the physical basis for leptons $l_i$ ($i=1,2,3$), sleptons $\til_X$
($X=1,..,6$), sneutrinos $\tinu_X$ ($X=1,2,3$),  neutralinos ${\tilde
  \chi}_A^0$ ($A=1,2,3,4$), charginos ${\tilde \chi}_A^\pm$ ($A=1,2$)
and Higgs bosons,  
$H_p\,\,\, (p=1,2,3)= h^0, H^0, A^0$,  one gets generically
non-vanishing couplings for intergenerational interactions
like, for instance: 
${\tilde \chi}_A^0 l_i {\tilde l}_X$, 
${\tilde \chi}_A^\pm l_i {\tilde \nu}_X$, 
$Z \til_X \til_Y$, $H_p \til_X \til_Y$ and  $H_p \tinu_X \tinu_Y$. 
When these interactions appear in loop-induced processes they
can then mediate LFV processes
involving leptons of different flavors $l_i$ and $l_j$, with $i\neq j$,
in the external states. The dependence of the LFV rates for these
processes on the previously introduced  $\deABij$ parameters then
appears both in the values of the physical slepton and sneutrino masses,
and in the values of these intergenerational couplings via the rotation
matrices  $R^{\til}$ and $R^{\tinu}$. For the present work, we use the
set of Feynman rules for these and other relevant couplings among mass
eigenstates, as summarized in
\citeres{Arganda:2005ji,Arganda:2007jw}.


\subsection{Selected LFV processes}

Our selection of LFV processes is driven by the requirement that we wish
to determine the constraints on all the slepton flavor mixing parameters
by studying different kinds of one-loop LFV vertices involving $l_i$ and
$l_j$ with $i \neq j$ in the external lines. In particular we want to
study the sensitivity to the $\deABij$'s in the most relevant
(three-point) LFV one-loop vertices, which are: the vertex with a
photon, $(l_il_j\gamma)_{\rm 1-loop}$, the vertex with a $Z$ gauge
boson, $(l_il_jZ)_{\rm 1-loop}$ and the vertices with the Higgs
bosons, 
$(l_il_jh^0)_{\rm 1-loop}$, $(l_il_jH^0)_{\rm 1-loop}$ and
$(l_il_jA^0)_{\rm 1-loop}$. This leads us to single out some specific
LFV processes where these one-loop generated LFV vertices play a
relevant role. We have chosen the following subset of LFV processes, all
together involving these particular LFV one-loop vertices:  
\begin{itemize}
\item[1.-] Radiative LFV decays: $\mu \to e \gamma$, $\tau \to e \gamma$ and
$\tau \to \mu \gamma$. These are sensitive to the $\deABij$'s via the
  $(l_il_j\gamma)_{\rm 1-loop}$ vertices with a real photon. 
\item[2.-] Leptonic LFV decays: $\mu \to 3 e$, $\tau \to 3 e$ and $\tau
  \to 3 \mu$. These are sensitive to the $\deABij$'s via the
  $(l_il_j\gamma)_{\rm 1-loop}$ vertices with a virtual photon, via
  the $(l_il_jZ)_{\rm 1-loop}$ vertices with a virtual $Z$, and via the
  $(l_il_jh^0)_{\rm 1-loop}$, $(l_il_jH^0)_{\rm 1-loop}$ and
  $(l_il_jA^0)_{\rm 1-loop}$ vertices with  virtual Higgs bosons. 
\item[3.-] Semileptonic LFV tau decays: $\tau \to \mu \eta$ and $\tau
  \to e \eta$. These are sensitive to the $\deABij$'s via $(\tau \mu
  A^0)_{\rm 1-loop}$ and $(\tau e A^0)_{\rm 1-loop}$ vertices,
  respectively, with a virtual $A^0$, and via  
$(\tau \mu Z)_{\rm 1-loop}$ and $(\tau e Z)_{\rm 1-loop}$ vertices,
  respectively with a virtual $Z$. 
\item[4.-] Conversion of $\mu$ into $e$ in heavy nuclei: These are
  sensitive to the $\deABij$'s via the $(\mu e\gamma)_{\rm 1-loop}$
  vertex with a virtual photon, via the $(\mu e Z)_{\rm 1-loop}$
  vertex with a virtual $Z$, and via the $(\mu e h^0)_{\rm 1-loop}$ and
  $(\mu e H^0)_{\rm 1-loop}$ vertices with a virtual $h^0$ and $H^0$
  Higgs boson, respectively.  
\end{itemize}  
%

\begin{figure}[ht!]
\begin{center}
\psfig{file=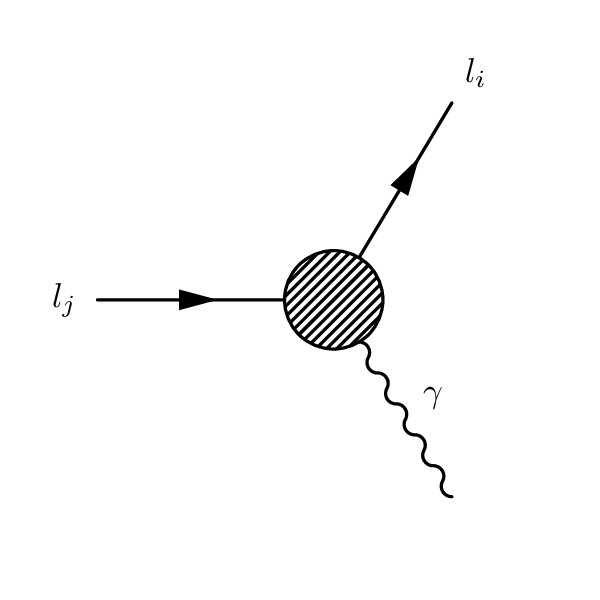 ,scale=0.77,clip=}\\
\psfig{file=figsLFV/lj3li_diagrams.epsi,scale=0.77,clip=}\\
\psfig{file=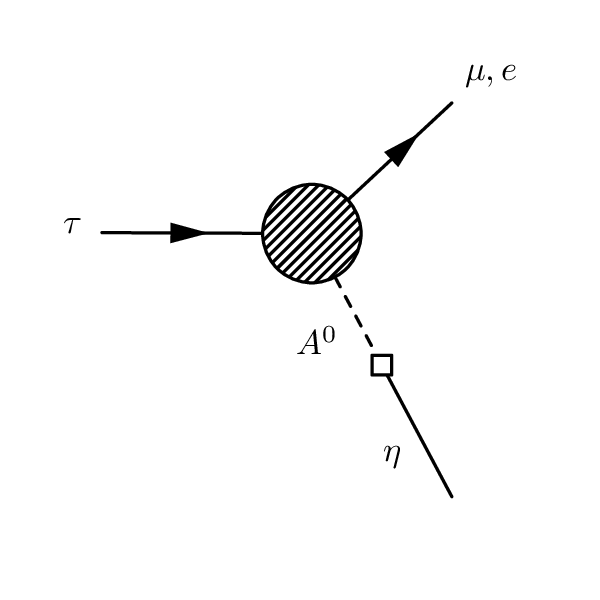,scale=0.77,clip=}
\psfig{file=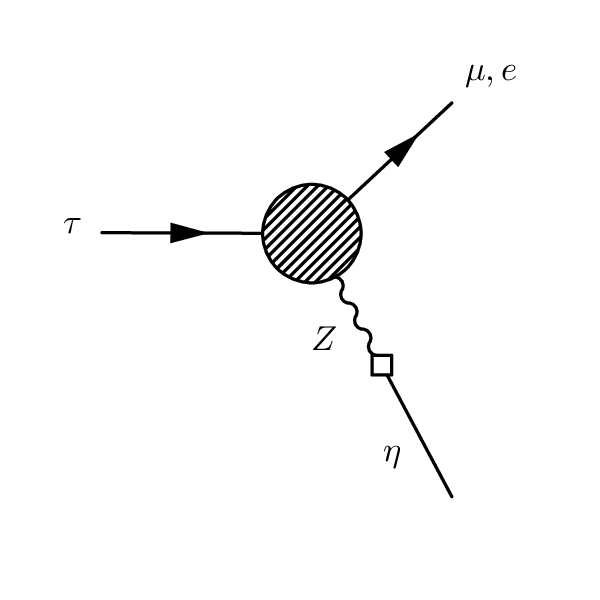,scale=0.77,clip=}\\
\psfig{file=figsLFV/CRmue_diagrams.epsi,scale=0.77,clip=} 
\end{center}
\caption{Generic one-loop diagrams contributing to LFV processes: 
1) $l_j \to l_i \gamma$; 
2) $l_j \to 3 l_i$, 
  mediated by $\gamma$ and $Z$ gauge bosons, by $H_p=h^0,H^0,A^0$ Higgs
  bosons and by boxes;   
3) $\tau \to \mu \eta$ and $\tau \to e \eta$, mediated by $A^0$ Higgs boson and by $Z$ gauge
  boson;
4) $\mu-e$ conversion in nuclei,
  mediated by $\gamma$, and $Z$ gauge bosons, by $H_p=h^0,H^0$ Higgs
  bosons, and by boxes.}  
\label{diagramsLFV}
\end{figure} 

The generic one-loop diagrams contributing to all the LFV processes above, are 
summarized in Fig.\ref{diagramsLFV}. These include the $\gamma$-mediated
diagrams, the $Z$-mediated diagrams, and the $h^0$, $H^0$ and
$A^0$-mediated diagrams. The generic one-loop box diagrams are also
shown in this figure. These also include the $\deABij$'s but their
sensitivities to these parameters are much lower than via the above
quoted three-point vertices. 
They are, however, included in our analytical results and in our
  numerical evaluation.

For our forthcoming numerical analysis of these LFV processes we have
implemented the full one-loop formulas into our private Fortran
  code. The analytical results are taken from various publications
  (with one of the authors as co-author):
\citere{Arganda:2005ji} for $\br(l_j \to 3 l_i)$ and  
$\br(l_j \to l_i \gamma)$, \citere{Arganda:2008jj} for $\br(\tau \to \mu
\eta)$ and $\br(\tau \to e \eta)$, and \citere{Arganda:2007jw} for  
the $\mu - e$ conversion rate in heavy nuclei, relative to the muon
capture rate $\CR(\mu-e, {\rm Nuclei})$.  
Following the same procedure of \cite{Arganda:2008jj} we use Chiral
Perturbation Theory for the needed hadronization of quark bilinears
involved in  the quark-level $\tau \to \mu q q'$ and $\tau \to e q q'$
decays that lead the $\eta$ particle in the final state. Our treatment
of the heavy nuclei and the proper approximations to go from the LFV
amplitudes at the parton level to the LFV rates at the nuclear level are
described in \cite{Arganda:2007jw}. For brevity, we omit to
explicit here all 
these needed formulas for the computation of the LFV rates and refer
the reader to the above quoted references for the details.  

The list of specific one-loop diagrams contributing to the relevant
$(l_il_j\gamma)_{\rm 1-loop}$, $(l_il_jZ)_{\rm 1-loop}$
$(l_il_jh^0)_{\rm 1-loop}$, $(l_il_jH^0)_{\rm 1-loop}$ and
$(l_il_jA^0)_{\rm 1-loop}$ vertices can also be found in
\citeres{Arganda:2005ji,Arganda:2007jw}. 
The main contributions come from the loops with
  charginos/sneutrinos and with neutralinos/sleptons. This will be
  relevant for the analytical interpretation of our results below.


\subsection{The MIA basic reference formulas} 
\label{sec:MIA}

For completeness, and in order to get a better understanding of the
forthcoming full one-loop results leading to the maximal allowed deltas
and their behavior with the relevant MSSM parameters, we include in this
section the main formulas for the LFV radiative decays within the Mass
Insertion Approximation (MIA) that we take from
\citere{Paradisi:2005fk}. These are simple formulas and illustrate
clearly the qualitative behavior of the LFV rates with all the deltas
and all the MSSM parameters.  The branching ratios of the radiative $l_j
\to l_i \gamma$ decays, with $ji=21$, $31$ and $32$, are:  
\BEA
{\rm BR}(l_j \to l_i \gamma) &=& \frac{\alpha}{4} 
(\frac{m_{l_j}^5}{\Gamma_{l_j}}) \left(|(A_{ij}^L)|^2+|(A_{ij}^R)|^2\right)
\label{BRs}
\EEA
where $\Gamma_{l_j}$ is the total ${l_j}$ width,  and the amplitudes, in the single delta insertion approximation, are given by~\cite{Paradisi:2005fk}:
\BEA
(A_{ij}^L)_{\rm MIA} &=& \frac{\alpha_2}{4 \pi} \Delta^{LL}_{ij}
\left[\frac{f_{1n}(a_{L2})+f_{1c}(a_{L2})}{m_{\tilde L}^4} + \frac{\mu M_2 \tb}{(M_2^2-\mu^2)} \frac{(f_{2n}(a_{L2},b_L)+f_{2c}(a_{L2},b_L))}{m_{\tilde L}^4}\right] \nonumber \\  
&+& \frac{\alpha_1}{4 \pi} \Delta^{LL}_{ij}
\left[\frac{f_{1n}(a_L)}{m_{\tilde L}^4}+\mu M_1 \tb \left(\frac{-f_{2n}(a_L,b_L)}
{m_{\tilde L}^4(M_1^2-\mu^2)} +\frac{2f_{2n}(a_L)}{m_{\tilde L}^4(m_{\tilde R}^2-m_{\tilde L}^2)}\right) \right] \nonumber \\ 
&+&\frac{\alpha_1}{4 \pi} \Delta^{LL}_{ij} \left[\frac{\mu M_1 \tb}{(m_{\tilde R}^2-m_{\tilde L}^2)^2}  
   \left(\frac{f_{3n}(a_R)}{m_{\tilde R}^2}-\frac{f_{3n}(a_L)}{m_{\tilde L}^2}\right) \right]
\nonumber \\ 
&+&\frac{\alpha_1}{4 \pi} \Delta^{LR}_{ij}\left[\frac{1}{(m_{\tilde L}^2-m_{\tilde R}^2)}
\left(\frac{M_1}{m_{l_j}}\right) \left(\frac{f_{3n}(a_R)}{m_{\tilde R}^2}-\frac{f_{3n}(a_L)}{m_{\tilde L}^2} \right)\right]
\EEA
and 
\BEA
(A_{ij}^R)_{\rm MIA} &=&
\frac{\alpha_1}{4 \pi} \Delta^{RR}_{ij}
\left[\frac{4f_{1n}(a_R)}{m_{\tilde R}^4}+\mu M_1 \tb \left(\frac{2f_{2n}(a_R,b_R)}
{m_{\tilde R}^4(M_1^2-\mu^2)} +\frac{2f_{2n}(a_R)}{m_{\tilde R}^4(m_{\tilde L}^2-m_{\tilde R}^2)}\right) \right] \nonumber \\ 
&+&\frac{\alpha_1}{4 \pi} \Delta^{RR}_{ij} \left[\frac{\mu M_1 \tb}{(m_{\tilde L}^2-m_{\tilde R}^2)^2}  
   \left(\frac{f_{3n}(a_L)}{m_{\tilde L}^2}-\frac{f_{3n}(a_R)}{m_{\tilde R}^2}\right) \right]
\nonumber \\ 
&+&\frac{\alpha_1}{4 \pi} \Delta^{RL}_{ij}\left[\frac{1}{(m_{\tilde R}^2-m_{\tilde L}^2)}
\left(\frac{M_1}{m_{l_j}}\right) \left(\frac{f_{3n}(a_L)}{m_{\tilde L}^2}-\frac{f_{3n}(a_R)}{m_{\tilde R}^2} \right)\right]~,
\EEA
where $\alpha_1=(5/3)(\alpha/\cos^2\theta_W)$,
$\alpha_2=(\alpha/\sin^2\theta_W)$, $a_{L2}=M_2^2/m_{\tilde L}^2$,
$a_L=M_1^2/m_{\tilde L}^2$, $a_R=M_1^2/m_{\tilde R}^2$,  
 $b_L=\mu^2/m_{\tilde L}^2$, $b_R=\mu^2/m_{\tilde R}^2$, 
 $\Delta^{AB}_{ij}=\delta^{AB}_{ij} m_{\tilde A}m_{\tilde B}$ and
$m_{\tilde L}$ and $m_{\tilde R}$ are the average slepton
masses in the ${\tilde L}$ and ${\tilde R}$ slepton sectors,
respectively. 
The $M_1$ and $M_2$ are the soft SUSY-breaking parameters in the
U(1) and SU(2) gaugino sector, respectively.
The $f_{in}$'s and $f_{ic}$'s are loop functions from
neutralinos and charginos contributions, respectively, given by: 
\BEA 
f_{1n}(a)&=&\frac{-17a^3+9 a^2+9a-1+6a^2(a+3)\ln a}{24(1-a)^5}~, \nonumber \\
f_{2n}(a)&=&\frac{-5a^2+4a+1+2a(a+2) \ln a}{4(1-a)^4}~,\nonumber \\
f_{3n}(a)&=&\frac{1+2a \ln a -a^2}{2(1-a)^3}~,\nonumber \\
f_{1c}(a)&=&\frac{-a^3-9a^2+9a+1+6a(a+1) \ln a}{6(1-a)^5}~,\nonumber \\
f_{2c}(a)&=&\frac{-a^2-4a+5+2(2a+1) \ln a}{2(1-a)^4}~, \nonumber \\
f_{2n}(a,b)&=&f_{2n}(a)-f_{2n}(b)~, \non \\ 
f_{2c}(a,b)&=&f_{2c}(a)-f_{2c}(b)~.
\label{functions}
\EEA
It is also very illustrative to compare the forthcoming results with
those of the MIA for the case of equal mass scales, $m_{\tilde
  L}=m_{\tilde R}=\mu=M_2=M_1 \equiv m_S$. From the previous formulas we
get: 
\BEA
(A_{ij}^L)_{\rm MIA}&=&\frac{\alpha_2}{4 \pi} \delta^{LL}_{ij} \left[ \frac{1}{240} \frac{1}{m_S^2}+\tb \frac{1}{15} \frac{1}{m_S^2} \right] \nonumber \\ 
&+&\frac{\alpha_1}{4 \pi} \delta^{LL}_{ij} \left[ \frac{-1}{80}\frac{1}{m_S^2}+
\tb \frac{1}{12}\frac{1}{m_S^2} \right] \nonumber \\ 
&+&\frac{\alpha_1}{4 \pi} \delta^{LR}_{ij} \left[\frac{1}{m_S m_{l_j}} \right]
\label{MIA-L}
\EEA
and 
\BEA
(A_{ij}^R)_{\rm MIA}&=& 
 \frac{\alpha_1}{4 \pi} \delta^{RR}_{ij} \left[ \frac{-1}{20}\frac{1}{m_S^2}-
\tb \frac{1}{60}\frac{1}{m_S^2} \right] \nonumber \\ 
&+&\frac{\alpha_1}{4 \pi} \delta^{RL}_{ij} \left[\frac{1}{m_S m_{l_j}} \right]~.
\label{MIA-R}
\EEA
In all these basic MIA formulas one can see clearly the scaling of
the BRs with all the deltas, in the single mass insertion approximation, and
with the most relevant parameters for the present study, namely, the
common/average SUSY mass $m_{S}$,  and $\tb$. These formulas will be
used below in the interpretation of the full numerical results.


\subsection{Experimental bounds on LFV}
\label{sec:expbounds}

So far, LFV has not been observed. 
The best present (90\% CL) experimental bounds on the previously
selected LFV processes are summarized in the following: 
 \BEA
\br(\mu \to e \gamma) < 5.7 \times 10^{-13} &\mbox{\cite{Adam:2013mnn}} \nonumber \\
\br(\tau \to \mu \gamma) < 4.4 \times 10^{-8} &\mbox{\cite{Aubert:2009ag}}   \nonumber\\
\br(\tau \to e \gamma) < 3.3 \times 10^{-8} &\mbox{\cite{Aubert:2009ag}} \nonumber\\
\br(\mu \to eee) < 1.0 \times 10^{-12}& \mbox{\cite{Bellgardt:1987du}} \nonumber\\
\br(\tau \to \mu\mu\mu) < 2.1 \times 10^{-8}&\mbox{\cite{Hayasaka:2010np}}\nonumber\\
\br(\tau \to e e e)< 2.7 \times 10^{-8}&\mbox{\cite{Hayasaka:2010np}}\nonumber\\
\CR(\mu-e, {\rm Au}) < 7.0 \times 10^{-13}&\mbox{\cite{Bertl:2006up}}\nonumber\\
\br(\tau \to \mu \eta) < 2.3\times10^{-8}&\mbox{\cite{Hayasaka:2010et}}\nonumber\\
\br(\tau \to e \eta) < 4.4\times10^{-8}&\mbox{\cite{Hayasaka:2010et}}
\EEA
 At present, the most constraining bounds are from
$\br(\mu \to e \gamma)$, which has been just improved by the MEG
collaboration, and from $\CR(\mu-e, {\rm Au})$, both being at the 
\order{10^{-13}} level. Therefore, the 12 slepton mixings are by far the
most constrained ones. 
All these nine upper bounds above will be applied next to extract the
maximum allowed $|\delta^{AB}_{ij}|$ values.


\subsection{MSSM scenarios}


Regarding our choice of MSSM parameters for our forthcoming numerical
analysis of the LFV processes, we have proceeded within two frameworks,
both compatible with present data, that we describe in the following.  

\subsubsection{Framework 1}

In the first framework, we have selected six specific points in the MSSM
parameter space, $S1,...,S6$, as examples of points that are allowed by
present data, including recent LHC searches and the measurements of the
muon anomalous magnetic moment. In \refta{tab:spectra} the
values of the various MSSM parameters as well as the values of the
predicted MSSM mass spectra are summarized. They were evaluated with
the program \fh~\cite{feynhiggs}. For simplicity, and to reduce the number of
independent MSSM input parameters we have assumed equal soft masses for
the sleptons of the first and second generations (similarly for the
squarks),  equal soft masses for the left and right slepton sectors
(similarly for the squarks, where $\tilde Q$ denotes the the
``left-handed'' squark sector, whereas $\tilde U$ and $\tilde D$ denote
the up- and down-type parts of the ``right-handed'' squark sector)
and also equal trilinear couplings for
the stop, $A_t$,  and sbottom squarks, $A_b$. In the slepton sector we
just consider the stau trilinear coupling, $A_\tau$. The other trilinear
sfermion couplings are set to zero value. Regarding the soft
SUSY-breaking parameters for the gaugino
masses, $M_i$ ($i=1,2,3$),  we assume an approximate GUT relation. The
pseudoscalar Higgs mass $\MA$, and the $\mu$ parameter are also taken as
independent input parameters. In summary, the six points $S1,..,S6$ are
defined in terms of the following subset of ten input MSSM parameters: 

\BEA
m_{\tilde L_1} &=& m_{\tilde L_2} \; ; \; m_{\tilde L_3} \; 
(\mbox{with~} m_{\tilde L_{i}} = m_{\tilde E_{i}}\,\,,\,\,i=1,2,3) \non \\
m_{\tilde Q_1} &=& m_{\tilde Q_2} \; ; \; m_{\tilde Q_3} \; 
(\mbox{with~} m_{\tilde Q_i} = m_{\tilde U_i} = m_{\tilde D_i}\,\,,\,\,i=1,2,3) 
                                                               \non \\
A_t&=&A_b\,\,;\,\,A_\tau \nonumber \\
M_2&=&2 M_1\, =\,M_3/4 \,\,;\,\,\mu \nonumber \\
\MA&\,\,;\,\, &\tb
\EEA

\begin{table}[h!]
\begin{tabular}{|c|c|c|c|c|c|c|}
\hline
 & S1 & S2 & S3 & S4 & S5 & S6 \\\hline
$m_{\tilde L_{1,2}}$& 500 & 750 & 1000 & 800 & 500 &  1500 \\
$m_{\tilde L_{3}}$ & 500 & 750 & 1000 & 500 & 500 &  1500 \\
$M_2$ & 500 & 500 & 500 & 500 & 750 &  300 \\
$A_\tau$ & 500 & 750 & 1000 & 500 & 0 & 1500  \\
$\mu$ & 400 & 400 & 400 & 400 & 800 &  300 \\
$\tb$ & 20 & 30 & 50 & 40 & 10 & 40  \\
$\MA$ & 500 & 1000 & 1000 & 1000 & 1000 & 1500  \\
$m_{\tilde Q_{1,2}}$ & 2000 & 2000 & 2000 & 2000 & 2500 & 1500  \\
$m_{\tilde Q_{3}}$  & 2000 & 2000 & 2000 & 500 & 2500 & 1500  \\
$A_t$ & 2300 & 2300 & 2300 & 1000 & 2500 &  1500 \\\hline
$m_{\tilde l_{1}}-m_{\tilde l_{6}}$ & 489-515 & 738-765 & 984-1018 & 474-802  & 488-516 & 1494-1507  \\
$m_{\tilde \nu_{1}}-m_{\tilde \nu_{3}}$& 496 & 747 & 998 & 496-797 & 496 &  1499 \\
$m_{{\tilde \chi}_1^\pm}-m_{{\tilde \chi}_2^\pm}$  & 375-531 & 376-530 & 377-530 & 377-530  & 710-844 & 247-363  \\
$m_{{\tilde \chi}_1^0}-m_{{\tilde \chi}_4^0}$& 244-531 & 245-531 & 245-530 & 245-530  & 373-844 & 145-363  \\
$M_{h}$ & 126.6 & 127.0 & 127.3 & 123.1 & 123.8 & 125.1  \\
$M_{H}$  & 500 & 1000 & 999 & 1001 & 1000 & 1499  \\
$M_{A}$ & 500 & 1000 & 1000 & 1000 & 1000 & 1500  \\
$M_{H^\pm}$ & 507 & 1003 & 1003 & 1005 & 1003 & 1502  \\
 $m_{\tilde u_{1}}-m_{\tilde u_{6}}$& 1909-2100 & 1909-2100 & 1908-2100 & 336-2000 & 2423-2585 & 1423-1589  \\
$m_{\tilde d_{1}}-m_{\tilde d_{6}}$ & 1997-2004 & 1994-2007 & 1990-2011 & 474-2001 & 2498-2503 &  1492-1509 \\
$m_{\tilde g}$ &  2000 & 2000 & 2000 & 2000 & 3000 &  1200 \\
\hline
\end{tabular}
\caption{Selected points in the MSSM parameter space (upper part)
and their corresponding spectra (lower part). 
All mass parameters and trilinear couplings are given in GeV.} 
\label{tab:spectra}
\end{table}

The specific values of these ten MSSM parameters in \refta{tab:spectra},
to be used in the forthcoming analysis of LFV, are chosen to provide
different  
patterns in the various sparticle masses, but all leading to rather
heavy spectra, thus they are naturally in agreement with the
absence of SUSY signals at LHC. In particular  
all points lead to rather heavy squarks and gluinos above $1200\gev$ and
heavy sleptons above $500\gev$ (where the LHC limits would also
  permit substantially lighter scalar leptons). 
The values of $\MA$ within the interval
$(500,1500)\gev$, $\tb$ within the interval $(10,50)$ and a large
$A_t$ within $(1000,2500)\gev$ are fixed such that a light Higgs boson
$h^0$ within the LHC-favoured range $(123,127)\gev$ is obtained%
\footnote{
The uncertainty takes into account experimental uncertainties as well as
theoretical uncertainties, where the latter would permit an even larger
interval. However, restricting to the chosen 
$\pm 2 \gev$ gives a good impression of the allowed parameter space.
}%
. It
should also be noted that the large chosen values of $\MA \ge 500$
GeV place the Higgs sector of our scenarios in the so called decoupling
regime\cite{Haber:1989xc},  
where the couplings of $h^0$ to gauge bosons and fermions are close to
the SM Higgs couplings, and the heavy $H^0$ couples like the
pseudoscalar $A^0$, and all heavy Higgs bosons are close in mass.
Increasing $\MA$  the heavy
Higgs bosons tend to decouple from low energy physics and the light
$h^0$ behaves like $H_{\rm SM}$. This type of MSSM Higgs sector seems
to be in good agreement with recent LHC data\cite{LHCHiggs,CMSpashig12050}. 
We have checked with the code {\it HiggsBounds}~\cite{higgsbounds}
  that the Higgs sector is in agreement with the LHC searches (where S3
  is right ``at the border'').
Particularly, the so far absence of gluinos at
LHC, forbids too low $M_3$ and, therefore, given the  assumed GUT
relation, forbids also a too low $M_2$. Consequently, the
values of $M_2$ and $\mu$ are fixed as to get gaugino masses compatible
with present LHC bounds. 
Finally, we have also required that all our points lead to a prediction
of the anomalous magnetic moment of the muon in the MSSM that can fill
the present discrepancy between the Standard Model prediction and the
experimental value. Specifically, we use \citeres{Bennett:2006fi} and 
\cite{Davier:2010nc} to extract the size of this discrepancy,
see also \citere{gm2-Jegerlehner}:
\begin{equation}
(g-2)_\mu^{\rm exp}-(g-2)_\mu^{\rm SM}= (30.2 \pm 9.0) \times 10^{-10}.
\label{gminus2}  
\end{equation}
We then require that the SUSY contributions from charginos and neutralinos in the MSSM to one-loop level, 
$(g-2)_\mu^{\rm SUSY}$  be within the interval defined by $3 \sigma$ around the central value in \refeq{gminus2}, namely:    
\BE
 (g-2)_\mu^{\rm SUSY} \in  (3.2 \times 10^{-10},57.2 \times 10^{-10}) 
\label{gminus2interval}  
\EE 
Our estimate of $(g-2)_\mu^{\rm SUSY}$ for the six $S1,..S6$ points with
the code {\it SPHENO}~\cite{Porod:2003um} is
(where \fh\ gives similar results), respectively,   
\BE
 (15.5 \, (\mbox{S1}), \, 13.8 \, (\mbox{S2}), \, 15.1 \, (\mbox{S3}),
  16.7 \, (\mbox{S4}), \, 6.1 \, (\mbox{S5}), \, 7.9 \, (\mbox{S6}))\, 
\times 10^{-10}
\EE 
which are clearly within the previous allowed interval.
The relatively low values are due to the relatively heavy slepton
spectrum that was chosen. However, they are well within the preferred
interval.


\subsubsection{Framework 2}
\label{sec:f2}

In the second framework, several possibilities for the
MSSM parameters have been considered, leading to simple patterns
of SUSY masses with specific 
relations among them and where the number of input parameters is strongly
reduced.  As in framework 1 the scenarios selected in
  framework~2 lead to predictions of $(g-2)_\mu$ and
$\Mh$ that are compatible with present data over a large part of
  the parameter space. To simplify the analysis of
the upper bounds of the deltas, we will focus in scenarios where the
mass scales that are relevant for the LFV processes  are all set 
relative to one mass scale, generically called here $m_{\rm SUSY-EW}$. 
This implies setting the slepton soft masses, the gaugino soft
masses, $M_2$ and $M_1$  and the $\mu$ parameter in terms of  this 
$m_{\rm SUSY-EW}$. It should also be noted that these same mass
parameters are the relevant ones for $(g-2)_\mu$. The remaining
relevant parameter in both LFV and $(g-2)_\mu$ is $\tb$, and the
analysis below is performed in the ($m_{\rm SUSY-EW}$, $\tb$) plane.
Our selected LFV observables to be analized in framework~2
are the radiative $l_j \to l_i \gamma$ decays.
As discussed before, these are expected to be the most constraining ones.   
On the other
hand, since we are  interested in choices of the MSSM parameters that
lead to a prediction of $\Mh$ that is compatible with LHC data, we also
have to set the corresponding relevant mass parameters for this
observable. These are mainly the squark soft masses and trilinear soft
couplings, with particular relevance of those parameters of the third
generation squarks. All these squark mass scales will be set, in our
framework 2, relative to one single mass scale, $m_{\rm SUSY-QCD}$.  
Since we wish to explore a wide range in $\tb$,
from 5 to 60, $\MA$ is fixed to $1000 \gev$ to ensure the agreement
with the present bounds in the $(\tb, \MA)$ plane from LHC
searches~\cite{CMSpashig12050}. 
 Finally, to reduce even further the number of
input parameters we will assume again an approximate GUT relation among
the gaugino soft masses, $M_2=2 M_1\, =\,M_3/4$ and the $\mu$ parameter
will be set equal to $M_2$. Regarding the trilinear couplings, they will
all be set to zero except those of the stop and sbottom sectors, being
relevant for $\Mh$, and that will be simplified to $A_t=A_b$. In
summary, our scenarios in framework 2 are set in terms of four input
parameters:  $m_{\rm SUSY-EW}$, $m_{\rm SUSY-QCD}$, $M_2$ and
$\tb$.
Generic scenarios in which the relevant parameters are fixed
independently are called ``phenomenological MSSM scenarios (pMSSM)'' 
in the literature (see, for instance,
\cite{Arbey:2012dq,Bechtle:2012jw}). We refer to our scenarios
here as ``pMSSM-4'', 
indicating the number of free parameters. These kind of scenarios have
the advantage of reducing considerably the number of input parameters
respect to the MSSM and, consequently, making easier the analysis of
their phenomenological implications.  
 
For the forthcoming numerical analysis we consider the following
specific pMSSM-4 mass patterns:   
\begin{itemize}
\item[{\bf (a)}]
\BEA
 m_{\tilde L}&=&m_{\tilde E}=m_{\rm SUSY-EW}\nonumber \\ 
 M_2&=& m_{\rm SUSY-EW} \nonumber \\ 
 m_{\tilde Q}&=&m_{\tilde U}=m_{\tilde D}=m_{\rm SUSY-QCD}\nonumber \\ 
 A_t&=&1.3 \, m_{\rm SUSY-QCD} \nonumber \\
 m_{\rm SUSY-QCD}&=& 2 \, m_{\rm SUSY-EW}
\label{Sa}
\EEA
\item[{\bf (b)}] 
\BEA
 m_{\tilde L}&=&m_{\tilde E}=m_{\rm SUSY-EW}\nonumber \\ 
 M_2&=& m_{\rm SUSY-EW}/5 \nonumber \\ 
 m_{\tilde Q}&=&m_{\tilde U}=m_{\tilde D}=m_{\rm SUSY-QCD}\nonumber \\ 
 A_t&=&m_{\rm SUSY-QCD} \nonumber \\
 m_{\rm SUSY-QCD}&=& 2 \, m_{\rm SUSY-EW}
\label{Sb}
\EEA
\item[{\bf (c)}] 
\BEA
 m_{\tilde L}&=&m_{\tilde E}=m_{\rm SUSY-EW}\nonumber \\ 
 M_2&=& 300 \,\,{\rm GeV} \nonumber \\ 
 m_{\tilde Q}&=&m_{\tilde U}=m_{\tilde D}=m_{\rm SUSY-QCD}\nonumber \\ 
 A_t&=&m_{\rm SUSY-QCD} \nonumber \\
 m_{\rm SUSY-QCD}&= & m_{\rm SUSY-EW}
\label{Sc}
\EEA 
\item[{\bf (d)}] 
\BEA
  m_{\tilde L}&=&m_{\tilde E}=m_{\rm SUSY-EW}\nonumber \\ 
 M_2&=& m_{\rm SUSY-EW}/3 \nonumber \\ 
 m_{\tilde Q}&=&m_{\tilde U}=m_{\tilde D}=m_{\rm SUSY-QCD}\nonumber \\ 
 A_t&=&m_{\rm SUSY-QCD} \nonumber \\
 m_{\rm SUSY-QCD}&=& m_{\rm SUSY-EW}
\label{Sd}
\EEA 
\end{itemize}
Where we have simplified the notation for the soft sfermion masses, by
using $m_{\tilde L}$ for 
$m_{\tilde L}=m_{\tilde L_{1}}=m_{\tilde L_{2}}=m_{\tilde L_{3}}$, etc.
In the forthcoming numerical analysis of the maximum allowed values of
the deltas within these scenarios, the most relevant parameters $m_{\rm
  SUSY-EW} \equiv m_{\rm SUSY}$ and $\tb$ will be varied within the
intervals: 
\BEA
500 \,\,{\rm GeV} \leq &m_{\rm SUSY}&\leq 2500 \,\,{\rm GeV} \nonumber \\
5 \leq &\tb& \leq 60
\EEA 
Due to the particular mass patterns chosen above, scenario (a) will deal
with approximately equally heavy sleptons and charginos/neutralinos and
with doubly heavy squarks; same for scenario (b) but with 1/5 lighter
charginos/neutralinos; scenario (c) with equally heavy sleptons and
squarks and charginos/neutralinos close to 300 GeV and scenario (d) with  
 1/3 lighter charginos/neutralinos.
The values of $\At$ have been selected to ensure that 
$\Mh \sim 125 - 126 \gev$ over large parts of the ($\msusy$, $\tb$) plane.


\subsection{Selected \boldmath{$\delta^{AB}_{ij}$} mixings}

Finally, for our purpose in this paper, we need to select the slepton
mixings and to set the range of values for the explored
$\delta^{AB}_{ij}$'s. First, we work in a complete basis, that is we
take into account the full set of twelve $\delta^{AB}_{ij}$'s.  
For simplicity, we will assume real values for these flavor slepton
mixing parameters, therefore we will not have to be concerned with the
Lepton Electric Dipole Moments (EDM). Concretely, the scanned interval
in our estimates of LFV rates will be:  
\BE
-1 \le \delta^{AB}_{ij} \le +1 
\EE
For each explored non-vanishing single delta, $\delta^{AB}_{ij}$, or
pair of deltas, $(\delta^{AB}_{ij}, \delta^{CD}_{kl})$, the
corresponding slepton and sneutrino physical masses,  the slepton and
sneutrino rotation matrices, as well as the LFV rates will be numerically
computed with our private Fortran code.


\section{Results and discussion} 
\label{sec:results}

\subsection{Results in framework 1}

\begin{figure}[ht!]
\begin{center}
\psfig{file=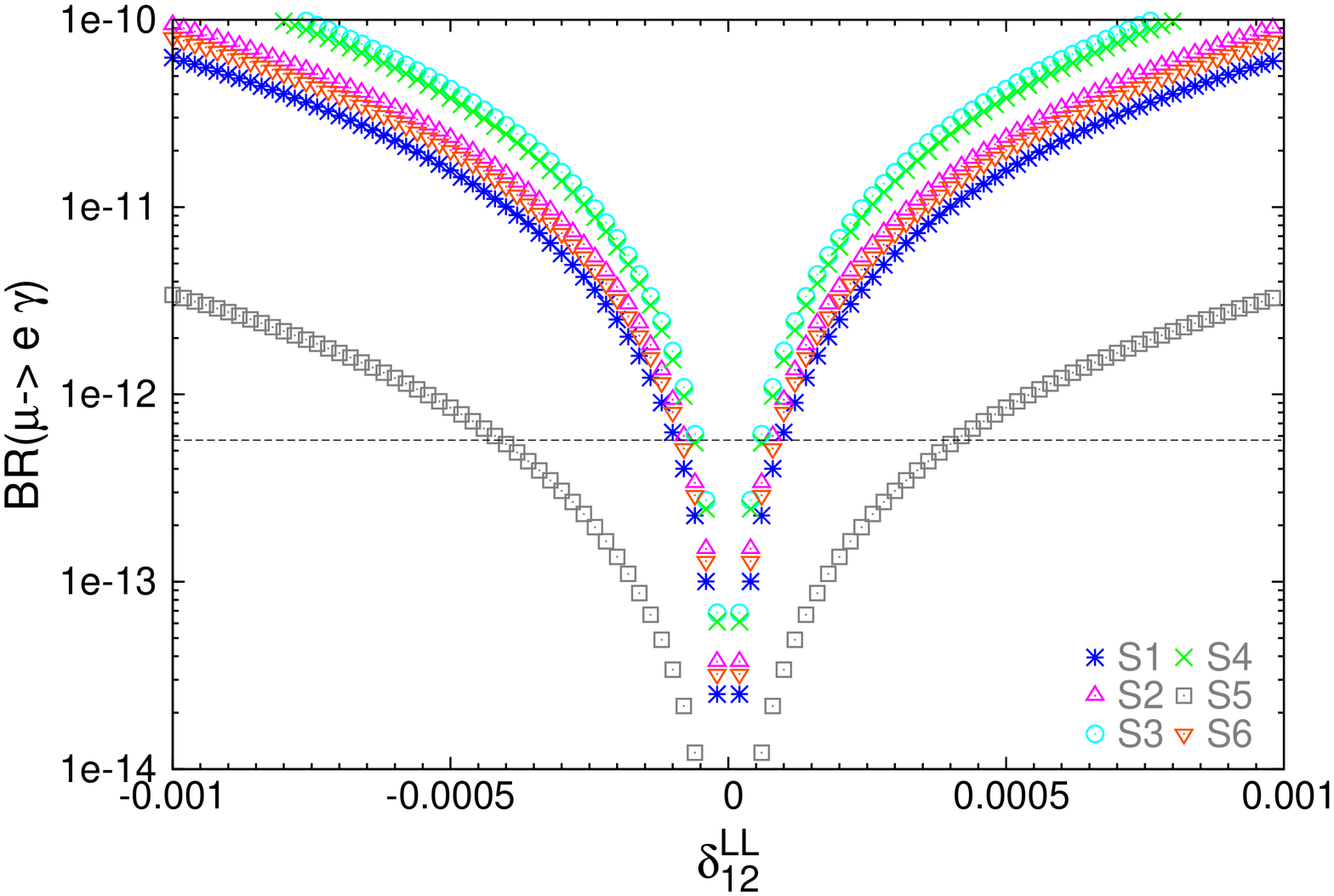   ,scale=0.45,angle=270,clip=}\\
\psfig{file=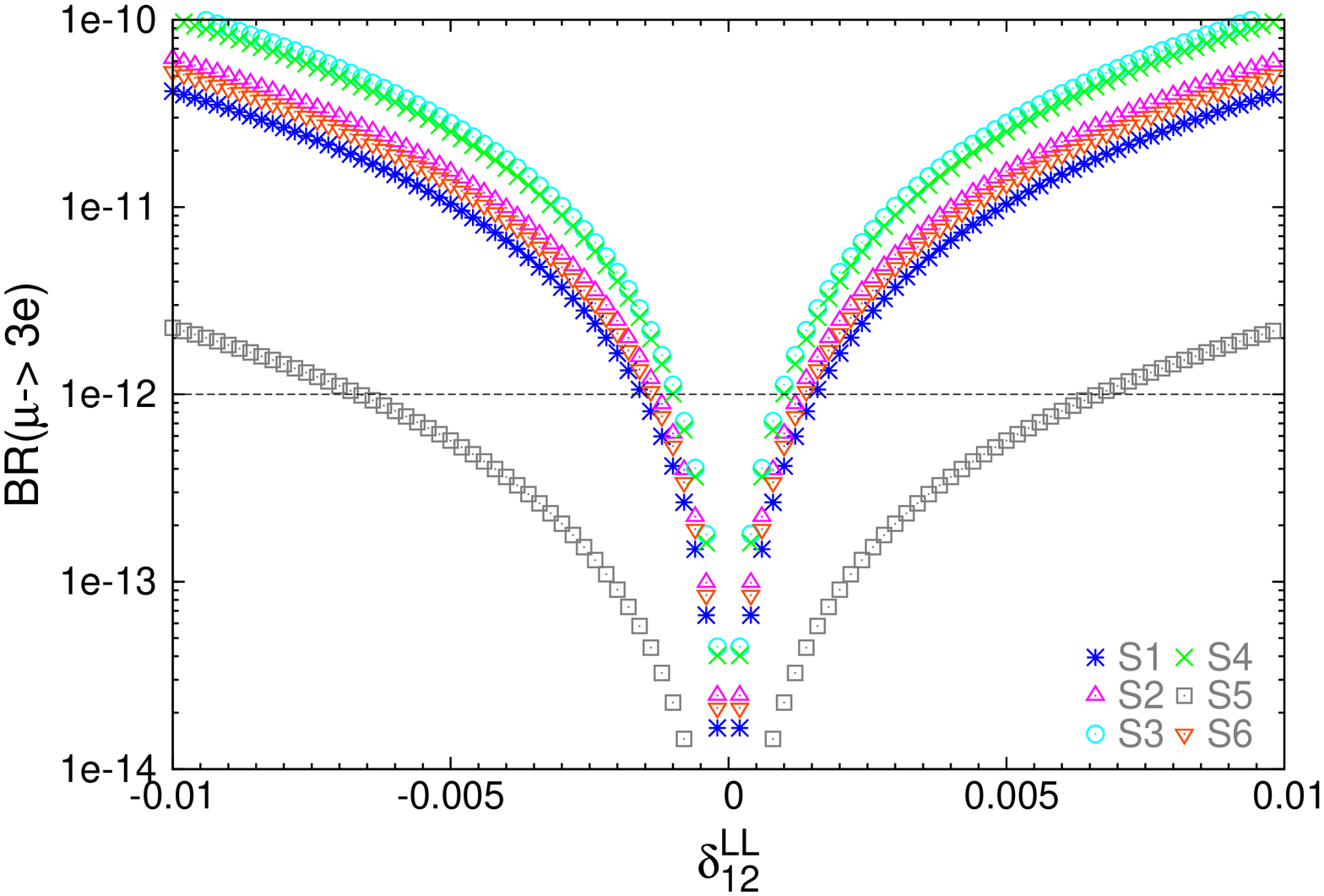     ,scale=0.30,angle=270,clip=}
\psfig{file=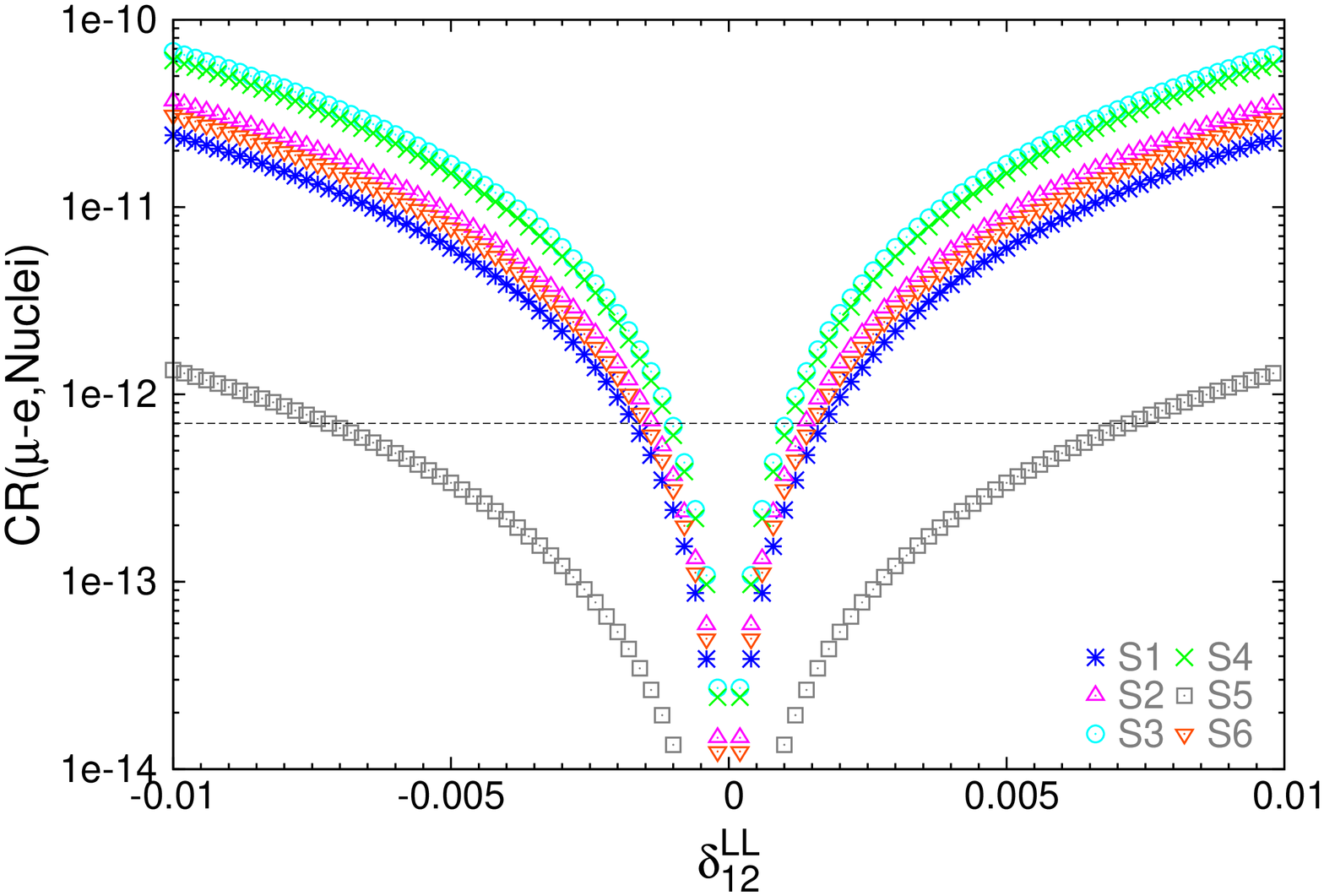    ,scale=0.30,angle=270,clip=}
\end{center}
\caption{LFV rates for $\mu-e$ transitions as a function of slepton
  mixing $\delta_{12}^{LL}$.}  
\label{mixing12LL}
\end{figure}

\begin{figure}[ht!]
\begin{center}
\psfig{file=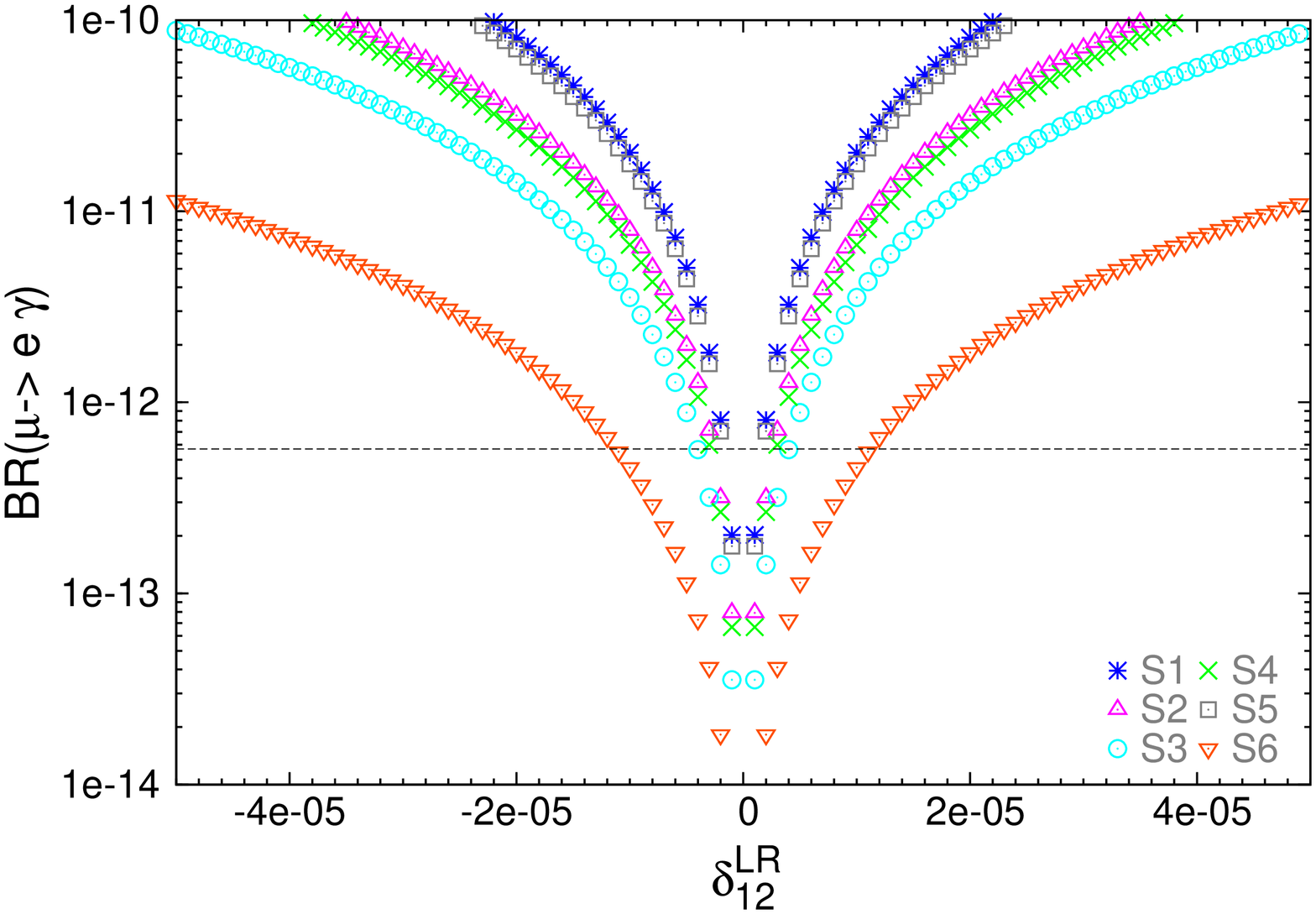   ,scale=0.45,angle=270,clip=}\\
\psfig{file=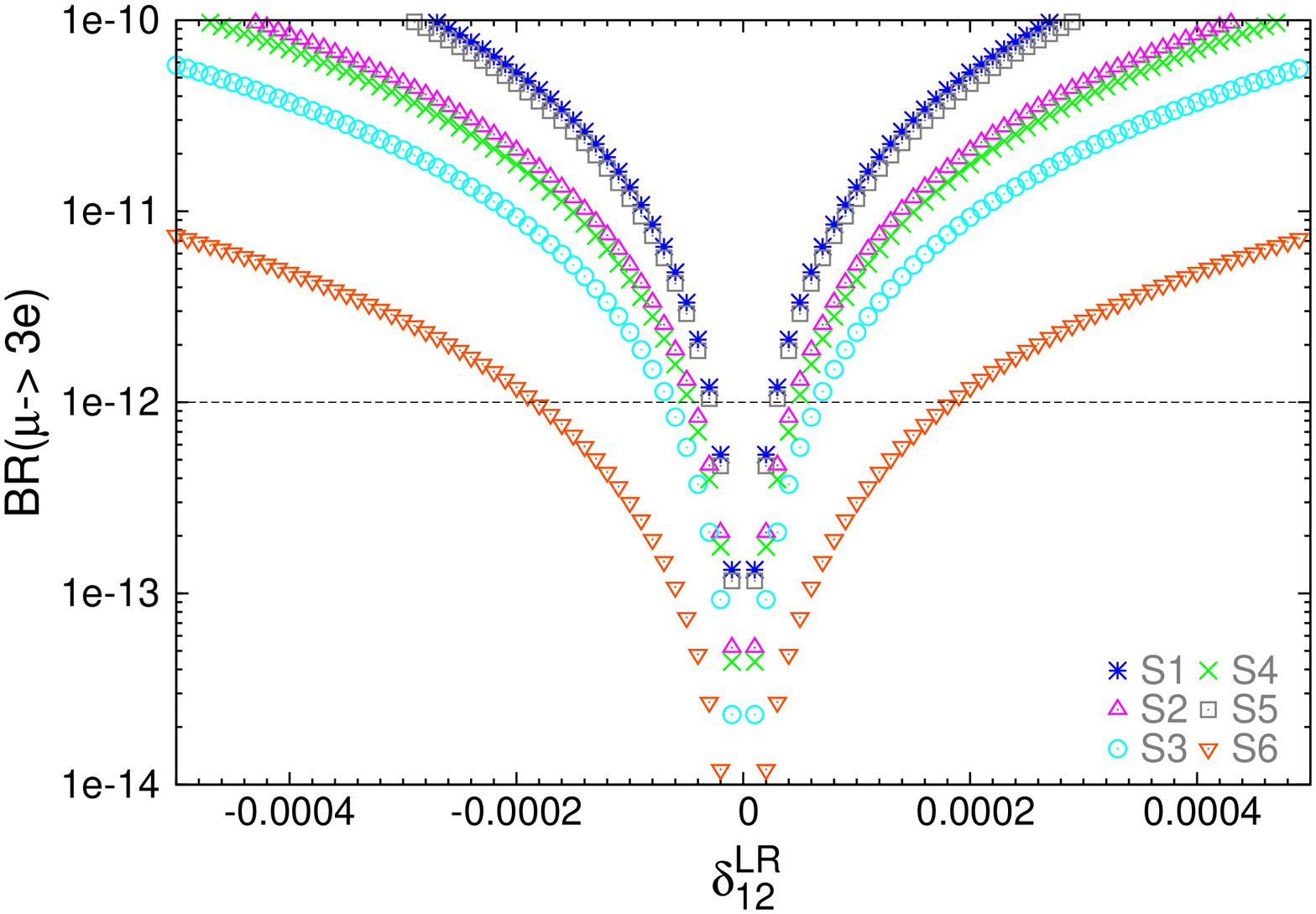     ,scale=0.30,angle=270,clip=}
\psfig{file=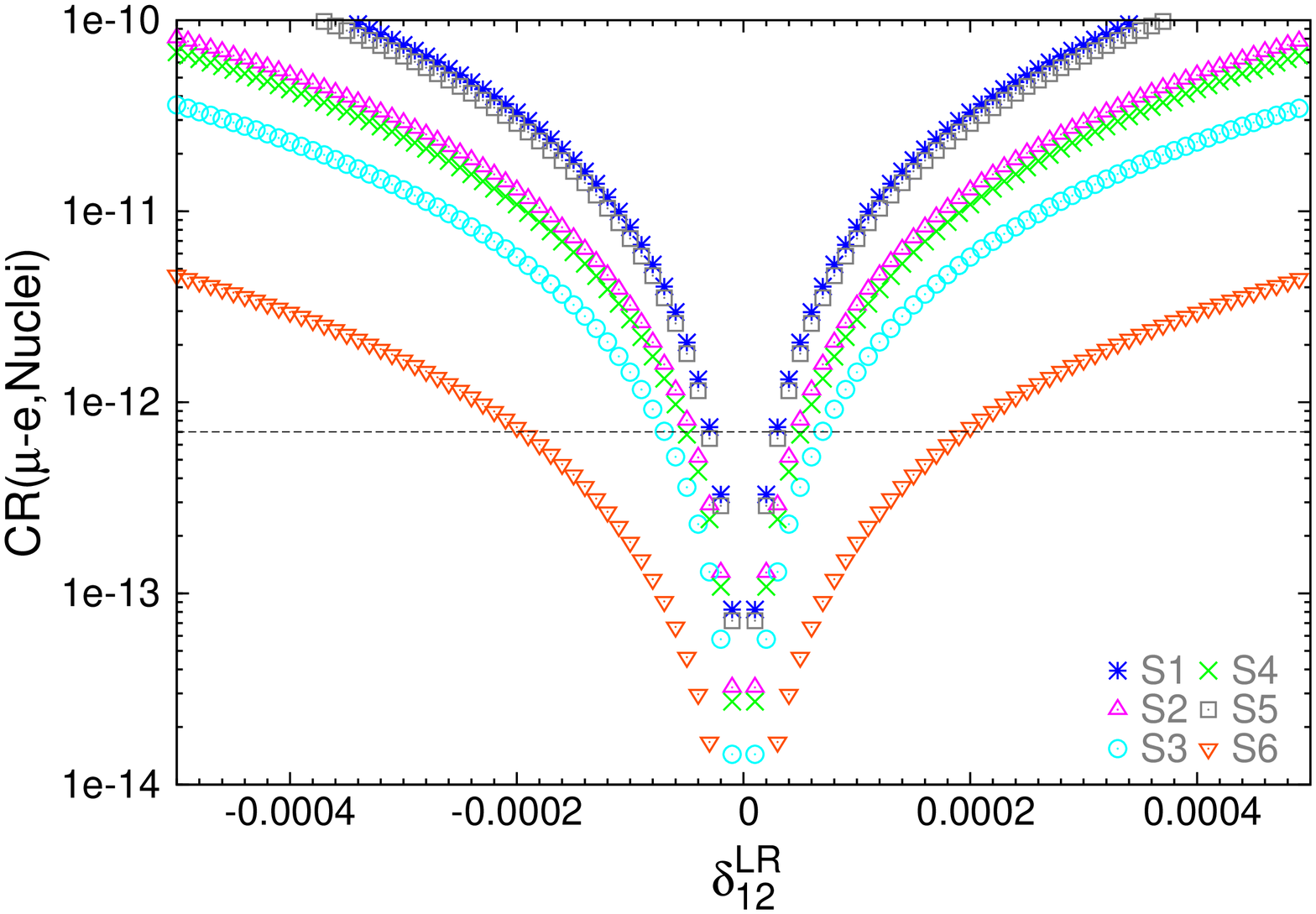    ,scale=0.30,angle=270,clip=}
\end{center}
\caption{LFV rates for $\mu-e$ transitions as a function of slepton
  mixing $\delta_{12}^{LR}$. The corresponding plots for
  $\delta_{12}^{RL}$, not shown here, are indistinguishable from these.}  
\label{mixing12LR}
\end{figure}

\begin{figure}[ht!]
\begin{center}
\psfig{file=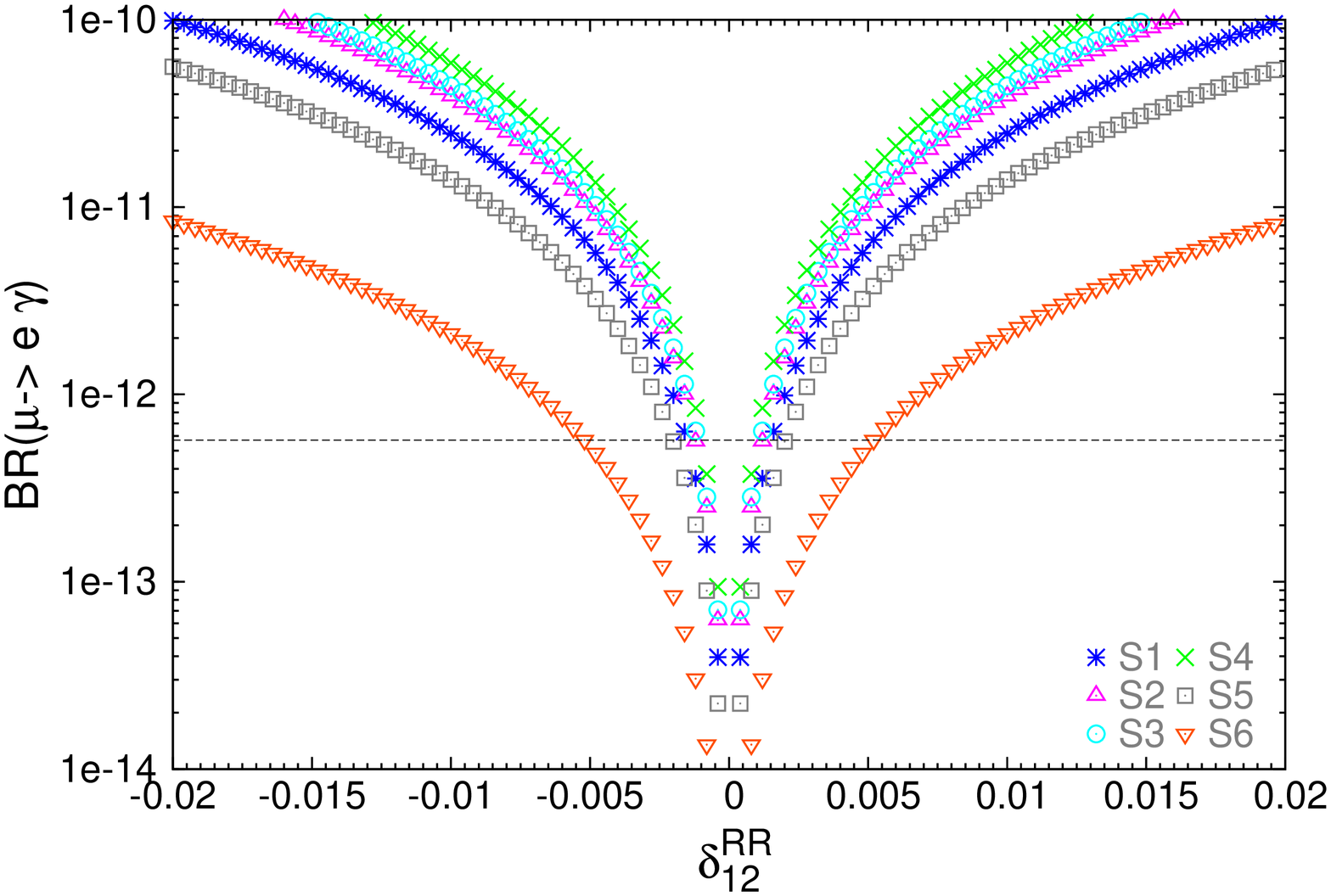   ,scale=0.45,angle=270,clip=}\\
\psfig{file=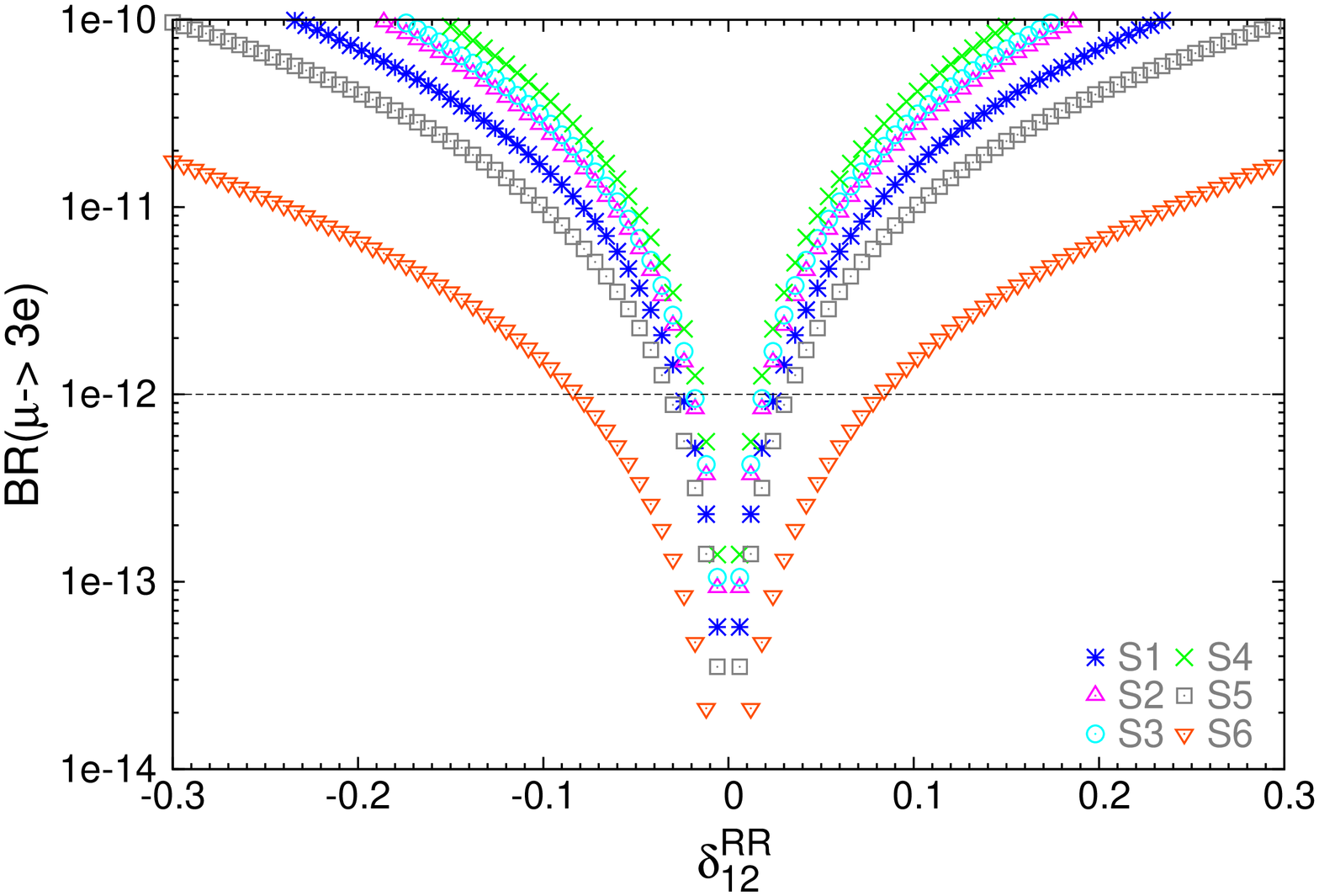     ,scale=0.30,angle=270,clip=}
\psfig{file=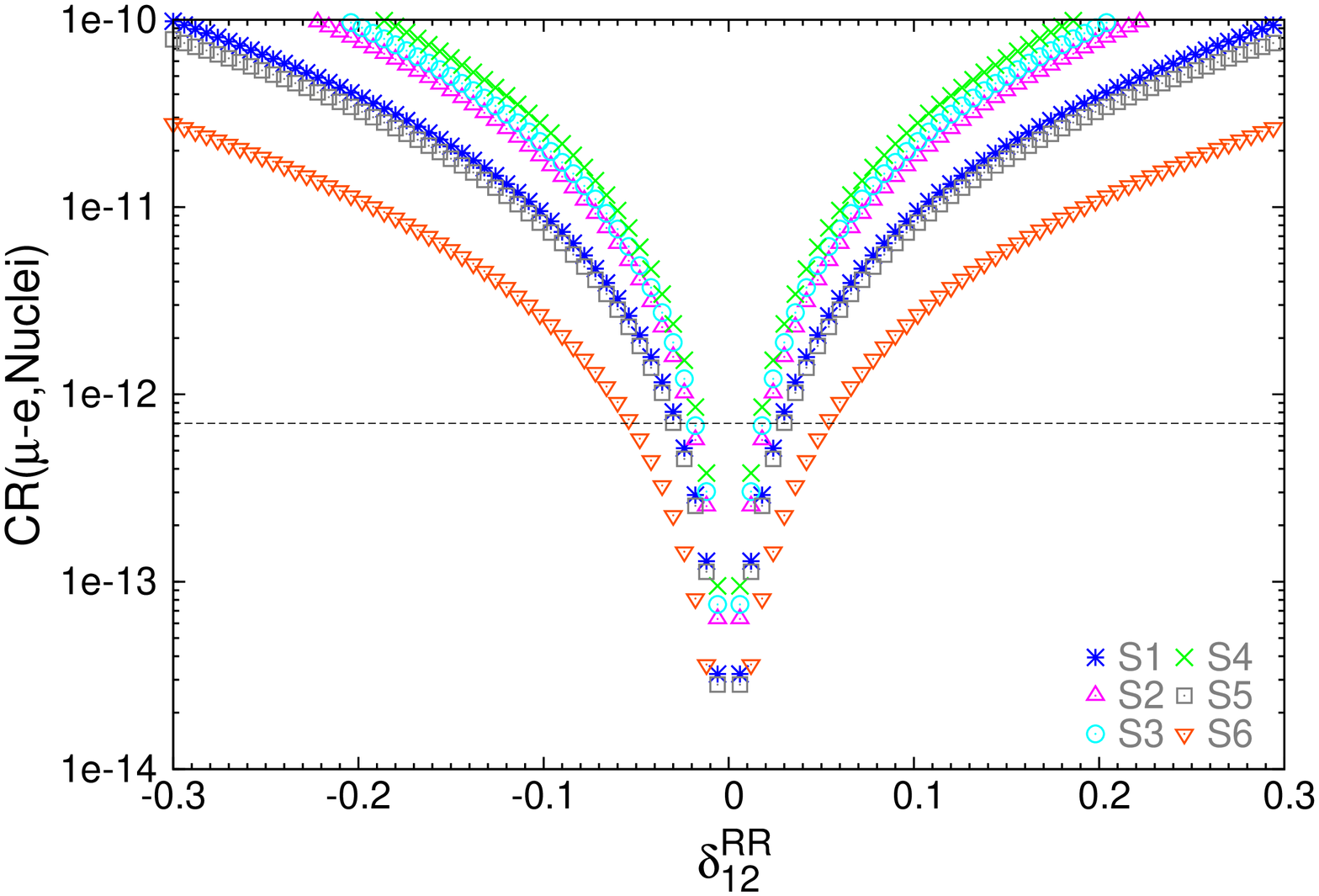    ,scale=0.30,angle=270,clip=}
\end{center}
\caption{LFV rates for $\mu-e$ transitions as a function of slepton
  mixing $\delta_{12}^{RR}$.}  
\label{mixing12RR}
\end{figure}

\begin{figure}[ht!]
\begin{center}
\psfig{file=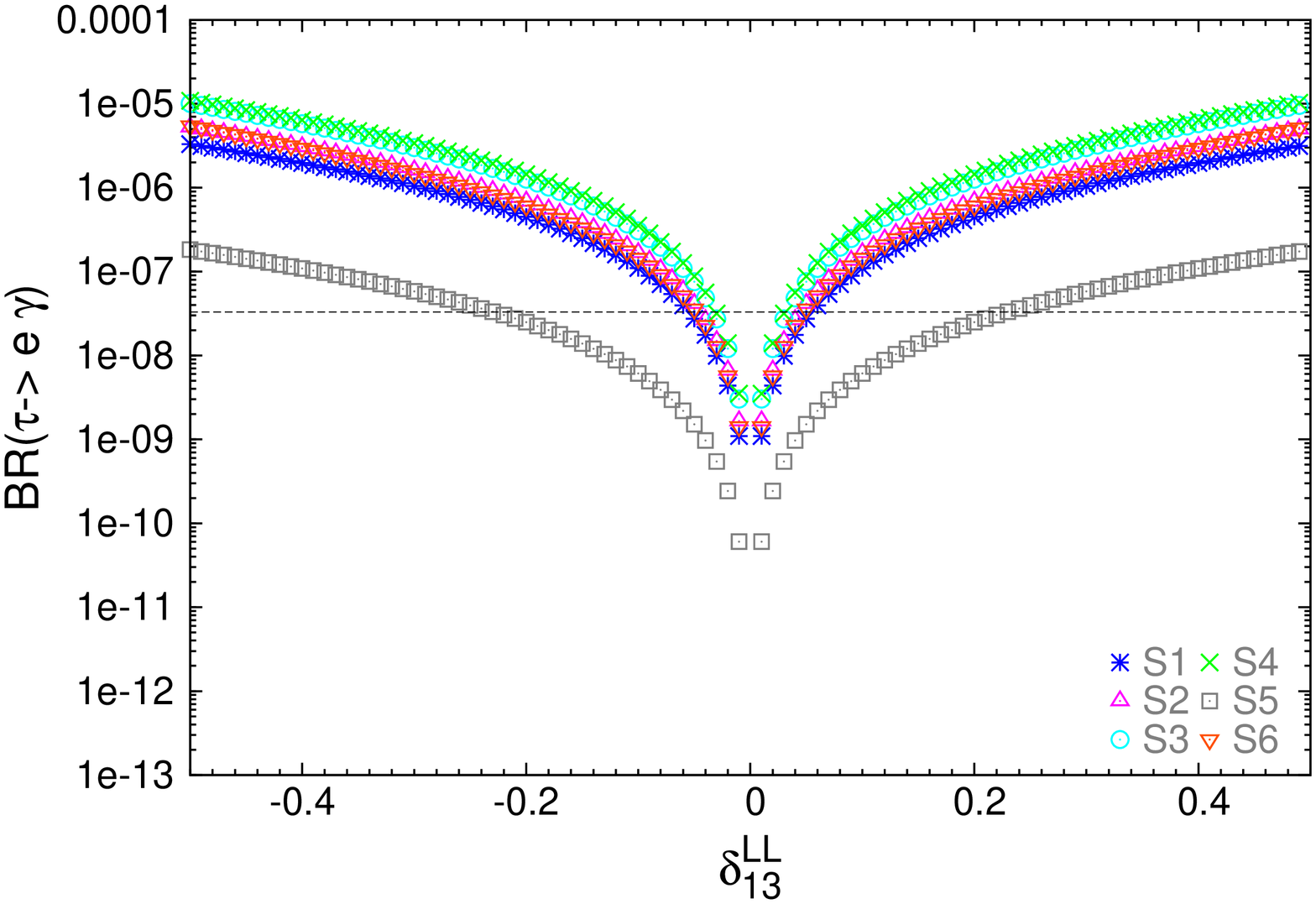  ,scale=0.45,angle=270,clip=}\\
\psfig{file=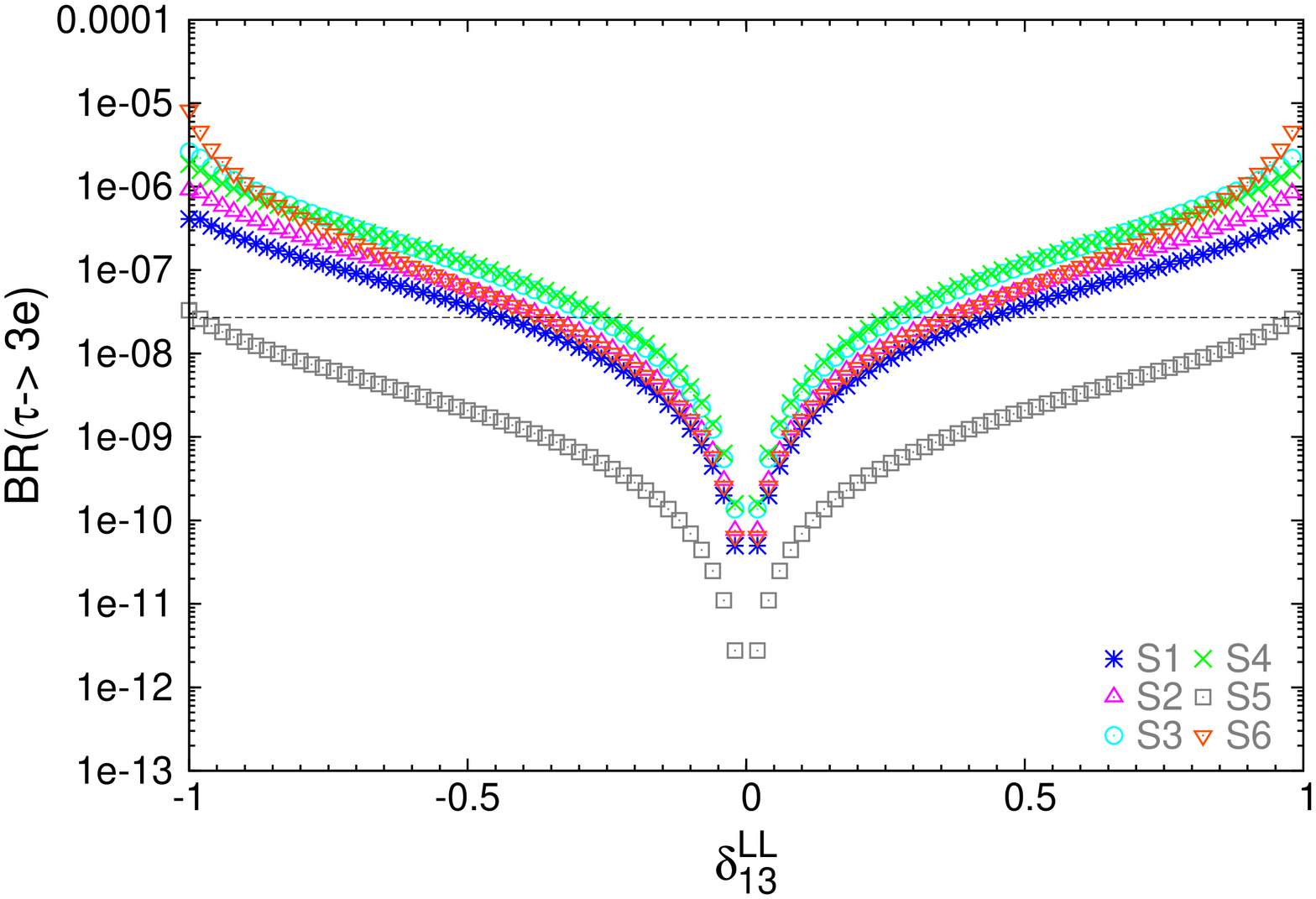    ,scale=0.30,angle=270,clip=}
\psfig{file=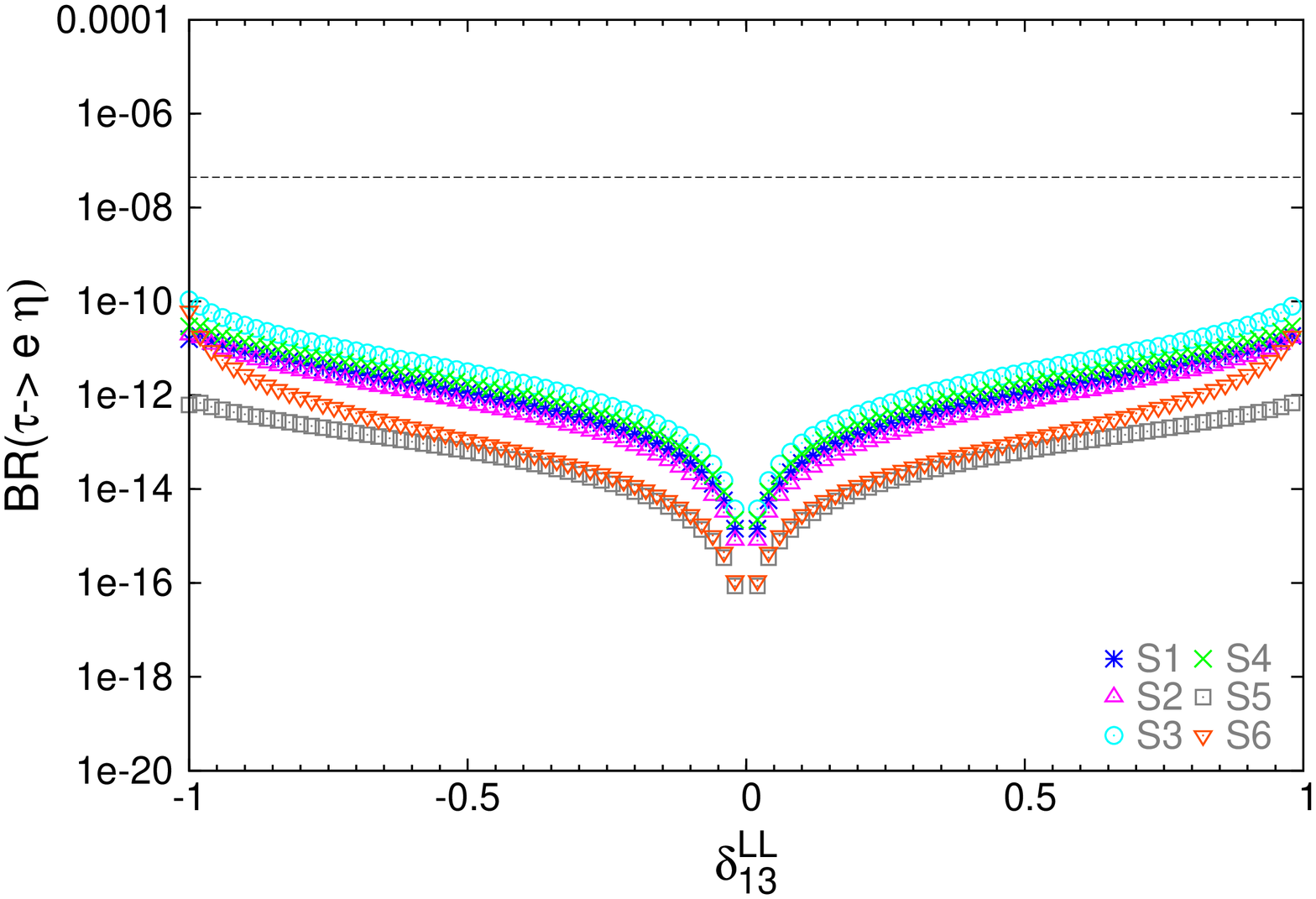   ,scale=0.30,angle=270,clip=}
\end{center}
\caption{LFV rates for $\tau-e$ transitions as a function of slepton 
mixing $\delta_{13}^{LL}$} 
\label{mixing13LL}
\end{figure}

\begin{figure}[ht!]
\begin{center}
\psfig{file=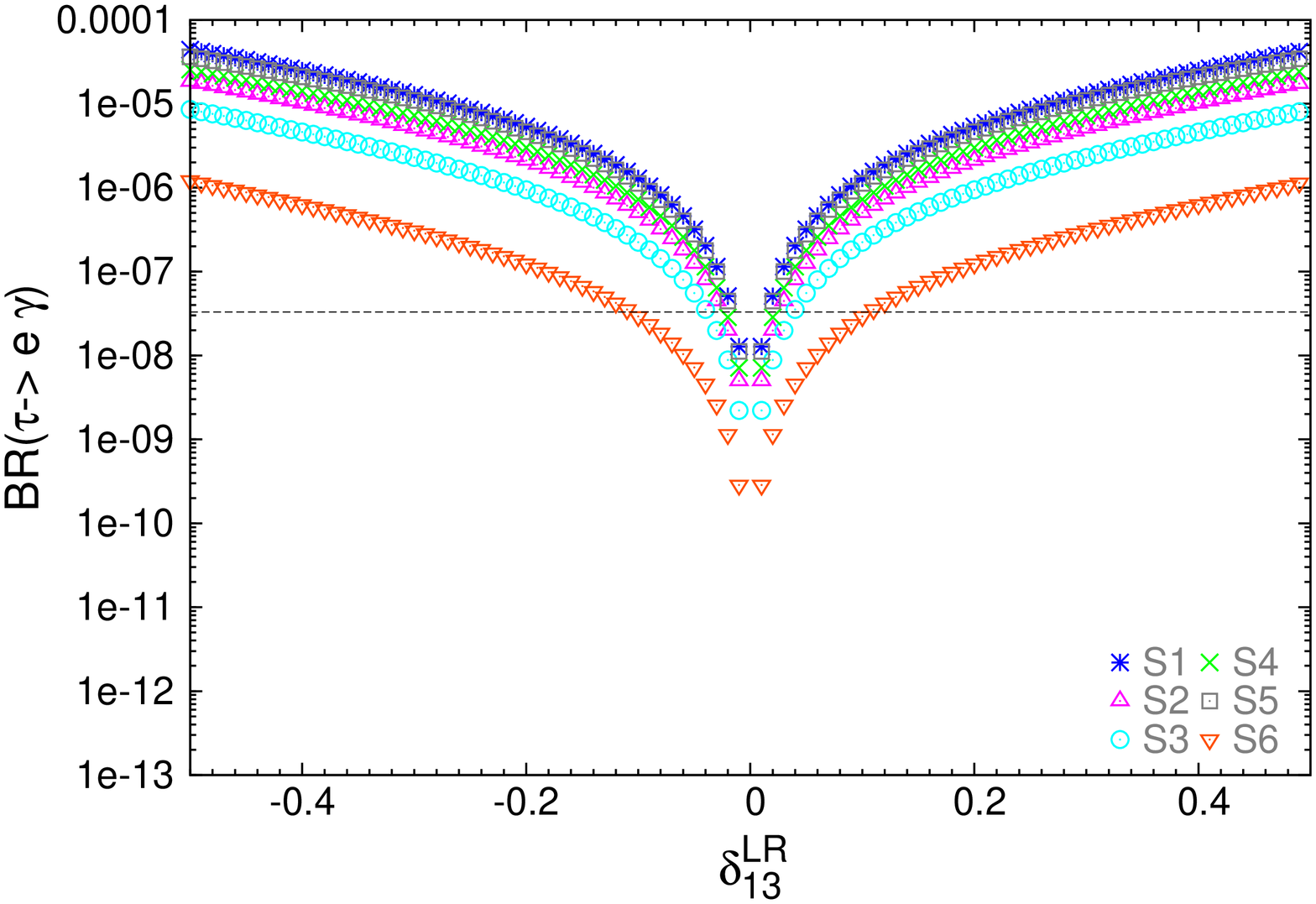  ,scale=0.45,angle=270,clip=}\\
\psfig{file=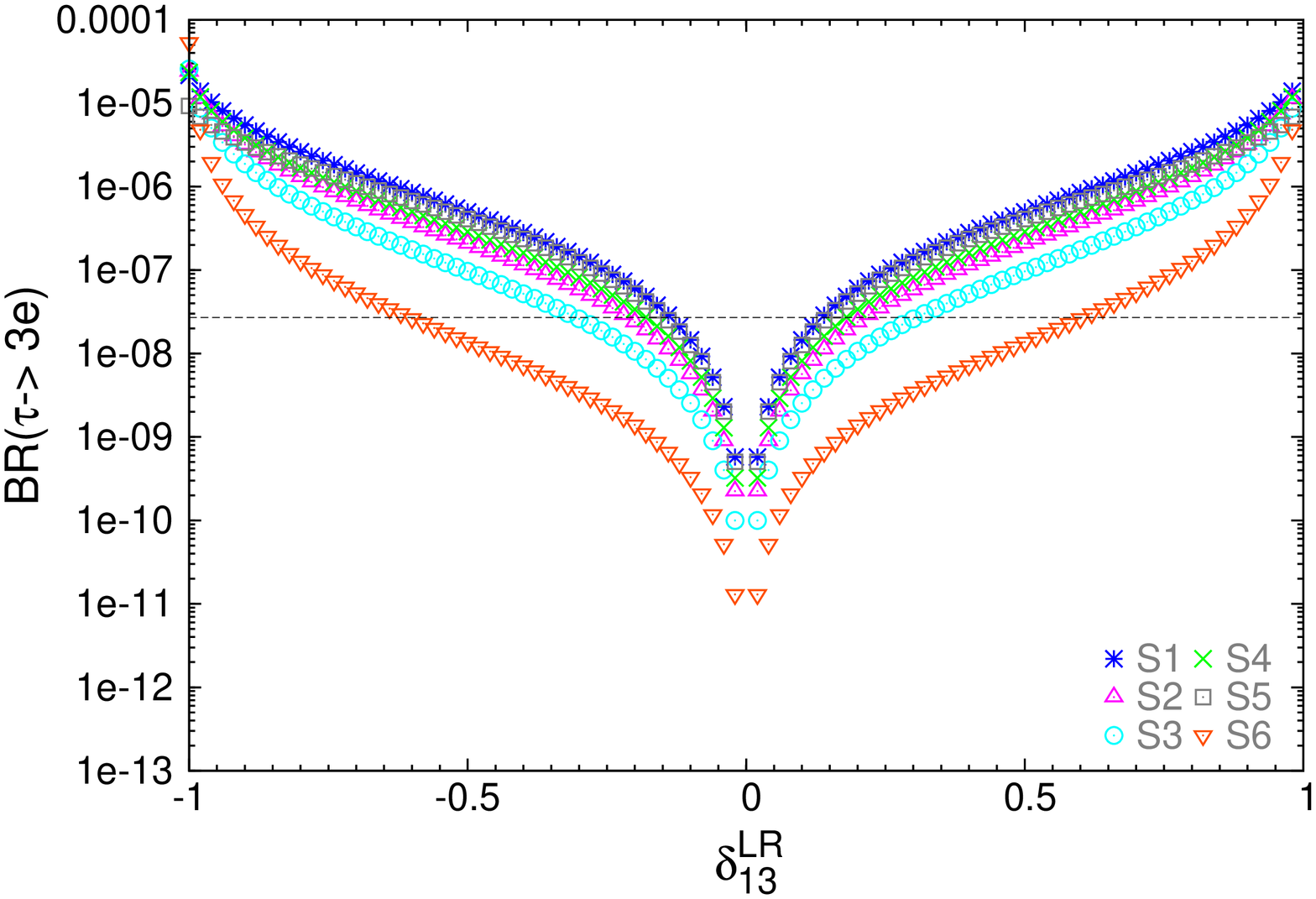    ,scale=0.30,angle=270,clip=}
\psfig{file=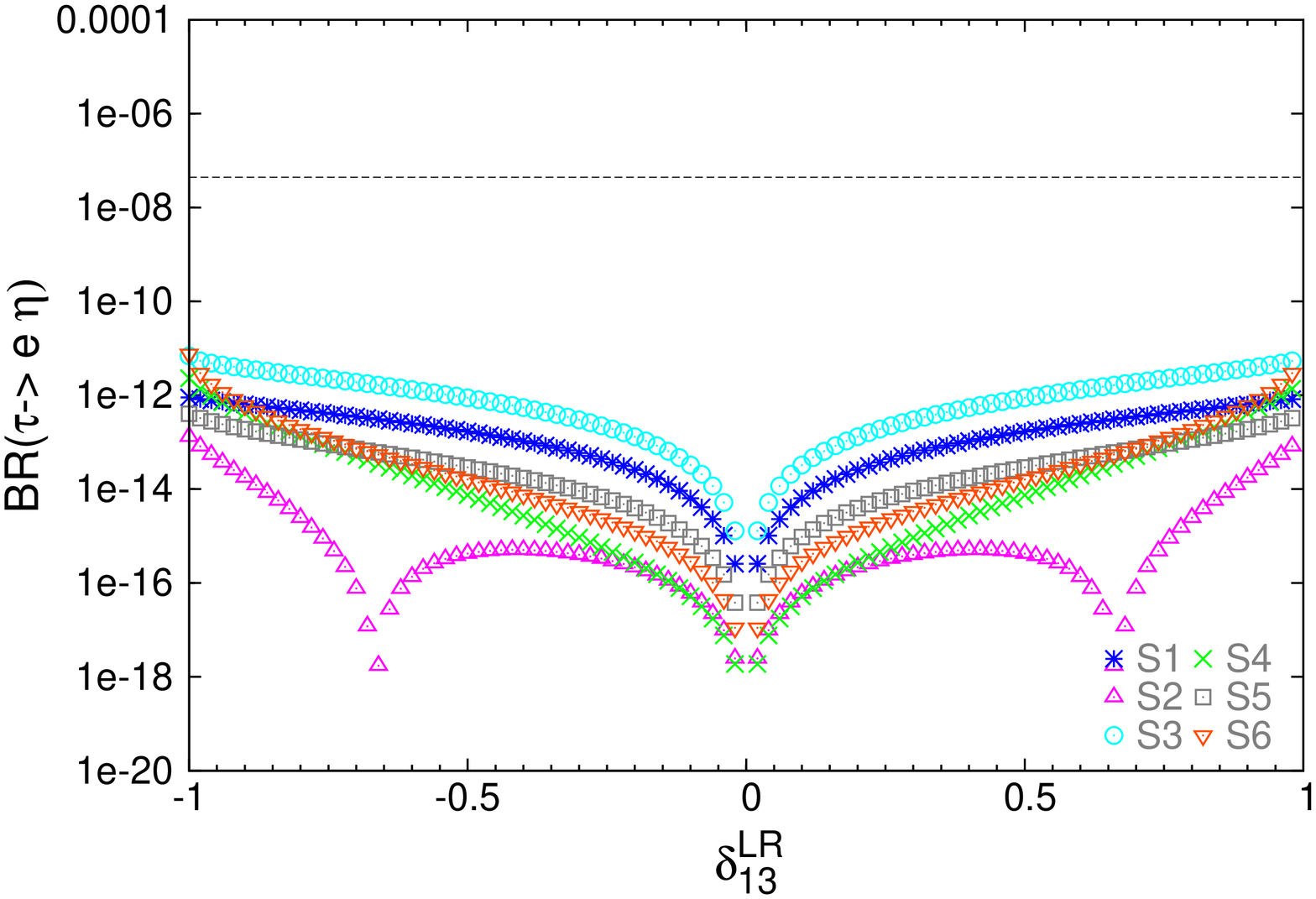   ,scale=0.30,angle=270,clip=}
\end{center}
\caption{LFV rates for $\tau-e$ transitions as a function of slepton
  mixing $\delta_{13}^{LR}$. The corresponding plots for
  $\delta_{13}^{RL}$, not shown here, are indistinguishable from these.}  
\label{mixing13LR}
\end{figure}

\begin{figure}[ht!]
\begin{center}
\psfig{file=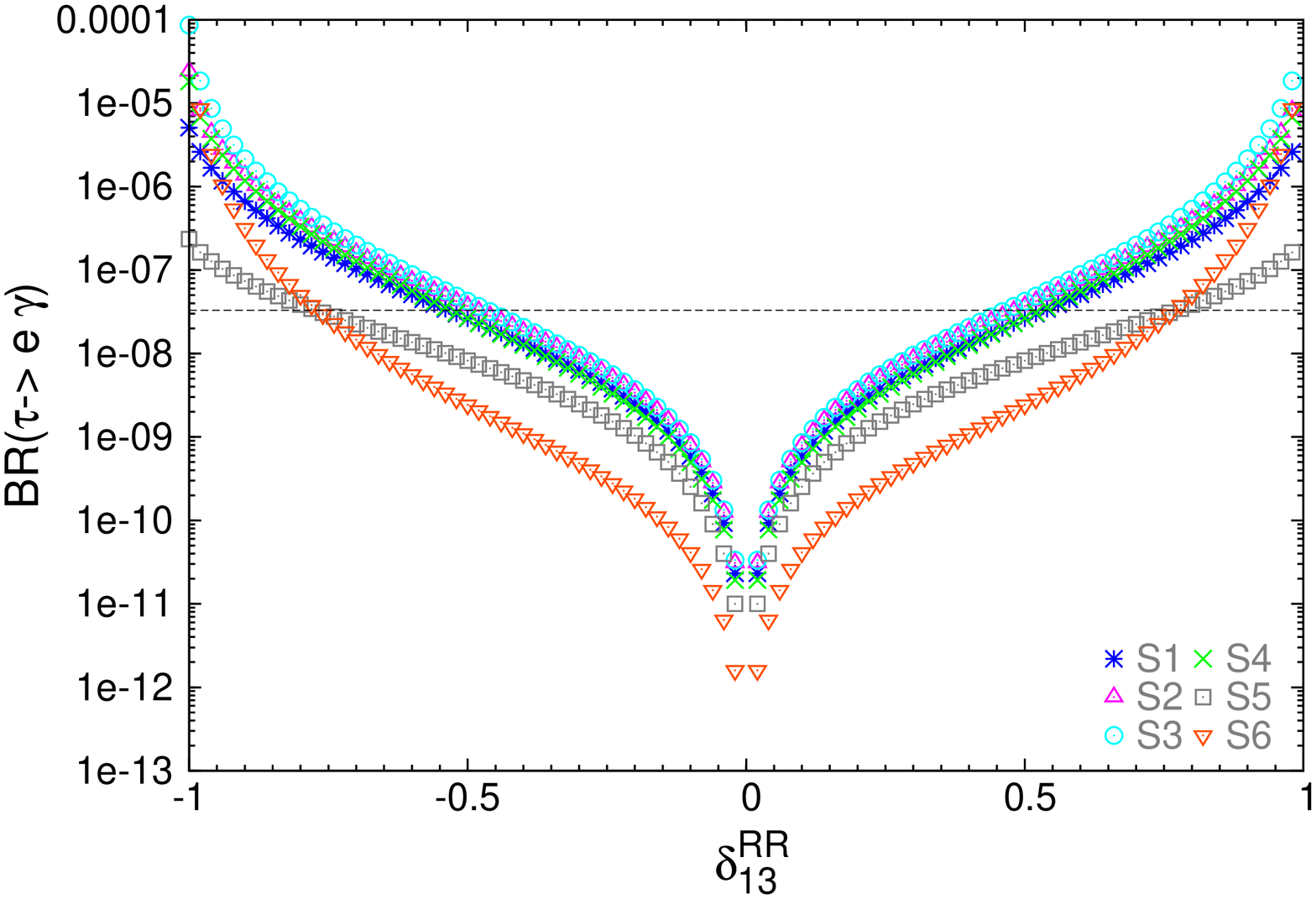  ,scale=0.45,angle=270,clip=}\\
\psfig{file=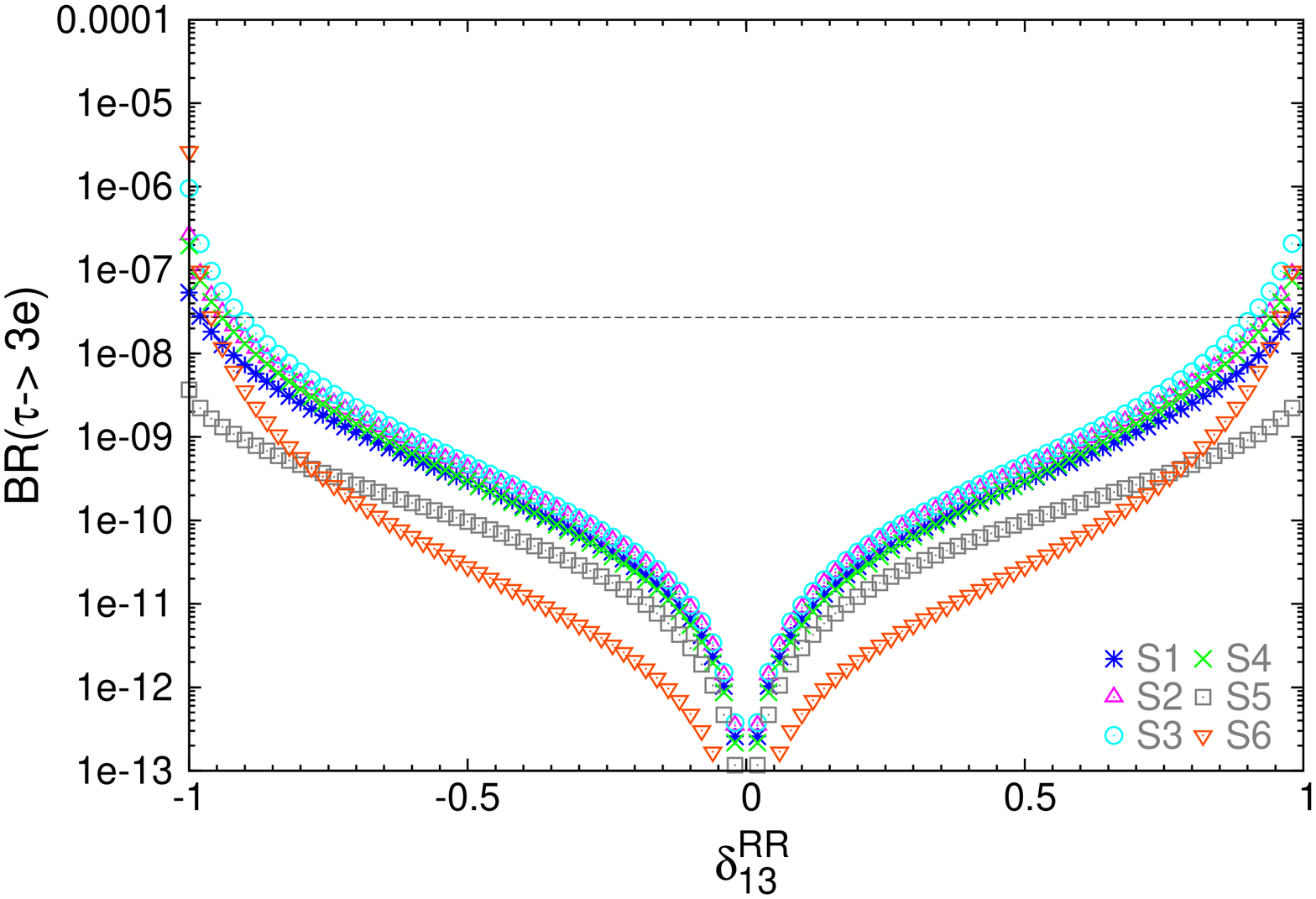    ,scale=0.30,angle=270,clip=}
\psfig{file=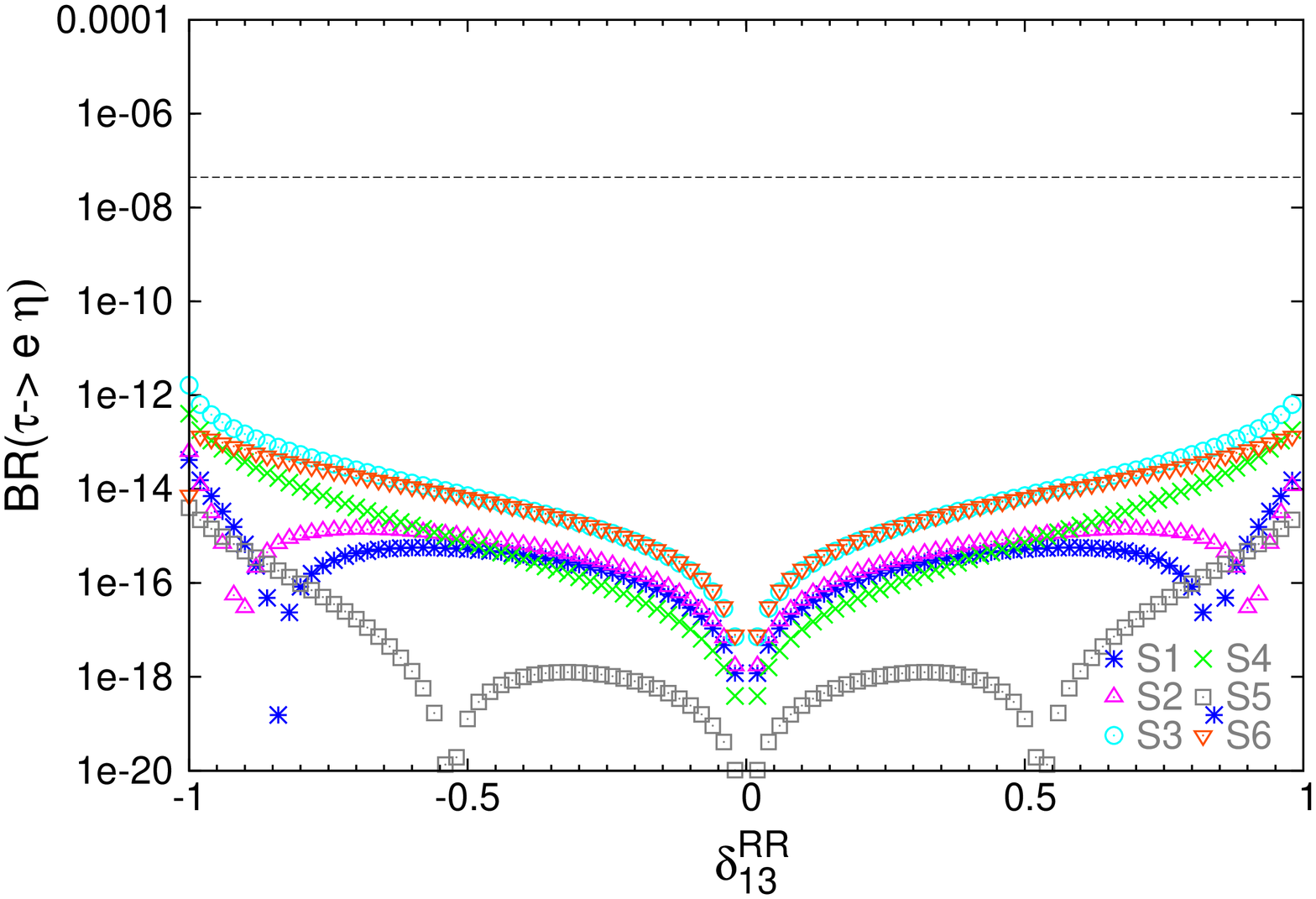   ,scale=0.30,angle=270,clip=}
\end{center}
\caption{LFV rates for $\tau-e$ transitions as a function of slepton
  mixing $\delta_{13}^{RR}$.}
\label{mixing13RR}
\end{figure}

\begin{figure}[ht!]
\begin{center}
\psfig{file=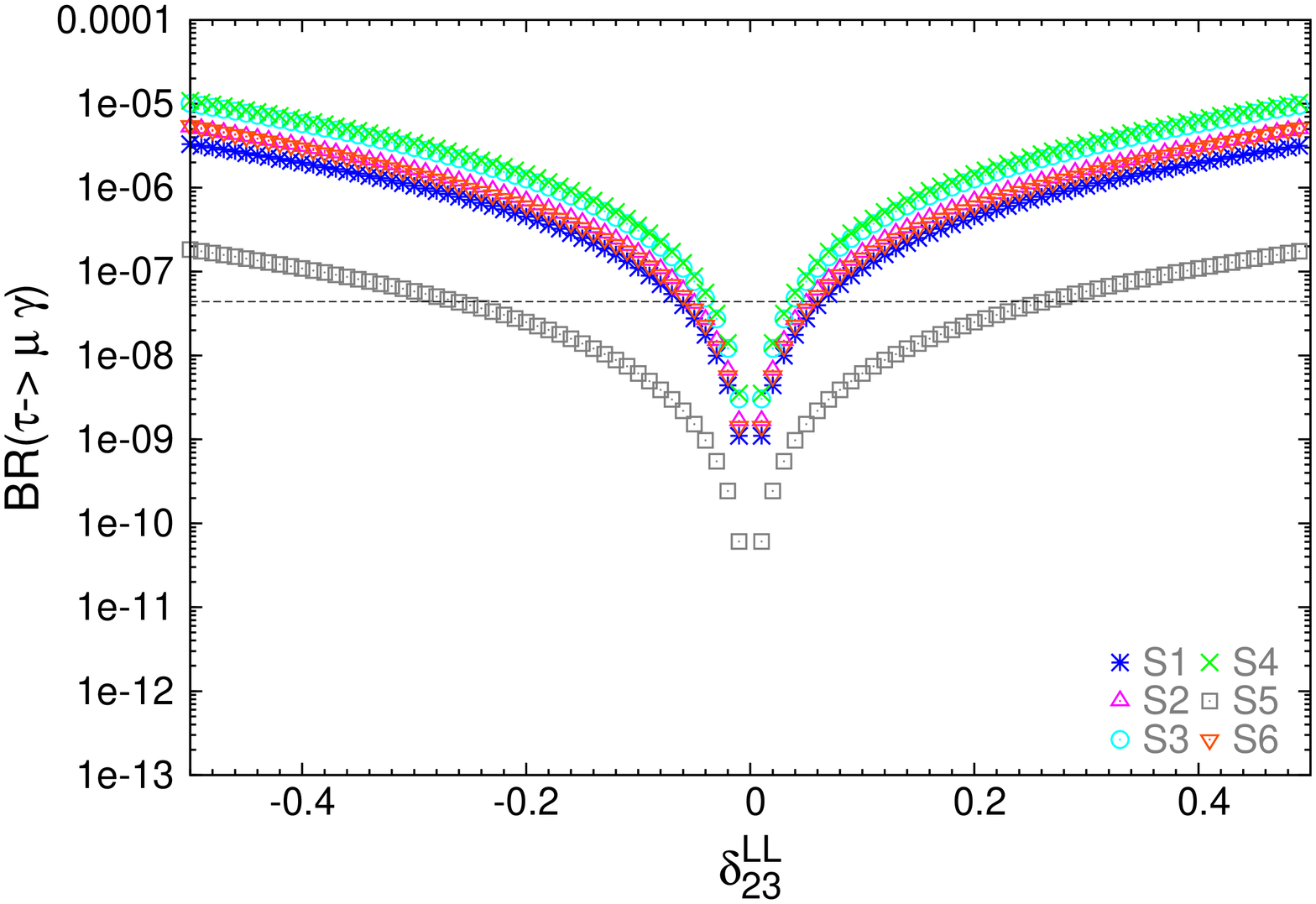  ,scale=0.45,angle=270,clip=}\\
\psfig{file=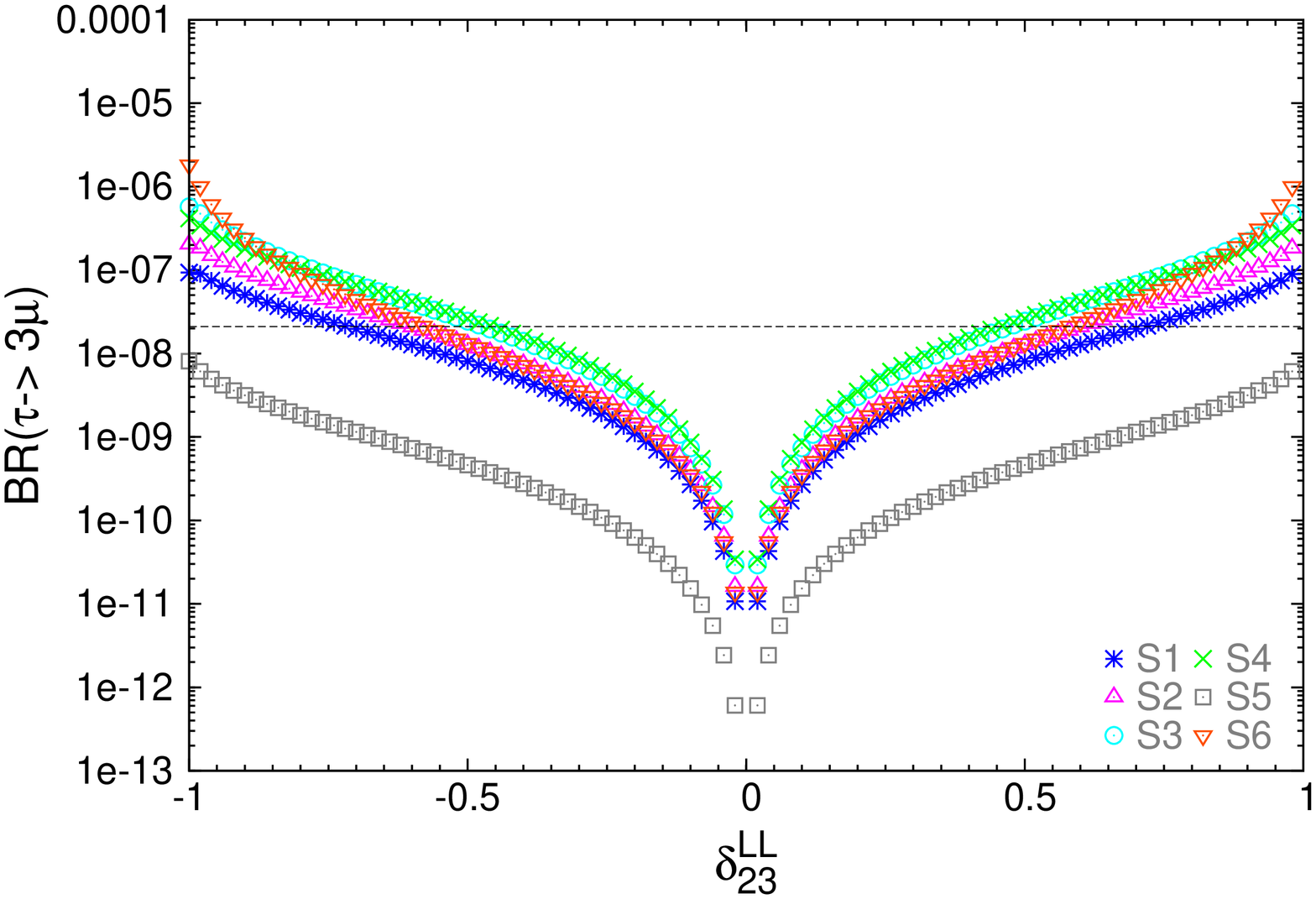    ,scale=0.30,angle=270,clip=}
\psfig{file=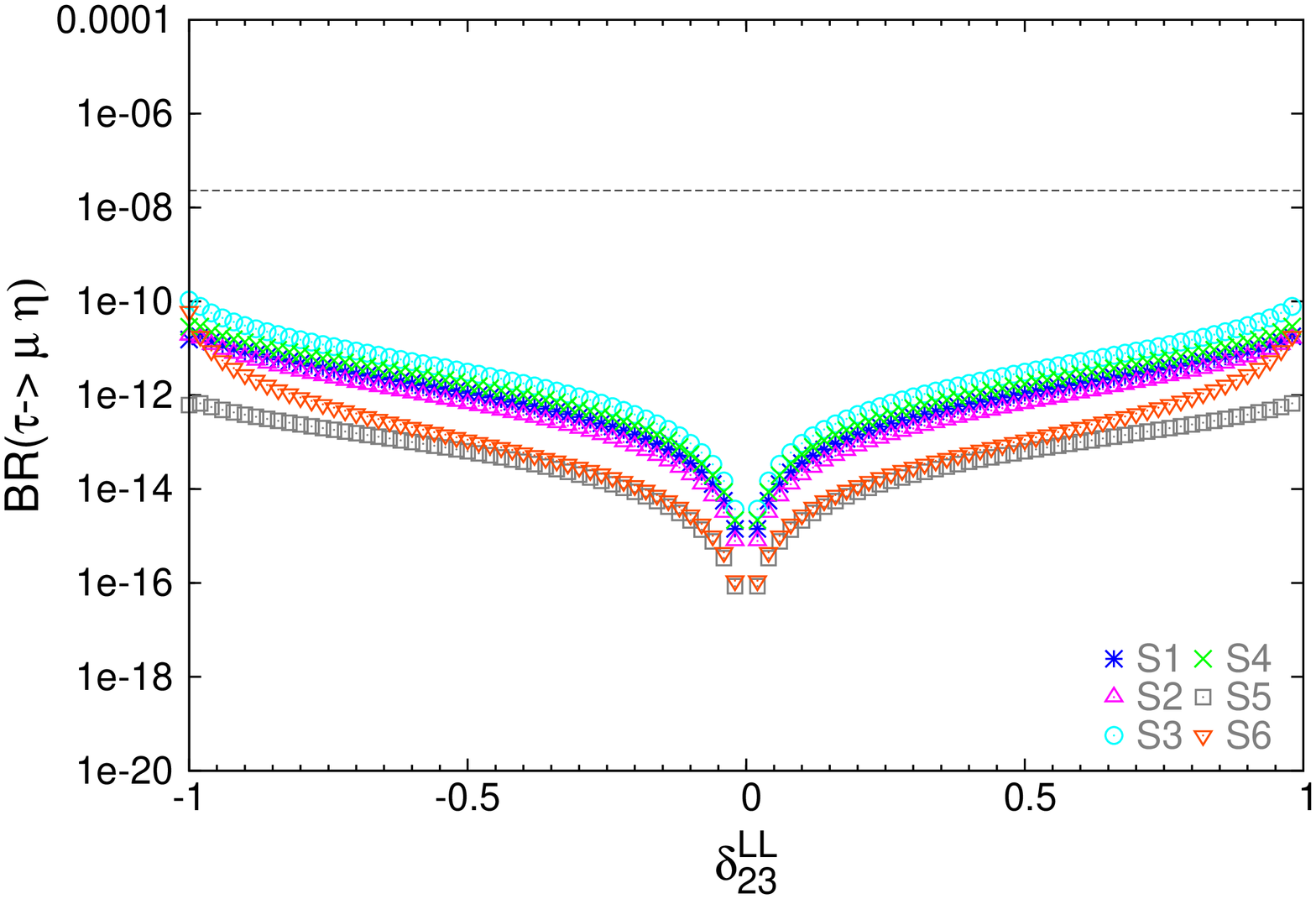   ,scale=0.30,angle=270,clip=}
\end{center}
\caption{LFV rates for $\tau-\mu$ transitions as a function of slepton mixing $\delta_{23}^{LL}$.} 
\label{mixing23LL}
\end{figure}

\begin{figure}[ht!]
\begin{center}
\psfig{file=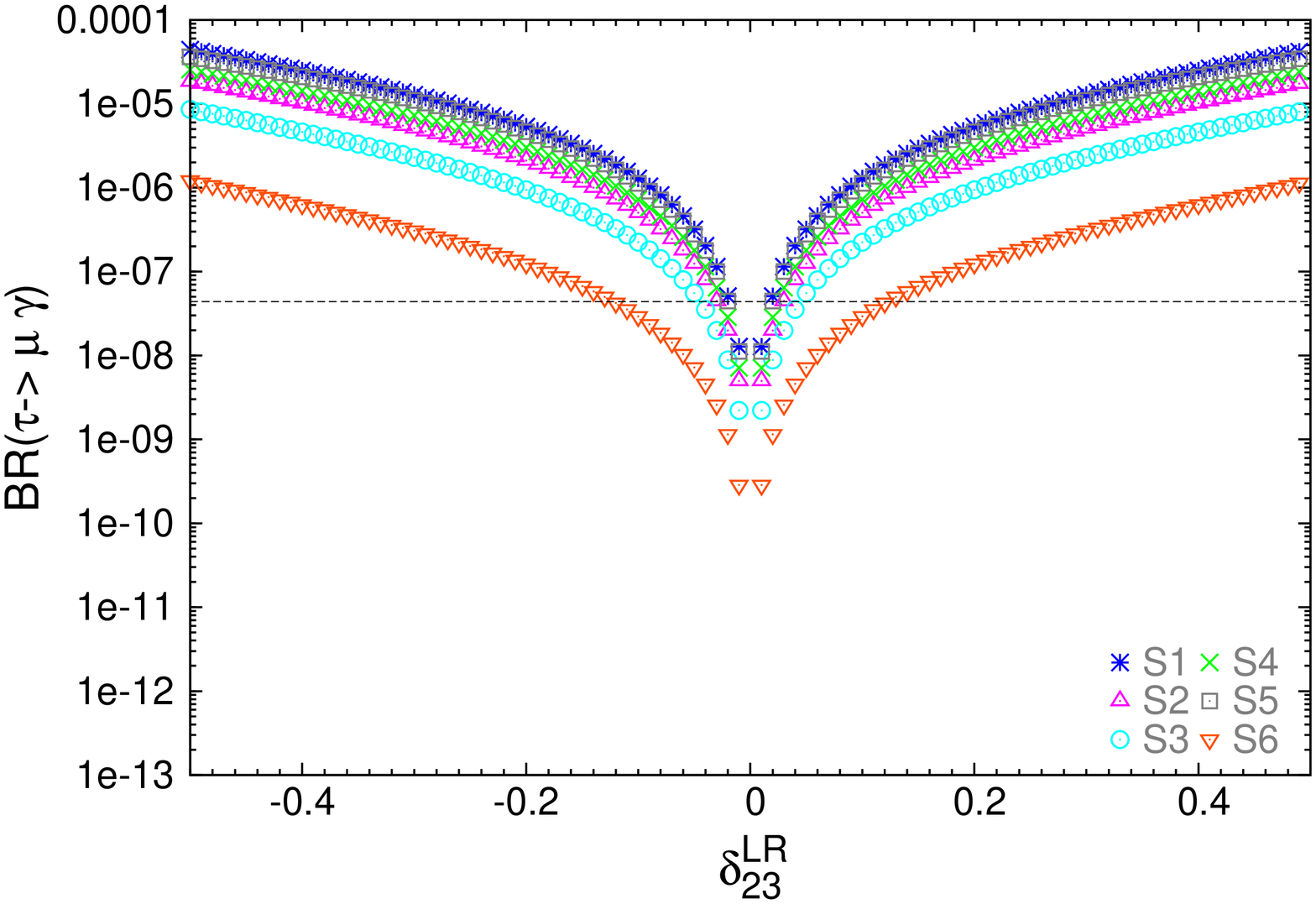  ,scale=0.45,angle=270,clip=}\\
\psfig{file=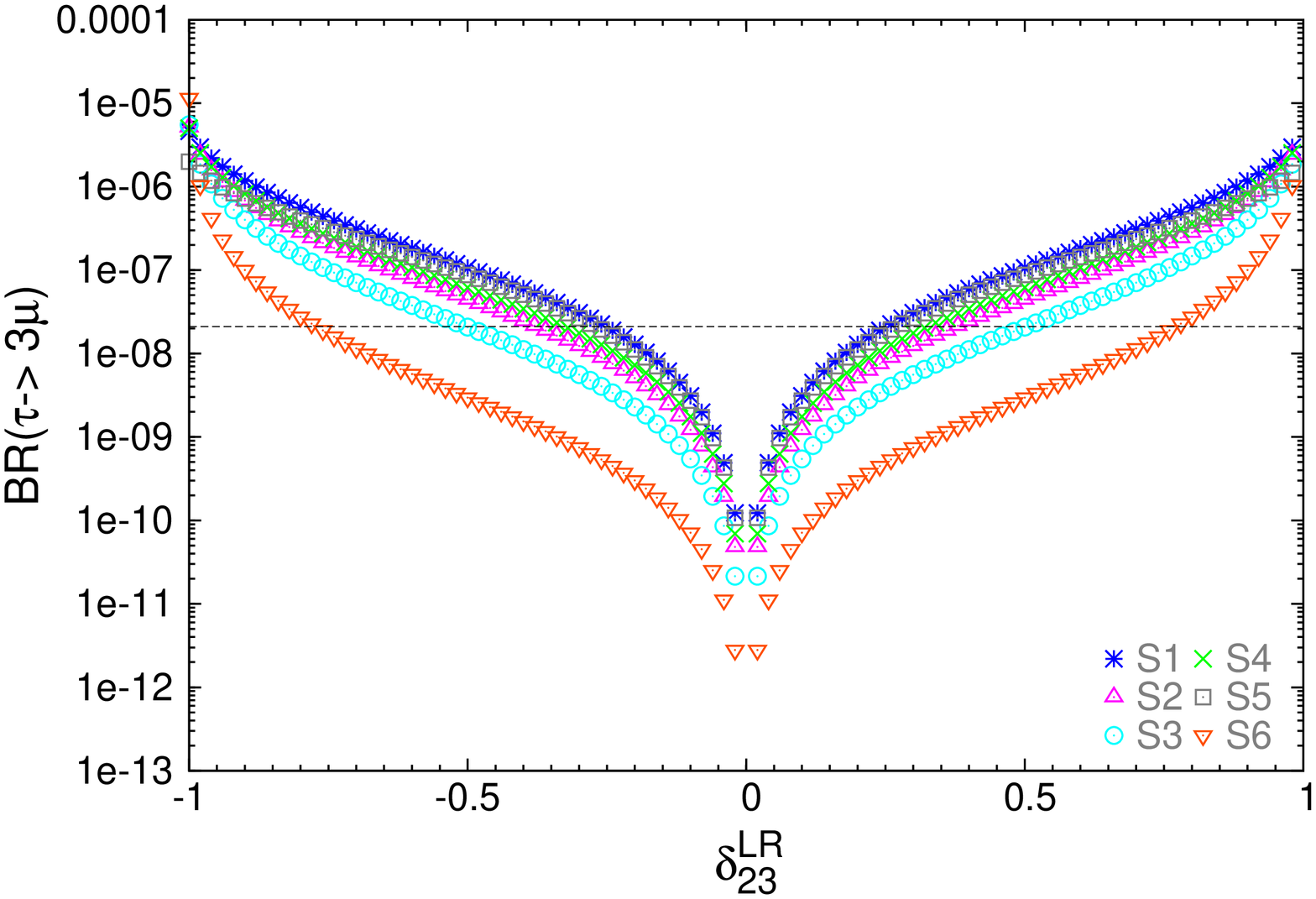    ,scale=0.30,angle=270,clip=}
\psfig{file=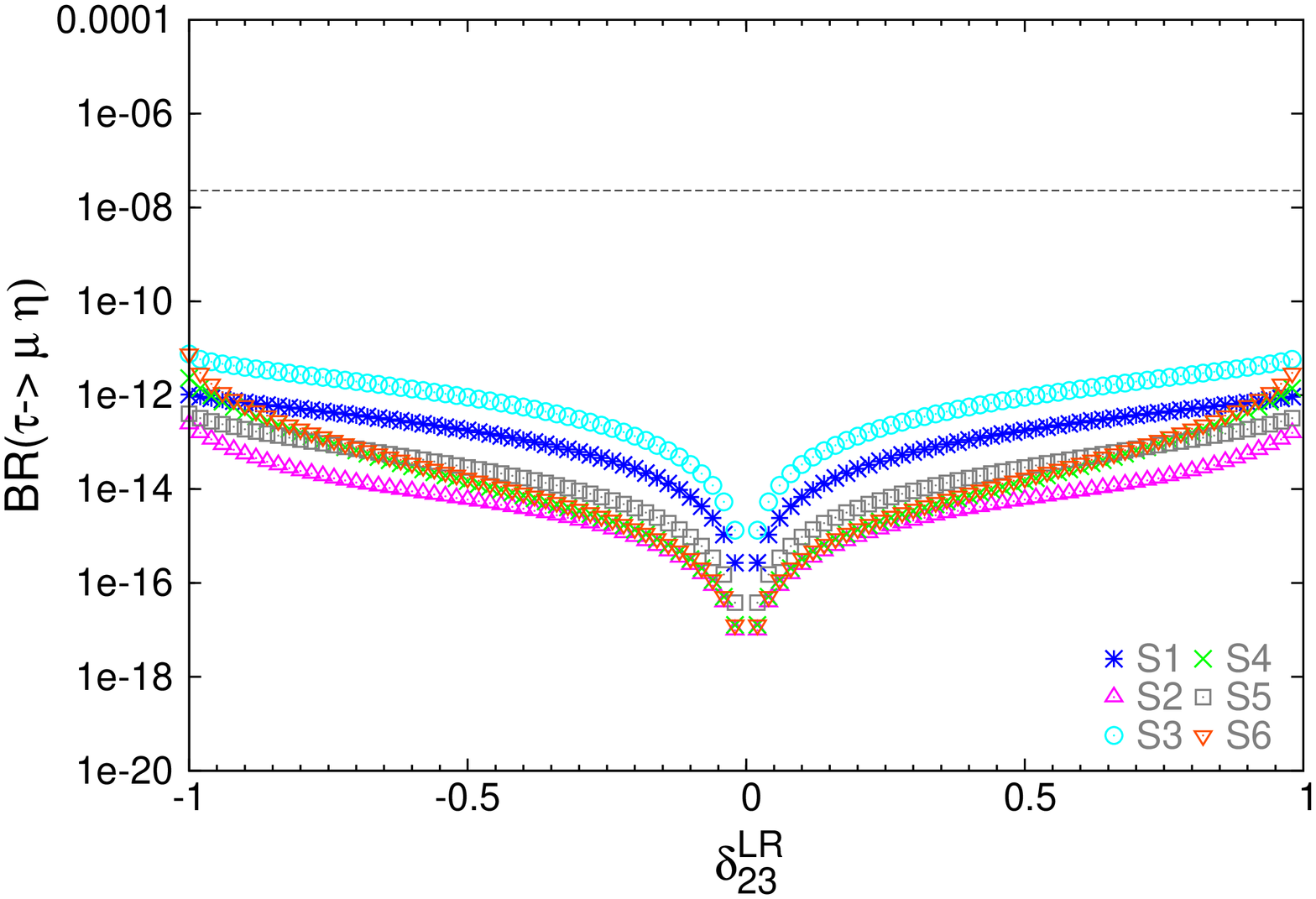   ,scale=0.30,angle=270,clip=}
\end{center}
\caption{LFV rates for $\tau-\mu$ transitions as a function of slepton
  mixing $\delta_{23}^{LR}$. The corresponding plots for
  $\delta_{23}^{RL}$, not shown here, are indistinguishable from these.}  
\label{mixing23LR}
\end{figure}

\begin{figure}[ht!]
\begin{center}
\psfig{file=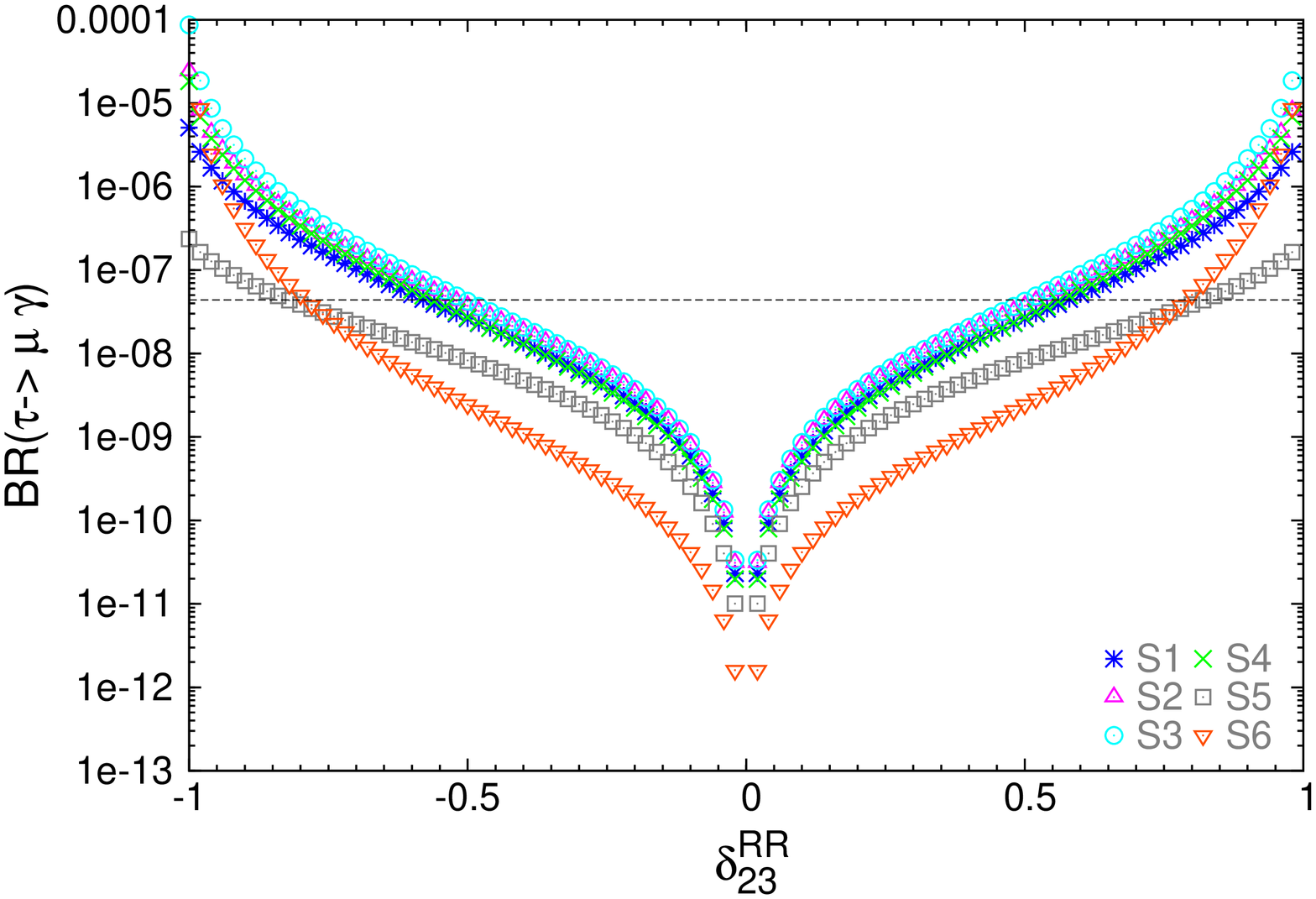  ,scale=0.45,angle=270,clip=}\\
\psfig{file=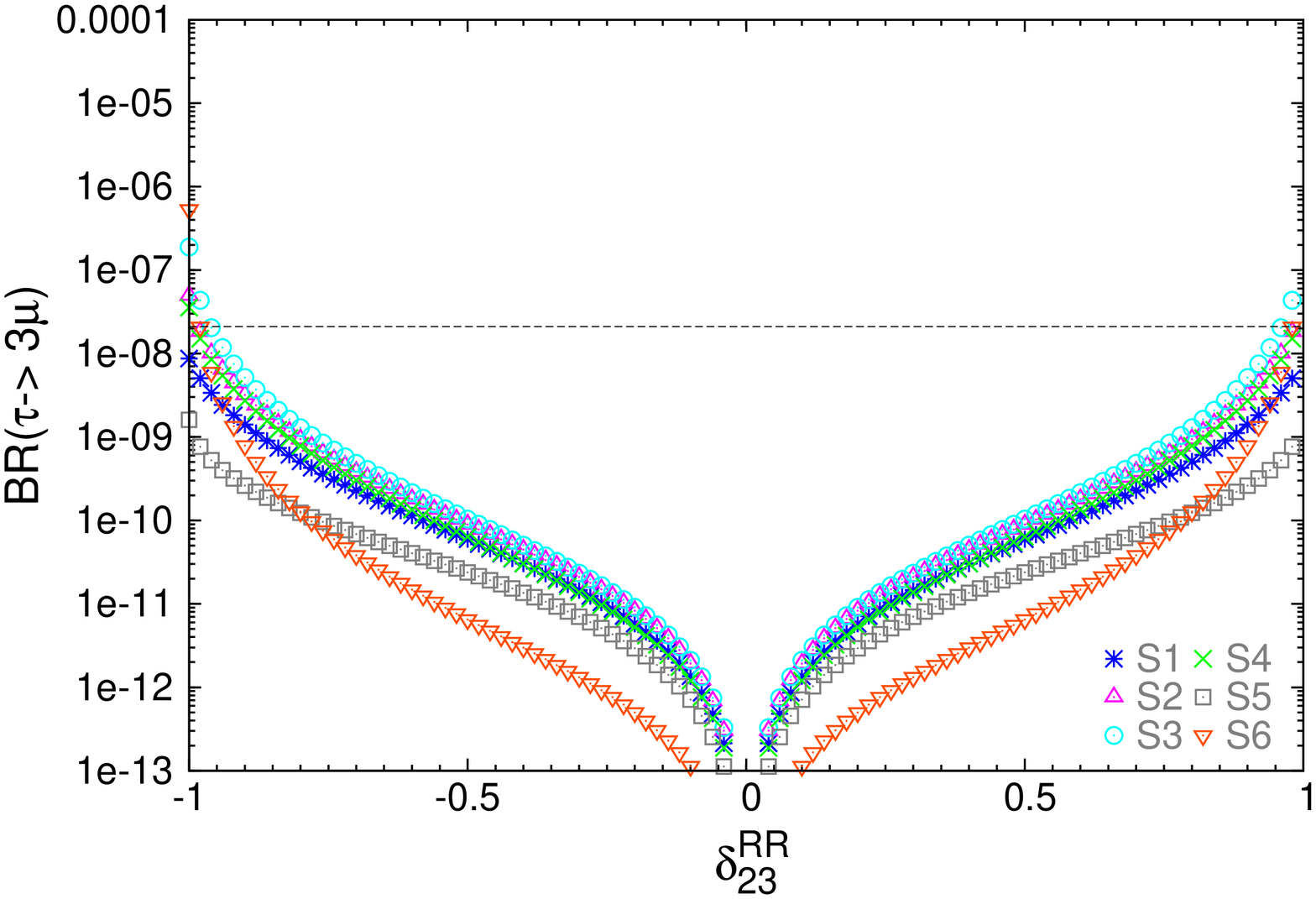    ,scale=0.30,angle=270,clip=}
\psfig{file=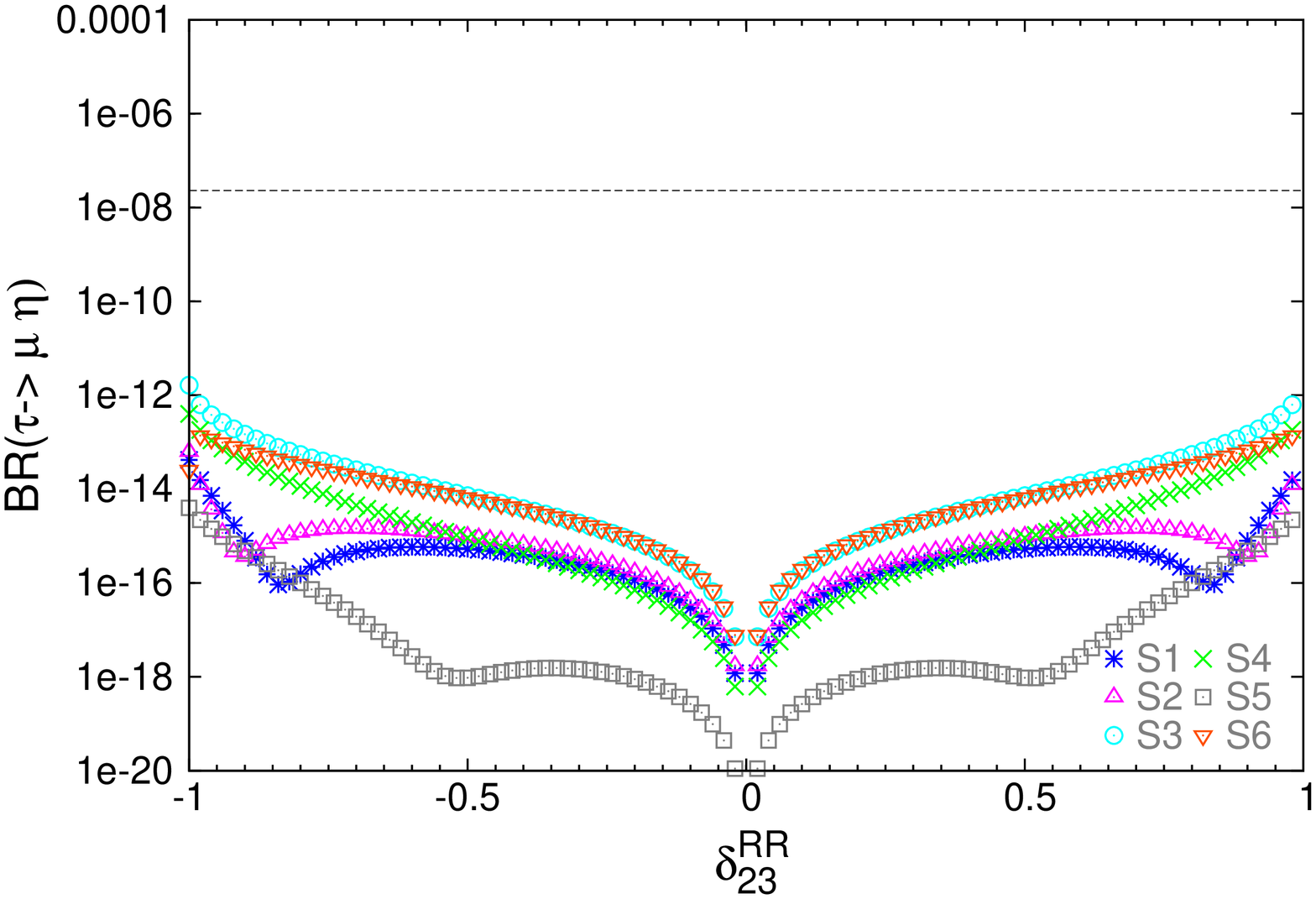   ,scale=0.30,angle=270,clip=}
\end{center}
\caption{LFV rates for $\tau-\mu$ transitions as a function of slepton
  mixing $\delta_{23}^{RR}$.}
\label{mixing23RR}
\end{figure}

The results of our numerical predictions of the branching ratios as
functions of the single deltas $\delta^{AB}_{ij}$, for the various
selected LFV processes and for the various scenarios S1 to S6 in
framework 1, are collected in figures \ref{mixing12LL} through
\ref{mixing23RR}, where a comparison with the corresponding present
upper experimental bound is also included, see \refse{sec:expbounds}. 
Figure \ref{mixing12LL}
summarizes the status of $\delta^{LL}_{12}$, \reffi{mixing12LR}
that of $\delta^{LR}_{12}$, \reffi{mixing12RR} that of
$\delta^{RR}_{12}$.
The analyzed experimental results are from $\br(\mu \to e \ga)$, 
$\br(\mu \to 3 e)$ and $\CR(\mu-e, \mbox{Nuclei})$.
Figure \ref{mixing13LL} depicts the results  of $\delta^{LL}_{13}$,  \reffi{mixing13LR} that of $\delta^{LR}_{13}$,
\reffi{mixing13RR} that of $\delta^{RR}_{13}$.
The analyzed experimental results are from $\br(\tau \to e \ga)$,
$\br(\tau \to 3 e)$ and $\br(\tau \to e \eta)$.
Figure \ref{mixing23LL} shows the results of 
$\delta^{LL}_{23}$, \reffi{mixing23LR} that of
$\delta^{LR}_{23}$, and \reffi{mixing23RR} that of
$\delta^{RR}_{23}$, where the experimental results are from 
$\br(\tau \to \mu \ga)$, $\br(\tau \to 3 \mu)$ and $\br(\tau \to \mu\eta)$.
The results for $\delta^{RL}_{ij}$ are indistinguishable from the corresponding ones for $\delta^{LR}_{ij}$, 
and consequently they have been omitted here.

 
A first look at these plots confirms the well
known result that the most stringent bounds are for the mixings between
the the first and the second slepton generations, 12. It is also evident
that the bounds for the mixings between the second and the third slepton
generations, 23,  are similar to the bounds for the mixings between the
first and the third generations, 13, and both are much weaker than the
bounds on the 12-mixings. 
As another general result one can observe that, 
whereas all the 12-mixings are constrained by
the three selected LFV processes,  
$\mu \to e \gamma$, $\mu \to 3 e$ and $\mu-e$ conversion in heavy (Au)
nuclei, the 23-mixings are not constrained, for the studied points, by
the semileptonic tau decay $\tau \to \mu \eta$. Similarly, the
13-mixings are not constrained either, by  $\tau \to e \eta$. The main
reason for this is that the studied points 
S1-S6 all have very heavy $A^0$ Higgs bosons, $M_{A}=500-1500\gev$ and
therefore the decay channel mediated by this $A^0$ is much
suppressed, even at large $\tb$, where the contribution from $A^0$ to 
$\br(\tau \to \mu \eta)$ and $\br(\tau \to e \eta)$, which is the dominant
one, grows as $(\tb)^6$ \cite{Arganda:2008jj,Herrero:2009tm}. It should also 
be noted the appearance of two symmetric minima in 
$\br(\tau \to \mu \eta)$ and $\br(\tau \to e \eta)$ of figs \ref{mixing23RR}
and \ref{mixing13RR} respectively, in the scenarios S5, S1 and S2. A similar 
feature can also be observed in
$\br(\tau \to e \eta)$ of fig \ref{mixing13LR} in scenario S2. For 
instance, in S5 these minima in $\br(\tau \to e \eta)$ appear at  
$\delta_{13}^{RR} \sim \pm 0.5$. We have checked that the origin of these minima
is due to the competing diagrams mediated by $A^0$ and $Z$ which give
contributions of similar size for  $\tb \lesssim 30$ but with opposite sign
, and 
this produces strong cancellations in the total rates. Similar comments
apply to $\br(\tau \to \mu \eta)$.
 Another general result, which confirms the known
literature for  particular models like SUSY-Seesaw models
\cite{Arganda:2005ji}, is the evident correlation between the $\br(l_j
\to 3 l_i)$ and  $\br(l_j \to l_i \gamma)$ rates. It should be
emphasized that we get these correlations in a model independent way and
without the use of any approximation, like the mass insertion
approximation or the large $\tb$ approximation. Since our
computation is full-one loop and has been performed in terms of
physical masses, our findings are valid for any  
value of $\tb$ and $\delta^{AB}_{ij}$'s. 
These correlations, confirmed in our plots, indicate that the general
prediction guided by the photon-dominance behavior in $\br(l_j \to 3 l_i)$  
indeed works quite well for all the studied $\delta^{AB}_{ij}$'s and all
the studied  
S1-S6 points. This dominance of the $\gamma$-mediated channel in the
$l_j \to 3 l_i$ decays allows to derive the following simplified
relation:
\begin{equation}\label{BRlj3li_approx}
\frac{\br(l_j \to 3 l_i)}{\br(l_j \to l_i \gamma)} =
\frac{\alpha}{3\pi}\left(\log\frac{m_{l_j}^2}{m_{l_i}^2}-\frac{11}{4}\right) \,,
\end{equation}
which gives the approximate values of
$\frac{1}{440}$, $\frac{1}{94}$ and $\frac{1}{162}$ for 
$(l_jl_i)= (\tau \mu), (\tau e)$ and  $(\mu e)$, respectively.
The ${\cal O}(\alpha)$ suppression in the predicted rates of  
$\br(l_j \to 3 l_i)$ versus $\br(l_j \to l_i \gamma)$ yields, 
despite the experimental sensitivities to the leptonic decays $l_j
\to 3 l_i$ have improved considerably in the last years, 
that the
radiative decays $l_j \to l_i \gamma$ are still the most efficient
decay channels in setting constraints to the slepton mixing
parameters. This holds for all the intergenerational mixings,
12, 13 and 23. 
As discussed in \cite{Arganda:2005ji}, in the context of SUSY, there
could be just a chance of departure from these ${\cal O}(\alpha)$
reduced ratios if the Higgs-mediated channels dominate the rates of the
leptonic decays, but this does not happen in our S1-S6 scenarios, with
rather heavy $H^0$ and $A^0$. We have checked that the
contribution from these Higgs channels are very small and can be safely
neglected, a scenario that is favored by the recent results from
the heavy MSSM Higgs boson searches at the LHC~\cite{CMSpashig12050}.

This same behavior can be seen in the comparison between the $\br(\mu
\to e  \gamma)$ and $\CR(\mu-e, {\rm Nuclei})$ rates. Again there is an
obvious correlation in our plots for these two rates that can be
explained by the same argument as above, namely,  the photon-mediated
contribution in $\mu-e$ conversion dominates the other contributions,
for all the studied cases, and therefore the corresponding rates are
suppressed by a ${\cal O}(\alpha)$ factor respect to the radiative decay
rates. These correlations are clearly seen in all our plots for all the
studied $\delta^{AB}_{ij}$'s and in all S1-S6 scenarios. The
relevance of 
$\CR(\mu-e, {\rm Nuclei})$ as compared to $\br(\mu \to 3 e)$ is
given by the fact that not only the
present  experimental bound is slightly better, but also
that the future perspectives for the expected sensitivities are
clearly more promising in the $\mu-e$ conversion case (see below). 
In general, as
can be seen in our plots, the present bounds for  $\delta^{AB}_{ij}$'s
as obtained from $\CR(\mu-e, {\rm Nuclei})$ and  $\br(\mu \to 3 e)$ are
indeed very similar. 

In summary, the best bounds that one can infer from our results in
figures \ref{mixing12LL} through \ref{mixing23RR} come from the
radiative $l_j \to l_i \gamma$ decays and we get the maximal allowed
values for all $|\delta^{AB}_{ij}|$'s that are collected in
\refta{boundsSpoints} for each of the studied scenarios S1 to S6. 
They give an overall idea of the size of the bounds with respect to
the latest experimental data.
When comparing the results in
this table for the various scenarios, we see that scenario S3 gives the
most stringent constraints to the  $\delta^{LL}_{ij}$ and
$\delta^{RR}_{ij}$ mixings, in spite of having rather heavy sleptons
with masses close to 1 TeV. The reason is well understood from the $\tb$
dependence of the BRs which enhances the rates in the case of  $LL$
and/or $RR$ single deltas at large $\tb$, in agreement with the simple
results of the MIA formulas in \refeqs{MIA-L} and (\ref{MIA-R}). 
Here it should be noted that within S3 we have $\tb = 50$, which
is the largest considered value in these S1-S6 scenarios. Something
similar happens in S4 with $\tb =40$. In contrast, the most stringent
constraints on the $\delta^{LR}_{ij}$ mixings occur in scenarios S1 and
S5. Here it is important to note that there are not enhancing
$\tb$ factors in the $\delta^{LR}_{ij}$ case. In fact, the contributions
from the $\delta^{LR}_{ij}$'s  to the most constraining LFV radiative
decay rates are $\tb$ independent, in agreement again with the MIA simple
expectations (see \refse{sec:MIA}). Consequently, the stringent
constraints on $\de^{LR}_{i,j}$ in S1 and S5 arise due to the relatively
light sleptons in these scenarios.

\begin{table}[h!]
\begin{tabular}{|c|c|c|c|c|c|c|}
\hline
 &  S1 &  S2 &  S3 &  S4 &  S5 & S6  
 \\ \hline
 & & & & & & \\
$|\delta^{LL}_{12}|_{\rm max}$ & $10 \times 10^{-5}$ & $7.5\times 10^{-5}$ &   $5 \times 10^{-5}$& $6 \times 10^{-5}$ & $42\times 10^{-5}$  &  $8\times 10^{-5}$  \\ 
& & & & & & \\
\hline
& & & & & & \\
$|\delta^{LR}_{12}|_{\rm max}$ & $2\times 10^{-6}$ & $3\times 10^{-6}$ &
$4\times 10^{-6}$  & $3\times 10^{-6}$ & $2\times 10^{-6}$  & $1.2\times 10^{-5}$   \\ 
& & & & & & \\
\hline
& & & & & & \\
$|\delta^{RR}_{12}|_{\rm max}$ & $1.5 \times 10^{-3}$& $1.2 \times 10^{-3}$ & 
$1.1 \times 10^{-3}$ & $1 \times 10^{-3}$ & $2 \times 10^{-3}$ & $5.2 \times 10^{-3}$   \\ 
& & & & & & \\
\hline
& & & & & & \\
$|\delta^{LL}_{13}|_{\rm max} $ &  $5 \times 10^{-2}$ & $5 \times 10^{-2}$ & 
$3 \times 10^{-2}$ &  $3 \times 10^{-2}$& $23 \times 10^{-2}$ & $5 \times 10^{-2}$   \\ 
& & & & & & \\
\hline
& & & & & & \\
 $|\delta^{LR}_{13}|_{\rm max}$& $2\times 10^{-2}$  & $3\times 10^{-2}$ & $4\times 10^{-2}$ & $2.5\times 10^{-2}$ & $2\times 10^{-2}$ & $11\times 10^{-2}$   \\ 
& & & & & & \\
 \hline
 & & & & & & \\
$|\delta^{RR}_{13}|_{\rm max}$ & $5.4\times 10^{-1}$  & $5\times 10^{-1}$ & 
 $4.8\times 10^{-1}$ &$5.3\times 10^{-1}$  & $7.7\times 10^{-1}$ & $7.7\times 10^{-1}$ 
  \\ 
 & & & & & & \\ 
  \hline
 & & & & & & \\ 
$|\delta^{LL}_{23}|_{\rm max}$ & $6\times 10^{-2}$  & $6\times 10^{-2}$ & 
 $4\times 10^{-2}$& $4\times 10^{-2}$ & $27\times 10^{-2}$ & $6\times 10^{-2}$ 
  \\ 
 & & & & & & \\ 
  \hline
 & & & & & & \\ 
$|\delta^{LR}_{23}|_{\rm max}$ & $2\times 10^{-2}$   & $3\times 10^{-2}$ & 
$4\times 10^{-2}$ & $3\times 10^{-2}$ & $2\times 10^{-2}$ & $12\times 10^{-2}$ 
  \\ 
 & & & & & & \\ 
  \hline
 & & & & & & \\ 
$|\delta^{RR}_{23}|_{\rm max}$ & $5.7\times 10^{-1}$  & $5.2\times 10^{-1}$ & 
 $5\times 10^{-1}$& $5.6\times 10^{-1}$ & $8.3\times 10^{-1}$ & $8\times 10^{-1}$ 
  \\ 
 & & & & & & \\  
  \hline
\end{tabular}
\caption{ Present upper bounds on the slepton mixing parameters $|\delta^{AB}_{ij}|$ for the selected S1-S6 MSSM points defined in \refta{tab:spectra}. The bounds for $|\delta^{RL}_{ij}|$ are similar 
to those of $|\delta^{LR}_{ij}|$.}
\label{boundsSpoints}
\end{table}

\bigskip

So far, we have studied the case where just one mixing delta is allowed
to be non-vanishing. However, it is known in the
literature\cite{Masina:2002mv,Paradisi:2005fk} that one can get more
stringent or more lose bounds in some particular cases if,
instead, two (or even more) deltas are 
allowed to be non-vanishing. In order to study the implications of these
scenarios with two deltas, we have analyzed the improved bounds on pairs
of mixings of the 13 and 23 type which are at present the less
constrained as long each delta is analyzed singly.

\begin{figure}[ht!]
\begin{center}
\psfig{file=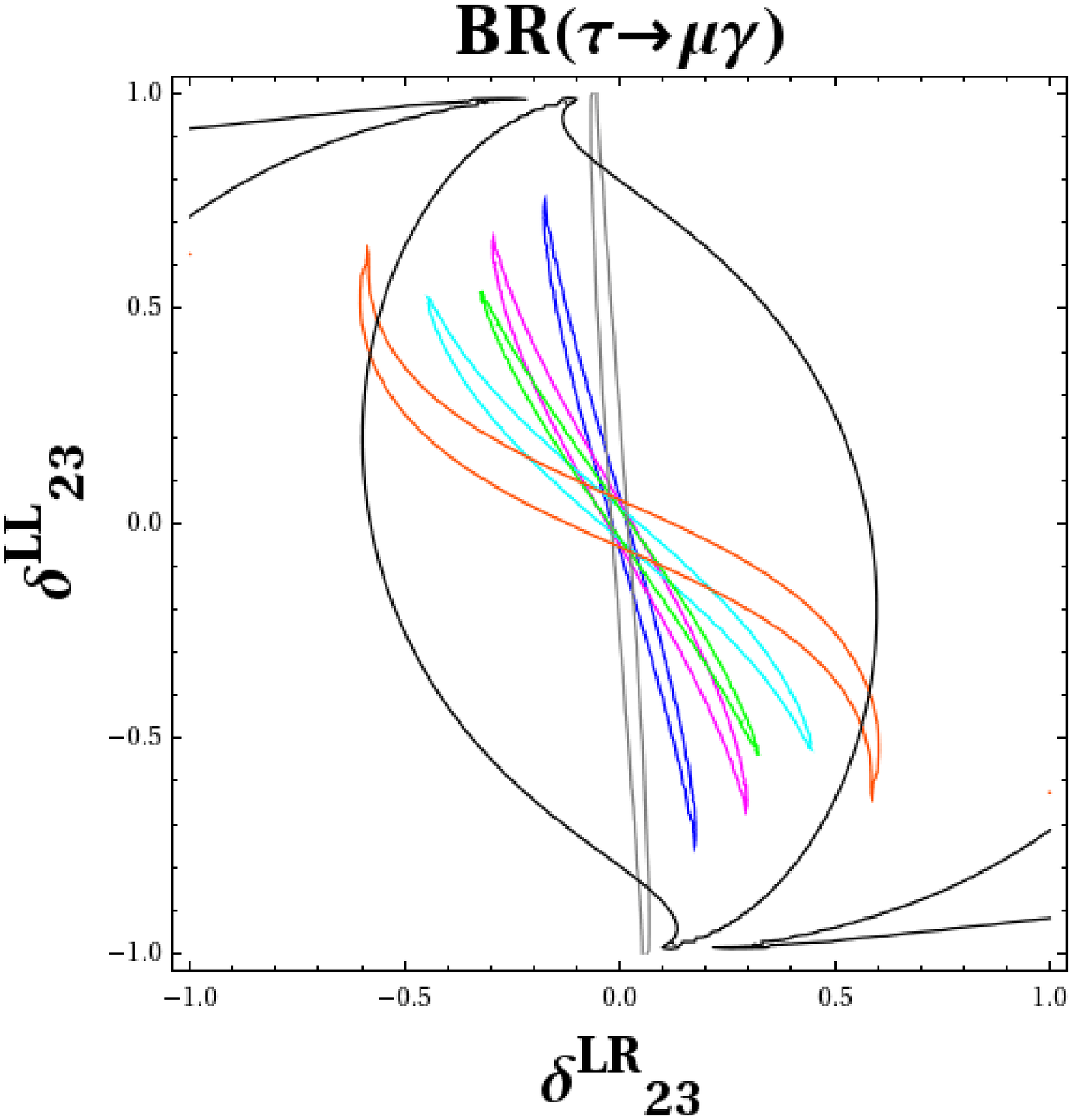,scale=0.70,clip=}
\end{center}
\caption{Maximum allowed values of  $(\delta^{LR}_ {23},\delta^{LL}_
  {23})$ in the scenarios S1 (dark blue), S2 (magenta), S3 (light blue),
  S4 (green), S5 (grey), S6 (orange) and S7 (black). The contourlines
  shown correspond to the present experimental upper limit: $\br(\tau
  \to \mu \gamma)_{\rm max}=4.4\times 10^{-8}$. For each scenario the
  allowed deltas are those inside the corresponding contourline.}  
\label{23LR-23LL}
\end{figure}

First we have looked into the various delta pairings of 23 type,
$(\delta^{AB}_{23},\delta^{CD}_{23})$, and we have found that some of
them lead to interesting interferences in the $\br(\tau \to \mu
\gamma)$ rates that can be either constructive or destructive, depending
on the relative delta signs, therefore leading to either a reduction or
an enhancement, respectively, in the  maximum allowed delta values as
compared to the one single delta case. More specifically, we have found
interferences in $\br(\tau \to \mu \gamma)$ for the case of
non-vanishing  $(\delta^{LR}_ {23},\delta^{LL}_ {23})$ pairs that are
constructive if these deltas are of equal sign, and destructive if they
are of opposite sign. Similarly, we have also found interferences in
$\br(\tau \to \mu \gamma)$ for the case of non-vanishing  $(\delta^{RL}_
{23},\delta^{RR}_ {23})$ pairs that are constructive if they are of
equal sign, and destructive if they are of opposite sign. However, in
this latter case the size of the interference is very small and does not
lead to very relevant changes with respect to the single delta case. The
numerical results for the most interesting case of $(\delta^{LR}_
{23},\delta^{LL}_ {23})$ are shown in \reffi{23LR-23LL}. We have
analyzed the six previous points, S1 through S6, and a new point S7 with
extremely heavy sleptons and whose relevant parameters for this analysis
of the 23 delta bounds are as follows: 
\BEA
S7& : &  m_{\tilde L_{1,2,3}}=m_{\tilde E_{1,2,3}}=10000\,\, {\rm GeV} \nonumber \\
&& \mu = 2000\,\,{\rm GeV}; \tb= 60  \nonumber \\
&& M_2=2000 \,\,{\rm GeV}; M_1=1000\,\,{\rm GeV}
\EEA  

This figure exemplifies in a clear way
that for some of the studied scenarios the destructive
interferences can be indeed quite relevant and produce new areas in the
$(\delta^{LR}_ {23},\delta^{LL}_ {23})$ plane with relatively large
allowed values of both $|\delta^{LR}_ {23}|$ and $|\delta^{LL}_ {23}|$
mixings. For instance, the orange contour which corresponds to the
maximum allowed values for scenario S6, leads to allowed mixings as
large as  $(\delta^{LR}_ {23},\delta^{LL}_ {23}) \sim (\pm 0.6, \mp
0.6)$. We also learn from this plot, that the relevance of this
$\delta^{LR}_ {23}-\delta^{LL}_ {23}$ interference grows in the
following order: Scenario S5 (grey contour) that has the smallest
interference effect, then S1, S2, S4, S3 and S6 that has the largest
interference effect. This growing interference effect is seen in the
plot as the contour being rotated anti-clockwise from
the most vertical one (S1) to the most inclined one
(S6). Furthermore, the size of the parameter space bounded by
these contours also grows, implying that 
``more'' parameter combinations are available for these two deltas.
It should be noted that, whereas the existence of the interference
effect can be already expected from the simple MIA formulas of
\refeqs{MIA-L} and (\ref{MIA-R}), the final found shape of these contours
in fig.\ref{23LR-23LL} and their quantitative relevance cannot be
explained by these simple formulas. The separation from the MIA
expectations are even larger in the new studied scenario S7, as can be
clearly seen in this figure. The big black contour, centered at zero,
contains a rather large allowed area in the  $(\delta^{LR}_
{23},\delta^{LL}_ {23})$ plane, allowing  values, for instance, of
$(\delta^{LR}_ {23},\delta^{LL}_ {23}) \sim (\pm 0.5, \mp
0.5)$. Furthermore, in this S7 there appear new allowed regions at the
upper left and lower right corners of the plot with extreme allowed
values as large as  $(\pm 0.9, \mp 0.9)$. 
These ``extreme'' solutions are only captured by a full one-loop
calculation
and cannot be explained by the simple MIA formulas.

\begin{figure}[ht!]
\begin{center}
\psfig{file=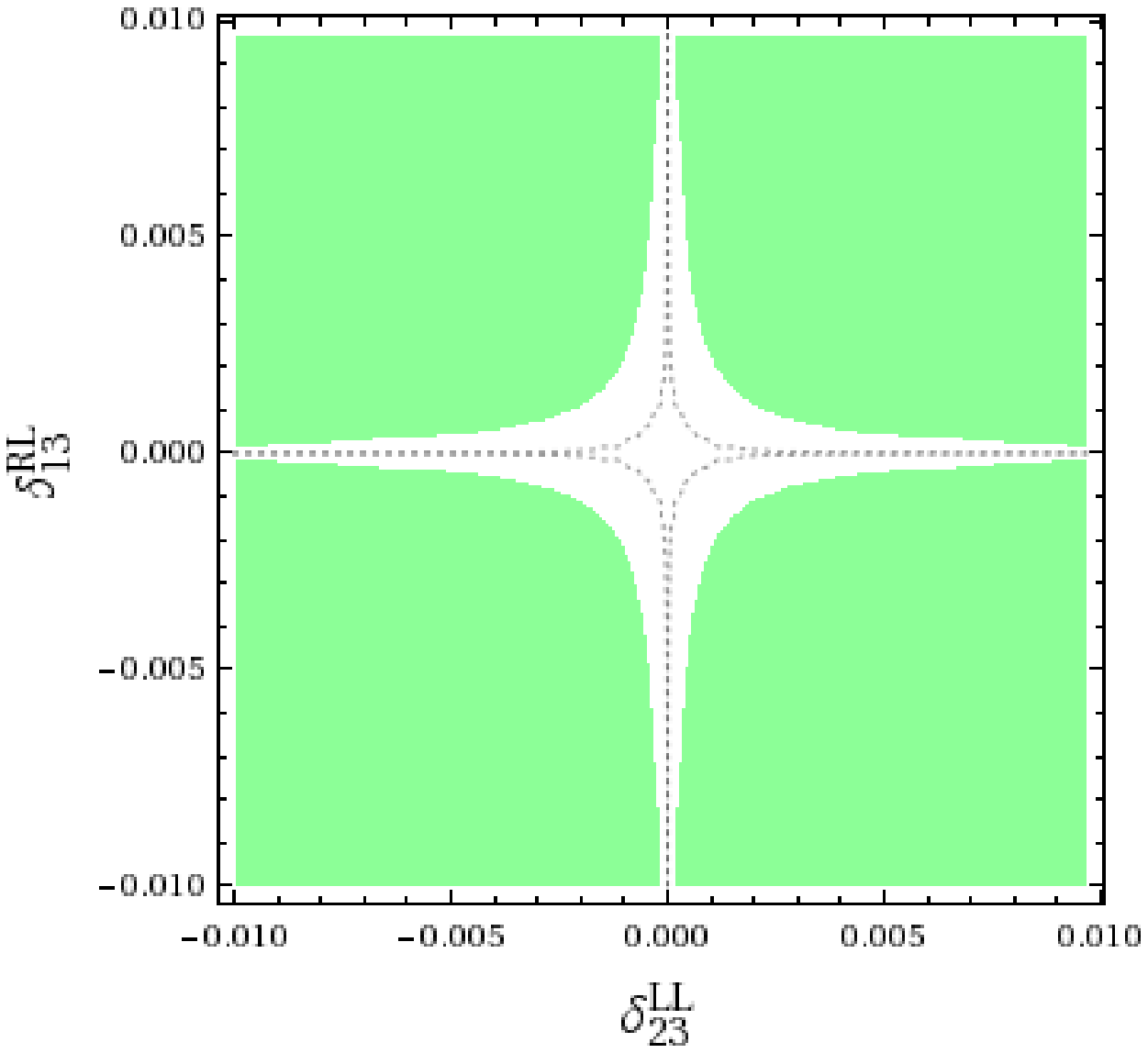     ,scale=0.65,clip=}
\psfig{file=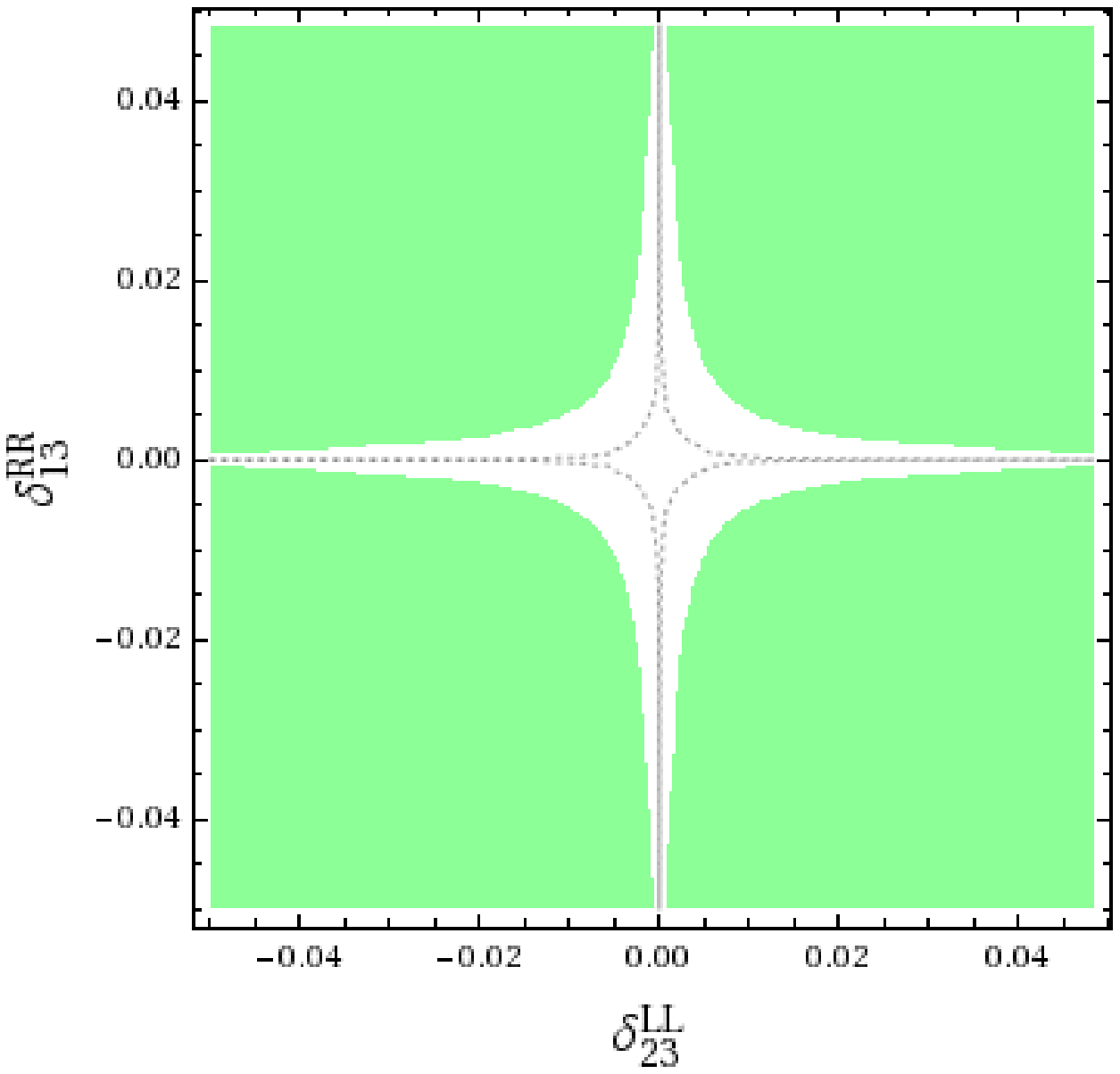   ,scale=0.615,clip=}\\
\psfig{file=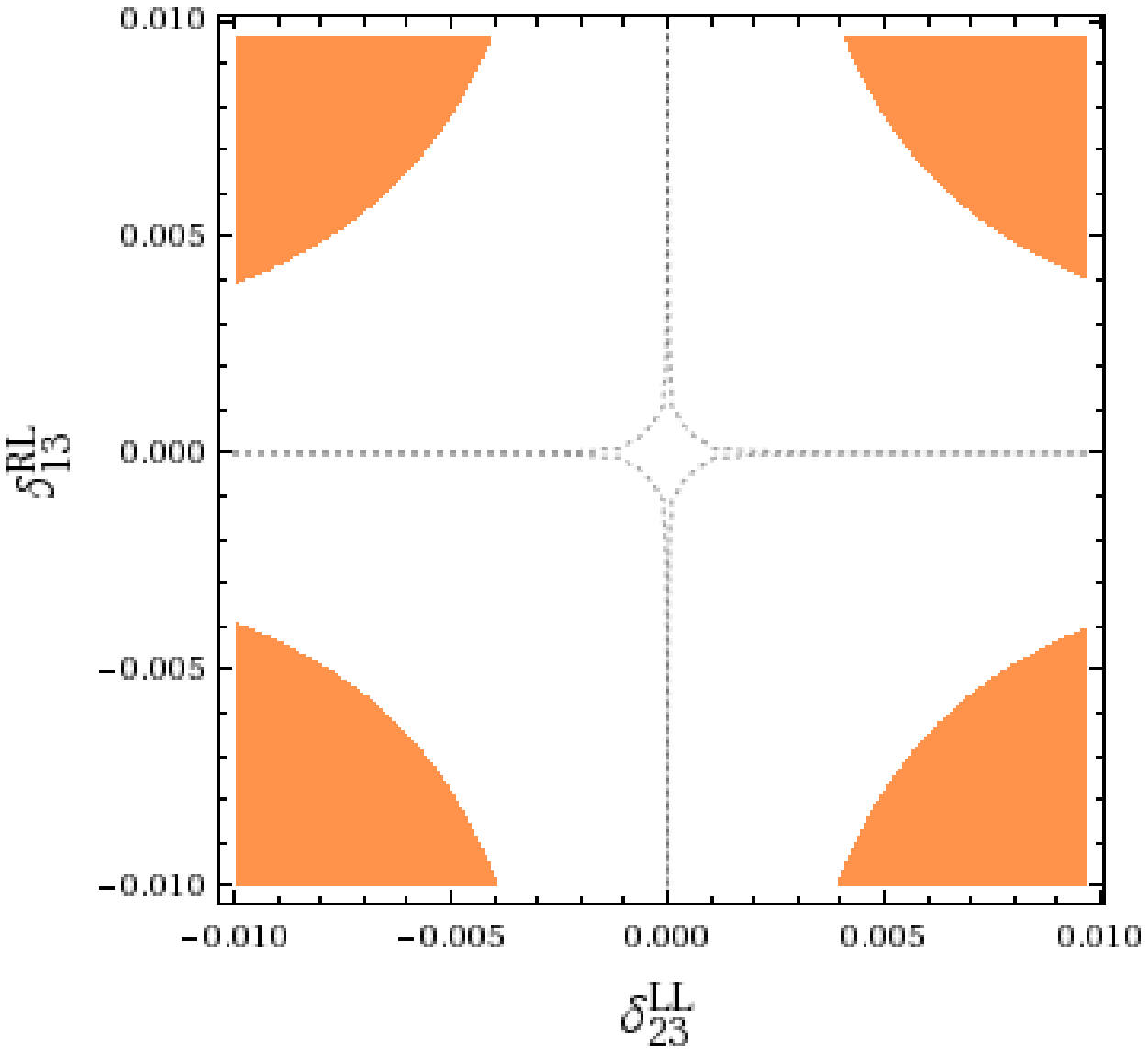
     ,scale=0.65,clip=}
\psfig{file=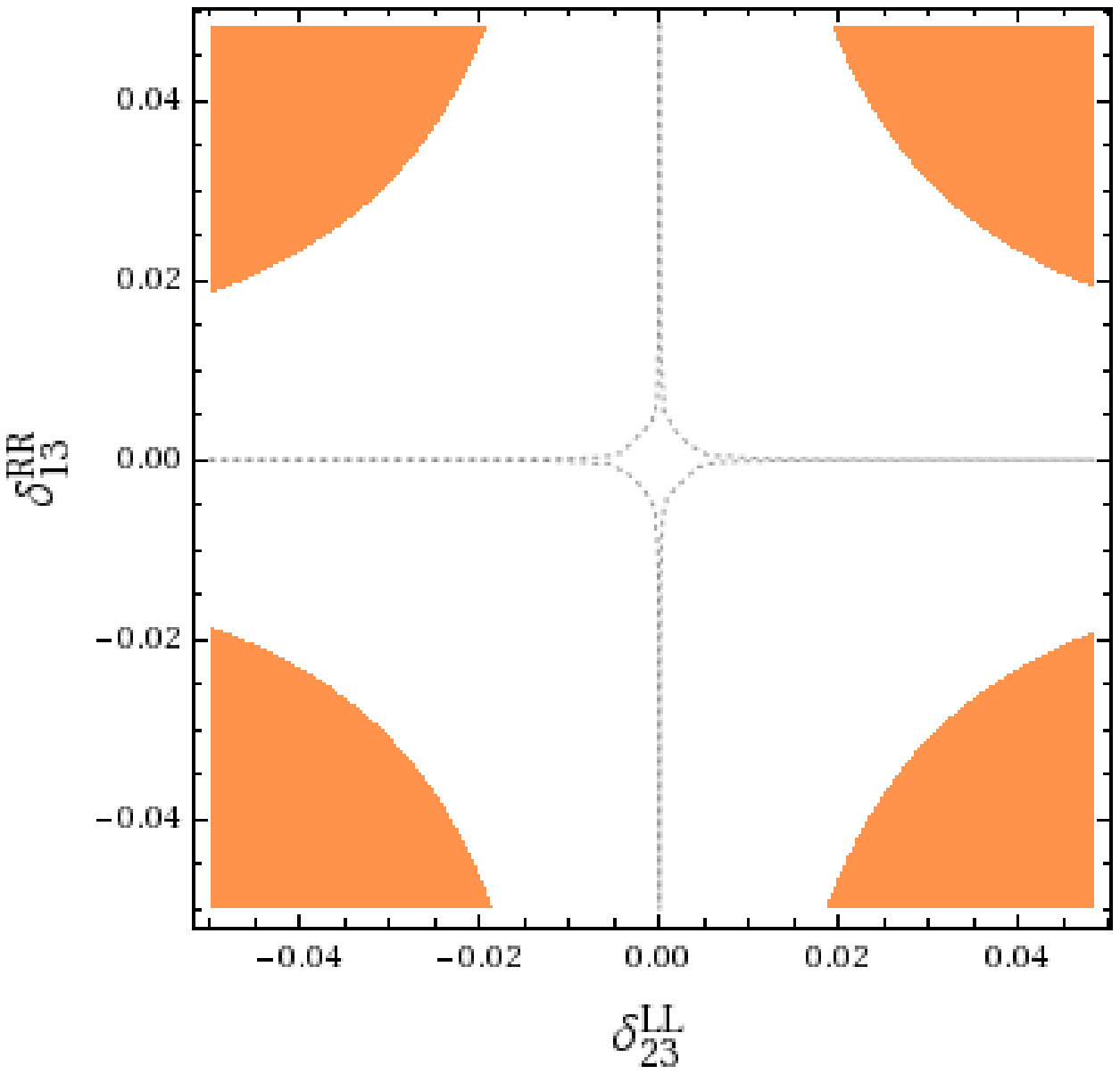    ,scale=0.615,clip=}
\end{center}
\caption{Bounds on pairs of slepton mixing parameters of (23,13) type
  for scenario S1: a) ($\delta_{23}^{LL}$,$\delta_{13}^{RL}$) in first
  column. Identical plots, not shown here, are found for:
  ($\delta_{23}^{RL}$,$\delta_{13}^{LL}$),
  ($\delta_{23}^{LR}$,$\delta_{13}^{RR}$), and
  ($\delta_{23}^{RR}$,$\delta_{13}^{LR}$); b)
  ($\delta_{23}^{LL}$,$\delta_{13}^{RR}$) in second column. Identical
  plots, not shown here, are found for:
  ($\delta_{23}^{RR}$,$\delta_{13}^{LL}$),
  ($\delta_{23}^{LR}$,$\delta_{13}^{RL}$), and
  ($\delta_{23}^{RL}$,$\delta_{13}^{LR}$). First row: Shaded regions (in
  green) are disallowed by the present upper experimental limit on 
  $\br(\mu \to e \gamma)$. 
  Second row: Shaded regions (in orange) are disallowed by the
  present upper experimental limit on $\CR(\mu - e, {\rm Nuclei})$. The
  allowed central areas in white will be shrinked by the future expected
  sensitivities in both $\mu \to e \gamma$ and $\mu-e$ conversion
  experimental searches (see text) to the small areas around the
  origin delimited 
  by the dotted lines.}  
\label{LL-RL}
\end{figure}

\begin{figure}[ht!]
\begin{center}
\psfig{file=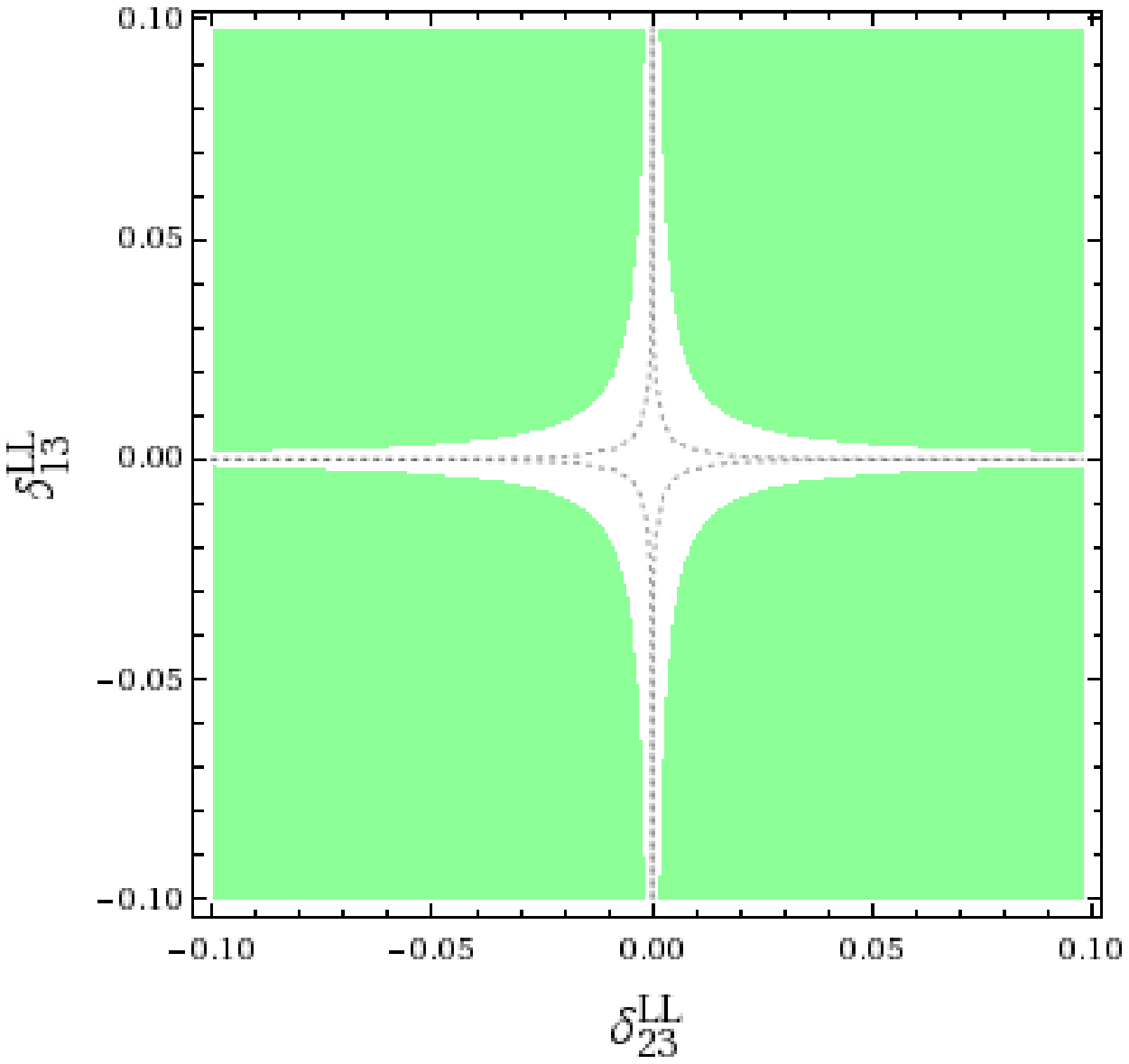     ,scale=0.645,clip=}
\psfig{file=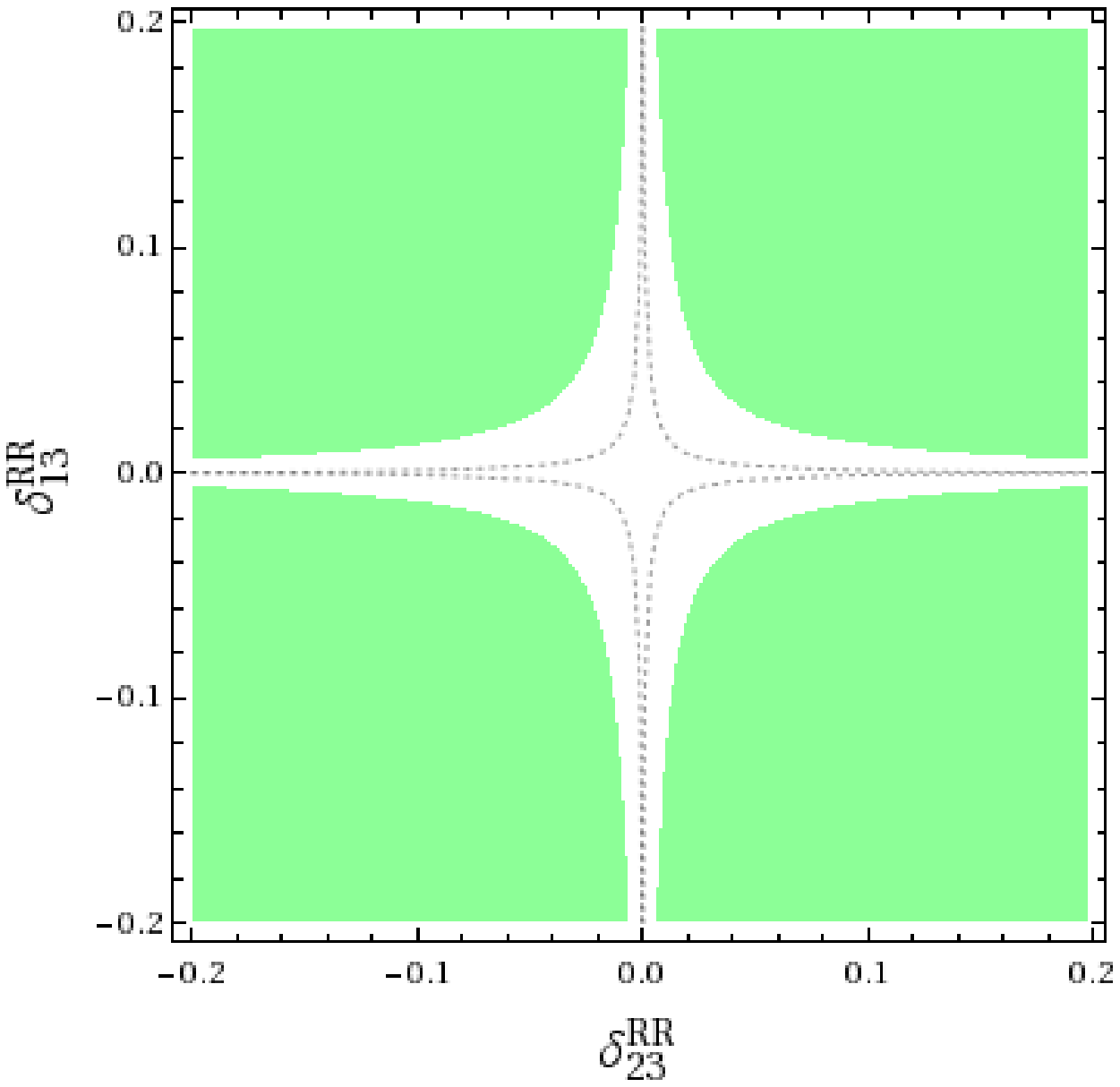   ,scale=0.62,clip=}\\
\psfig{file=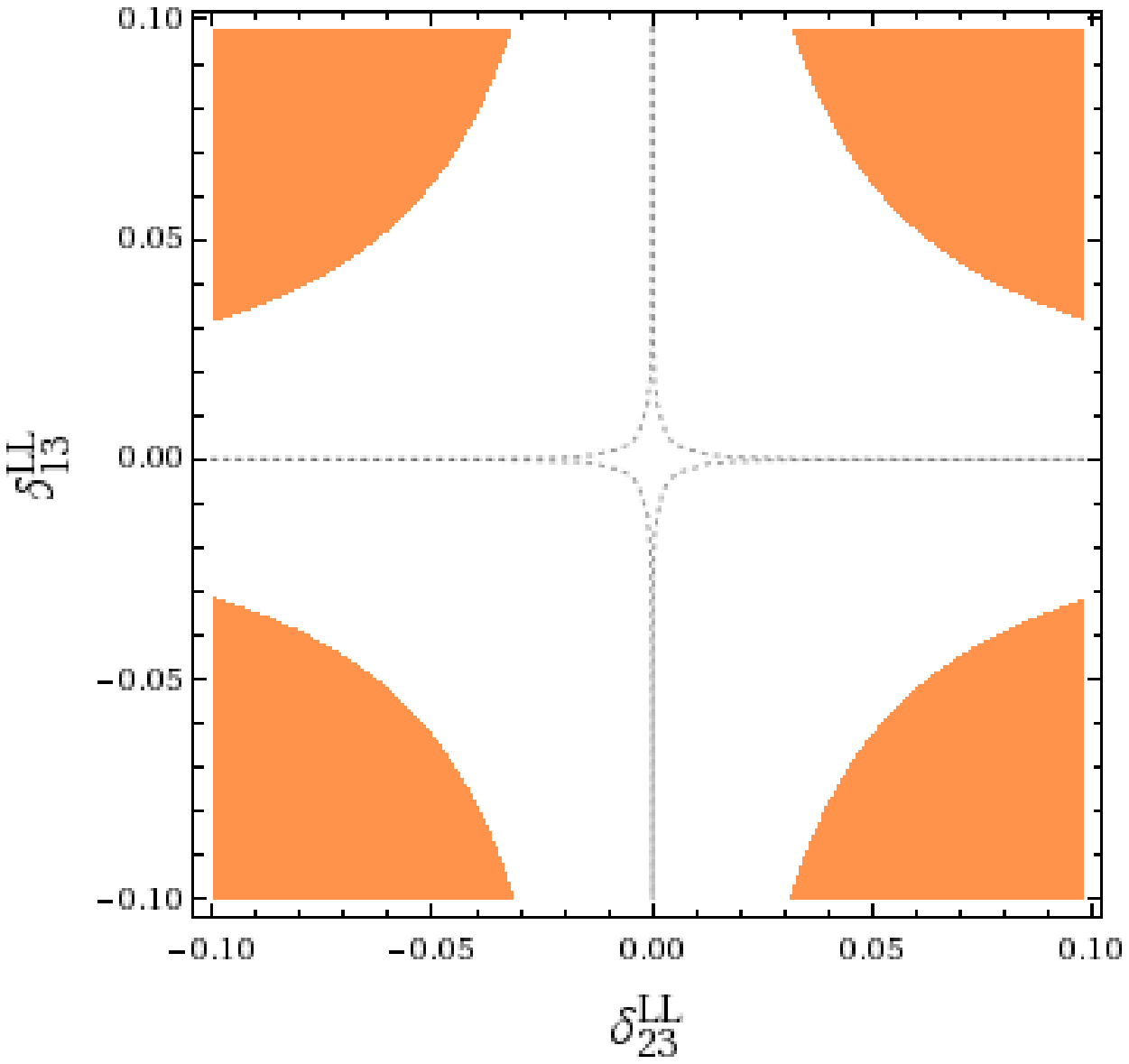
     ,scale=0.645,clip=}
\psfig{file=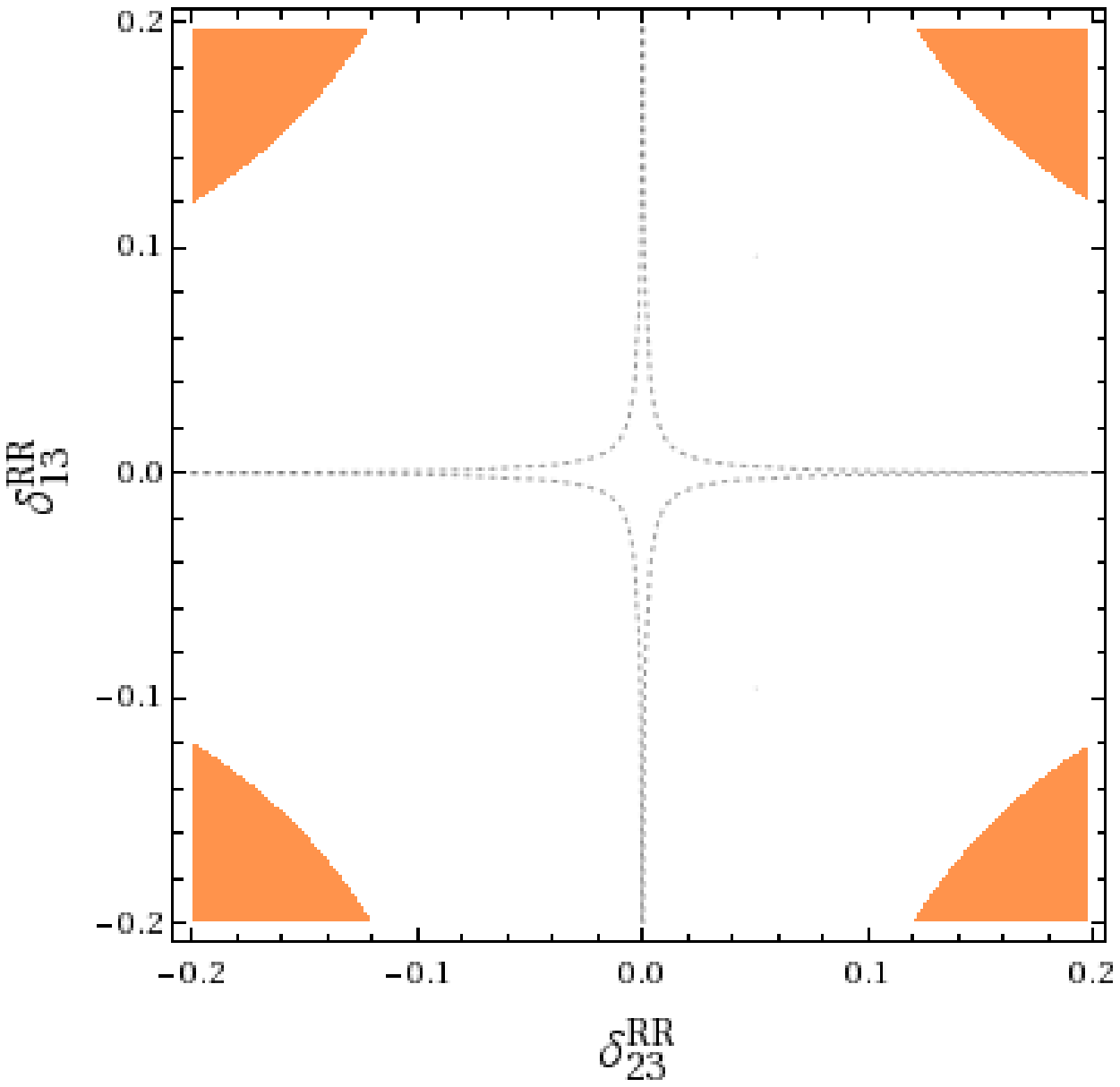    ,scale=0.62,clip=}
\end{center}
\caption{Bounds on pairs of slepton mixing parameters of (23,13) type
  for scenario S1: a)  ($\delta_{23}^{LL}$,$\delta_{13}^{LL}$) in first
  column; b)  ($\delta_{23}^{RR}$,$\delta_{13}^{RR}$) in second
  column. First row: Shaded regions (in green) are disallowed by 
  $\br(\mu \to e \gamma)$. 
  Second row: Shaded regions (in orange) are disallowed
  by $\CR(\mu - e, {\rm Nuclei})$. All inputs and explanations are
  as in \reffi{LL-RL}.}  
\label{LL-LLandRR-RR}
\vspace{-3em}
\end{figure}

\medskip
We now turn to examples in which more stringent bounds on 
  combinations of two deltas are derived. In particular,
 we have explored the restrictions that are obtained on the 
(13,23)  mixing pairs from the present bounds on $\br(\mu \to e \gamma)$
and $\CR(\mu-e, {\rm nuclei})$. In figures \ref{LL-RL} and
\ref{LL-LLandRR-RR} we show the results of this analysis for the S1
point. We have only selected  the pairs where we have found  improved
bounds respect to the previous single delta analysis. From \reffi{LL-RL}
we conclude that, for S1,  the maximal allowed values  
by present $\mu \to e \gamma$ (($ \mu-e$ conversion)) searches are
(given specifically here for equal input deltas):  
\BEA
(|\delta^{LL}_{23}|_{\rm max},|\delta^{RL}_{13}|_{\rm max})
&=&(0.0015,0.0015)\,\,((0.0062,0.0062))\,\,  
\label{double1}
\EEA
These numbers can be understood as follows: if, for instance, 
$\de^{RL}_{13} = 0.0015$ then $|\de^{LL}_{23}| < 0.0015$. If, on the
other hand, one delta goes to zero the bound on the other delta
disappears (from this particular observable).
We find equal bounds as in \refeq{double1} for: 
$(|\delta^{RL}_{23}|_{\rm max},|\delta^{LL}_{13}|_{\rm max})$, 
$(|\delta^{LR}_{23}|_{\rm max},|\delta^{RR}_{13}|_{\rm max})$ and   
$(|\delta^{RR}_{23}|_{\rm max},|\delta^{LR}_{13}|_{\rm max})$.  

Other pairings of deltas give less stringent bounds than
\refeq{double1} but still more stringent than the ones from the
single delta analysis. In particular, we get: 
\BEA
 (|\delta^{LL}_{23}|_{\rm max},|\delta^{RR}_{13}|_{\rm max}) 
&=&(0.0073,0.0073)\,\,((0.031,0.031))\,\,
\label{double2}
\EEA
And we get equal bounds as in \refeq{double2} for:
$(|\delta^{RR}_{23}|_{\rm max},|\delta^{LL}_{13}|_{\rm max})$, $(|\delta^{LR}_{23}|_{\rm max},|\delta^{RL}_{13}|_{\rm max})$ and $(|\delta^{RL}_{23}|_{\rm max},|\delta^{LR}_{13}|_{\rm max})$.

Finally, from  \reffi{LL-LLandRR-RR} we get:
\BEA
(|\delta^{LL}_{23}|_{\rm max},|\delta^{LL}_{13}|_{\rm max}) &=&(0.013,0.013)\,\,((0.056,0.056))\,\,
\label{double3}
\EEA
and 
\BEA
(|\delta^{RR}_{23}|_{\rm max},|\delta^{RR}_{13}|_{\rm max}) &=&(0.036,0.036)\,\,((0.16,0.16))\,\,
\label{double4}
\EEA
We have also studied the implications of the future expected
sensitivities in both 
$\br(\mu \to e \gamma) < 10^{-14}$~\cite{Mihara:2012zz} and 
$\CR(\mu - e, {\rm Nuclei}) < 2.6 \times 10^{-17}$~\cite{mueconvfuture},
 which are anticipated from future searches.
From our results in 
\reffis{LL-RL} and \ref{LL-LLandRR-RR} we conclude that the previous bounds
in \refeqs{double1}, (\ref{double2}), (\ref{double3}) and (\ref{double4})
will be improved (for both $\mu \to e
\gamma$ and  $\mu-e$ conversion)  to  
$(0.0005,0.0005)$, $(0.0025,0.0025)$, $(0.005,0.005)$ and $(0.01,0.01)$,
respectively. 

\clearpage
\newpage

\subsection{Results in framework 2}

The main goal of this part is to investigate how the upper
bounds for the slepton mixing deltas that we have found previously could
change for different ranges of the MSSM parameter space, 
that go beyond the selected S1-S6 points.  

In order to explore the variation of these bounds for different choices
in the MSSM parameter space, we investigate the four qualitatively
different pMSSM-4 scenarios (a), (b), (c) and (d) defined in \refeqs{Sa},
(\ref{Sb}), (\ref{Sc}) and (\ref{Sd}), respectively. 
As explained above, the idea is to explore generic scenarios that are
compatible with present data, in particular with the measurement of
a Higgs boson mass, which we interpret as the mass of the light
$\cp$-even Higgs boson in the MSSM, and the present
experimental measurement of $(g-2)_\mu$. Taking these experimental
results into account, we have re-analyzed the full set of bounds for
the single deltas that are 
extracted from the most restrictive LFV processes as a function of the
two most relevant parameters in our framework 2: the generic
SUSY mass scale $m_{\rm SUSY}$ $(\equiv m_{\rm SUSY-EW})$ and
$\tb$. In order to find $\Mh$ around $\sim 125 - 126 \gev$ the
scale $m_{\rm SUSY-QCD}$ as well as the trilinear couplings have been
chosen to sufficiently high values, see \refse{sec:f2}.
For the analysis in this framework 2, we
use the bounds on the radiative decays, $l_j \to l_i \gamma$ which, as
we have already shown, are at present the most restrictive ones in the
case of one single non-vanishing delta. And to simplify the analysis in
this part of the work, we use the mass insertion approximation (MIA)
formulas of \refeqs{BRs} through (\ref{functions}) to evaluate the
$\br(l_j \to l_i \gamma)$ rates. We have checked that these simple MIA
formulas provide a sufficiently accurate estimate of the
LFV rates in the case of single 
deltas, in agreement with \citere{Paradisi:2005fk}. 
 
\begin{figure}[ht!]
\begin{center}
\psfig{file=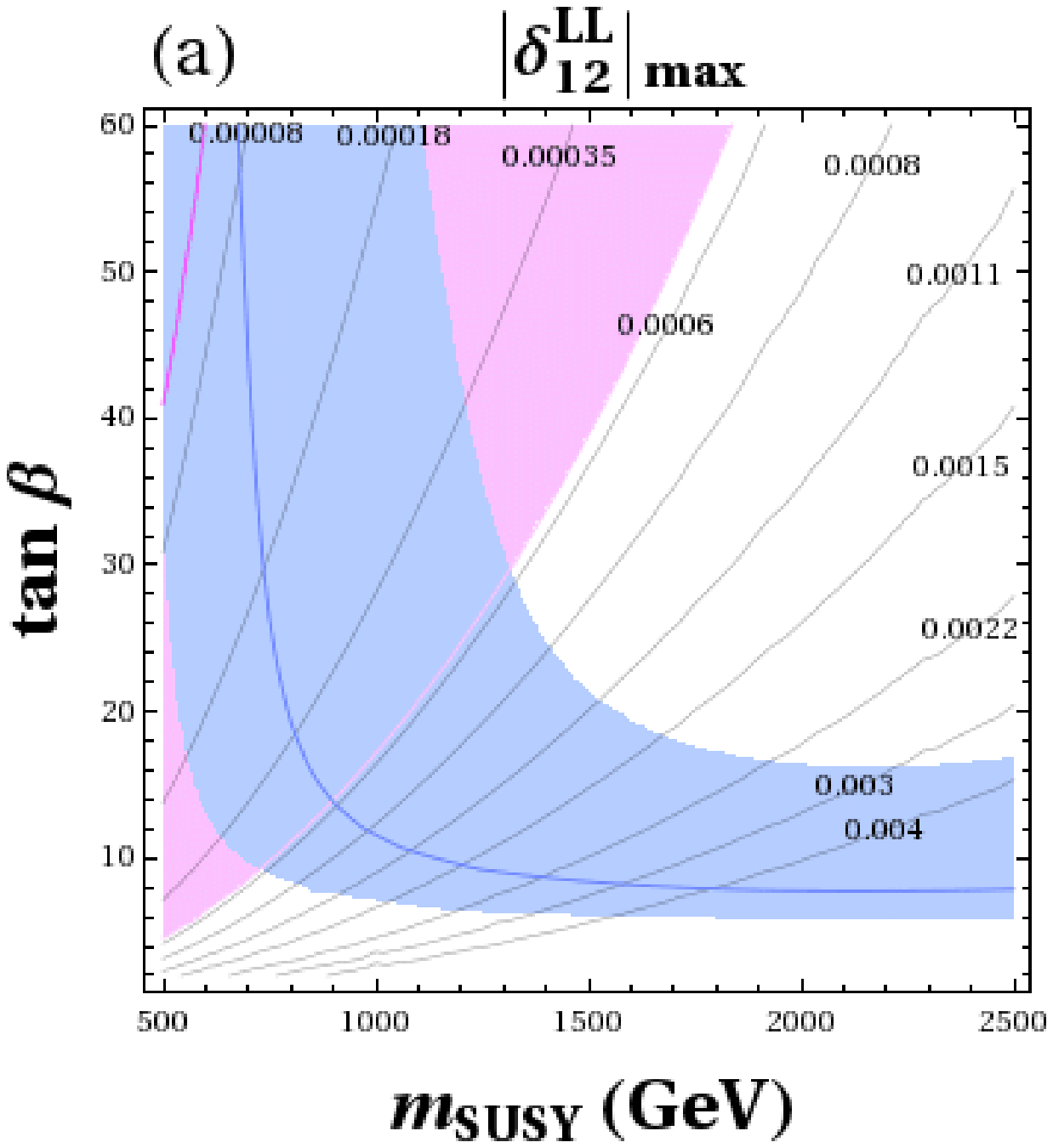,scale=0.60,clip=}
\psfig{file=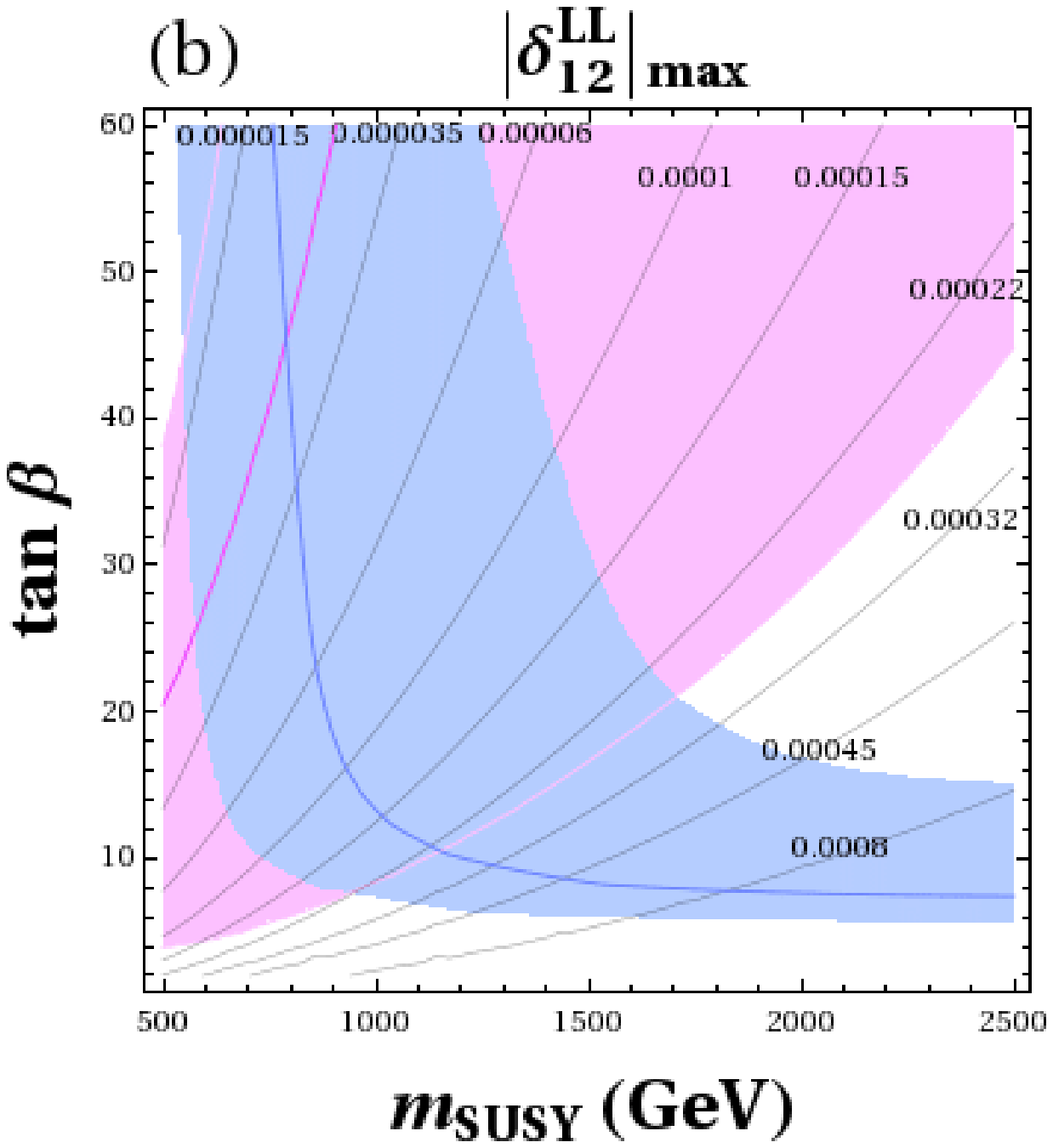,scale=0.60,clip=}\\
\psfig{file=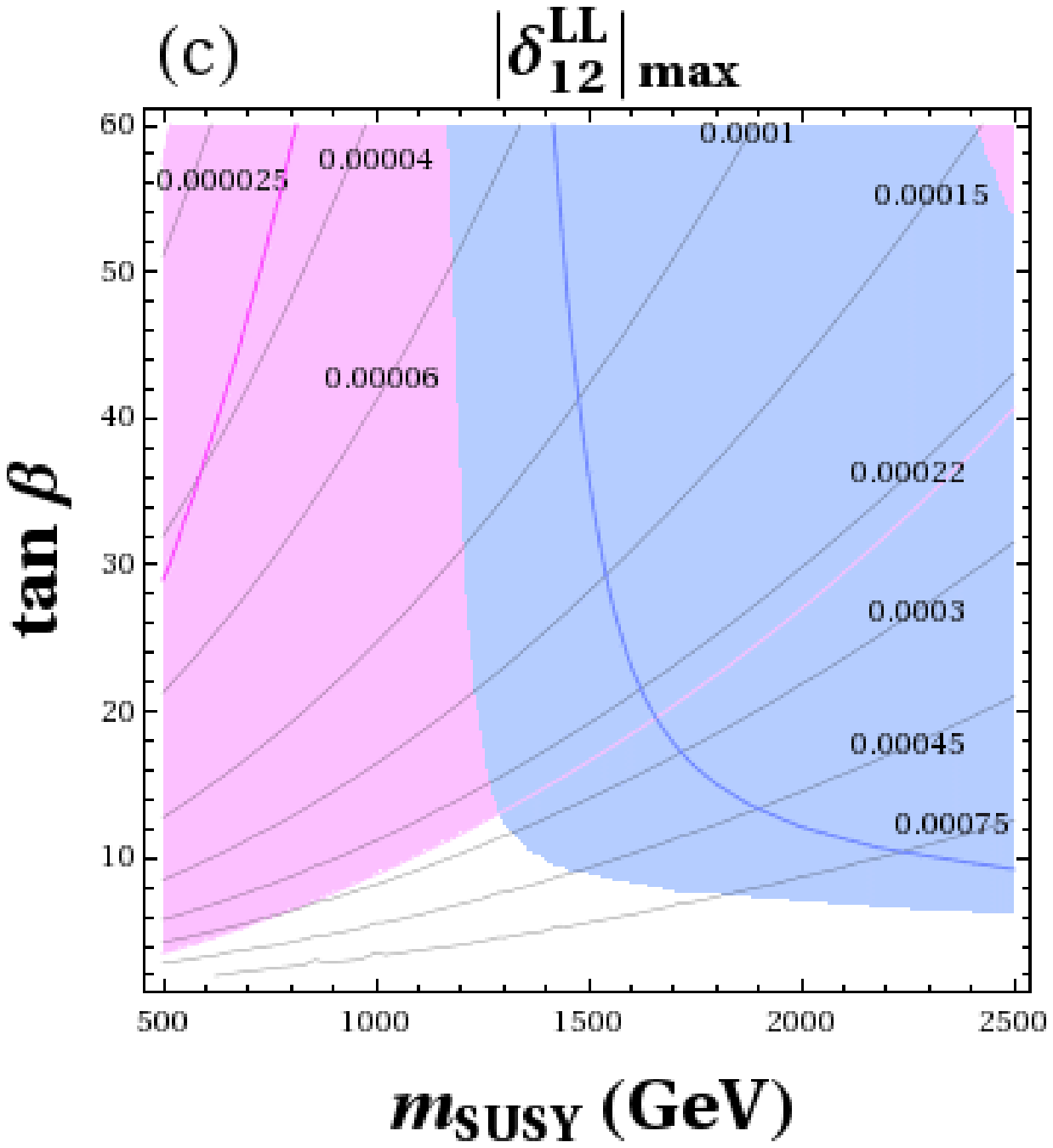,scale=0.60,clip=}
\psfig{file=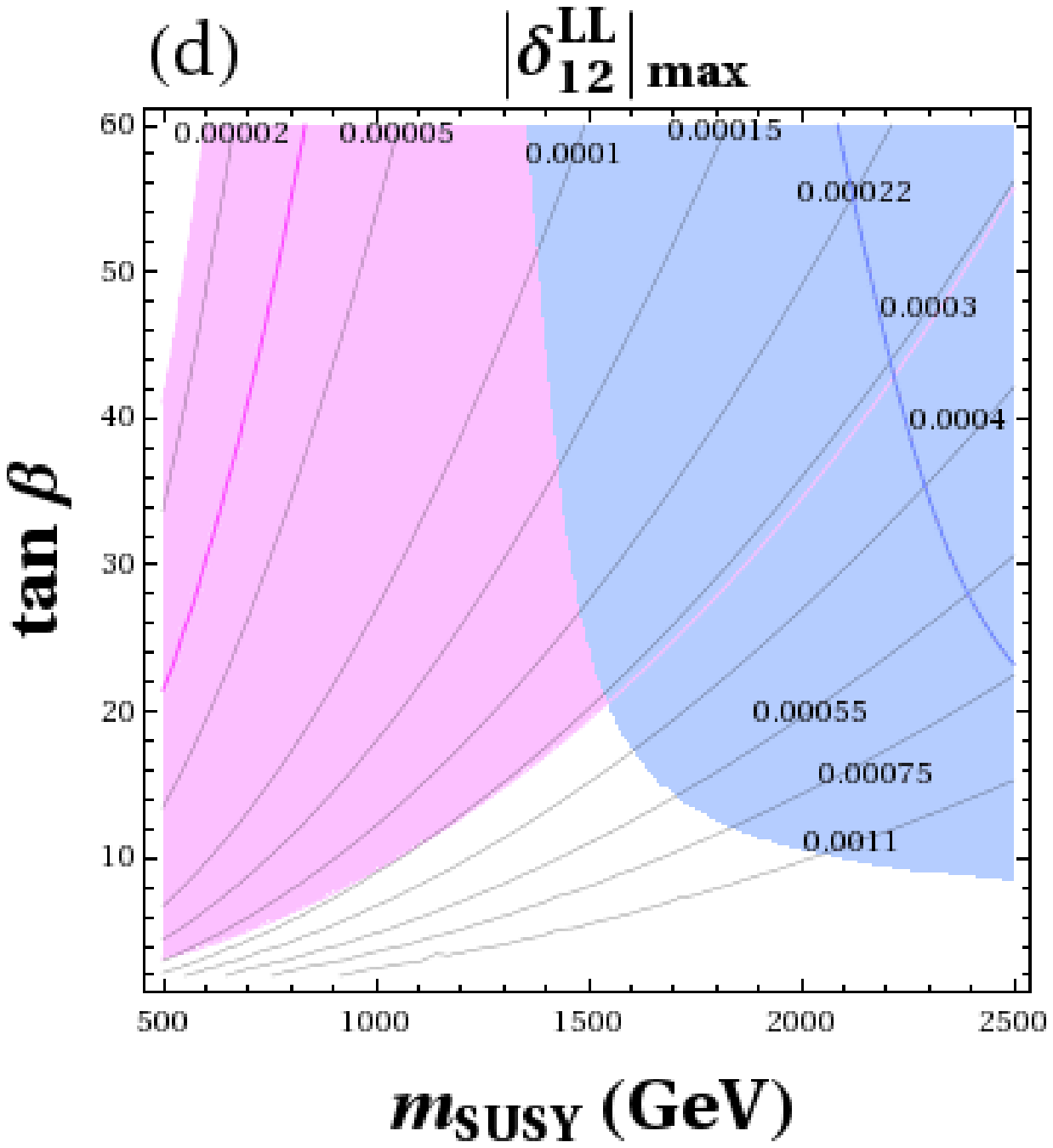,scale=0.60,clip=}
\end{center}
\caption{ Contourlines in the 
($m_{\rm SUSY}$, $\tb$) plane of maximum slepton mixing
  $|\delta_{12}^{LL}|_{\rm max}$ that are allowed by LFV searches in
  $\mu \to e \gamma$ for scenarios {\bf (a)}, {\bf (b)}, {\bf (c)} and
  {\bf (d)} of our framework 2, defined in section 3.4.2. The shaded
  areas in pink are the regions leading to a $(g-2)_\mu^{\rm SUSY}$
  prediction in the $(3.2,57.2) \times 10^{-10}$ interval. The interior
  pink contourline (without number) corresponds to setting $(g-2)_\mu^{\rm SUSY}$ exactly
  at the central value of the discrepancy $(g-2)_\mu^{\rm
    exp}-(g-2)_\mu^{\rm SM}=30.2 \times 10^{-10}$ . 
The shaded overimposed areas in blue are the regions leading to a
$\Mh$ prediction within the  $(123,127)\gev$ interval. The
interior blue contourline (without number) corresponds to the particular 
$\Mh=125\gev$ value.}
\label{msusytb-LL12}
\vspace{1em}
\end{figure}

\begin{figure}[ht!]
\vspace{4em}
\begin{center}
\psfig{file=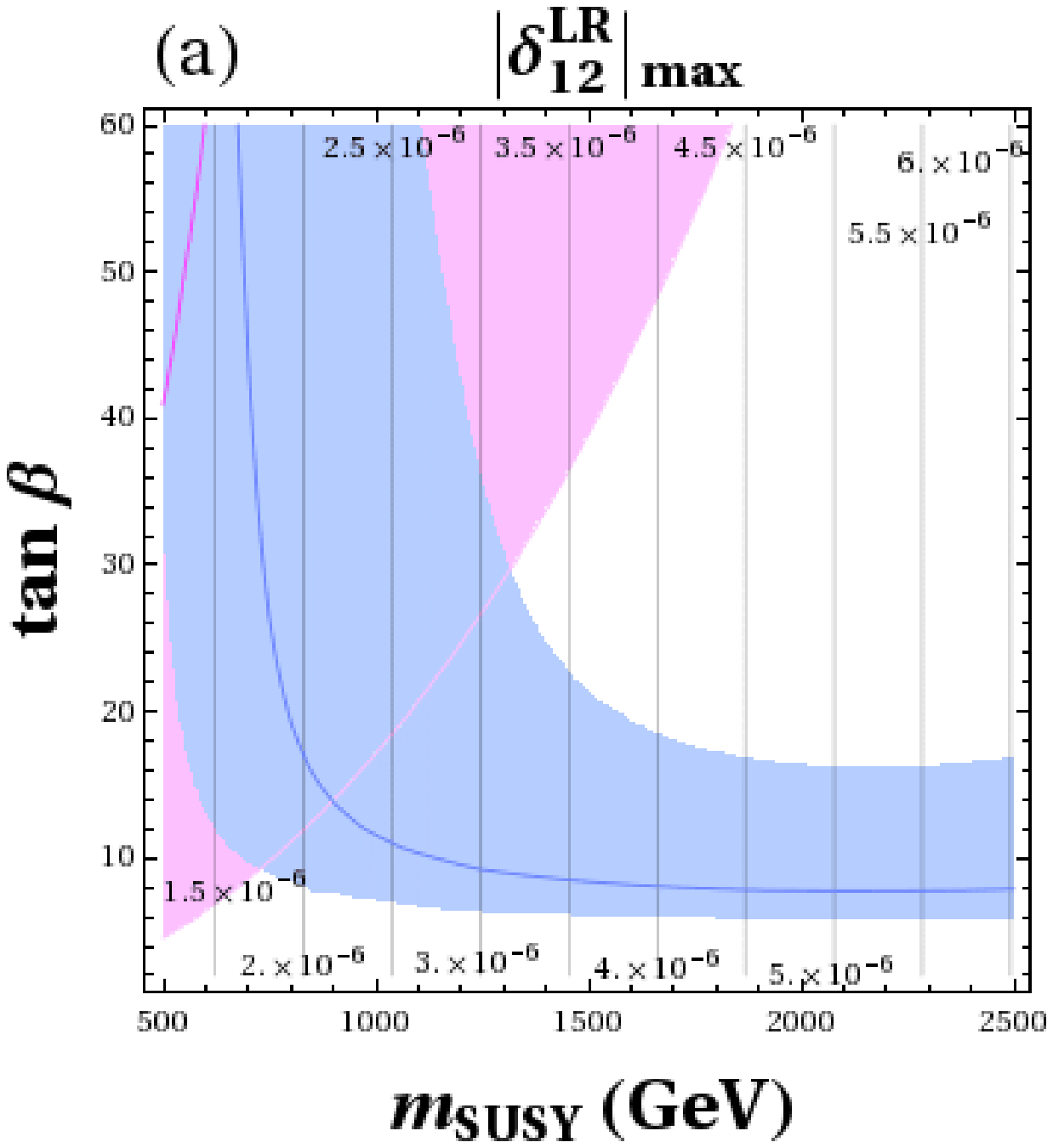,scale=0.60,clip=}
\psfig{file=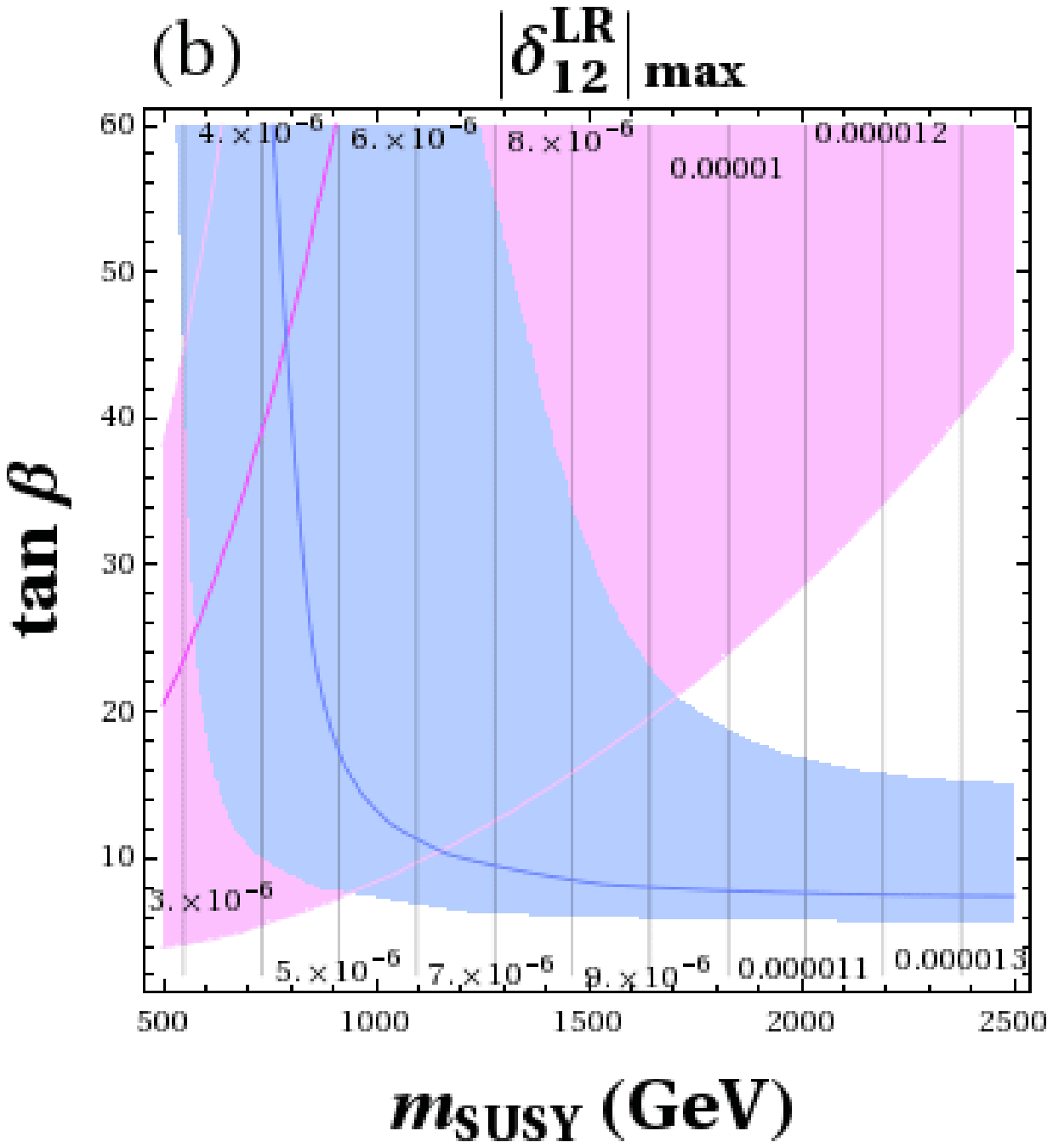,scale=0.60,clip=}\\
\psfig{file=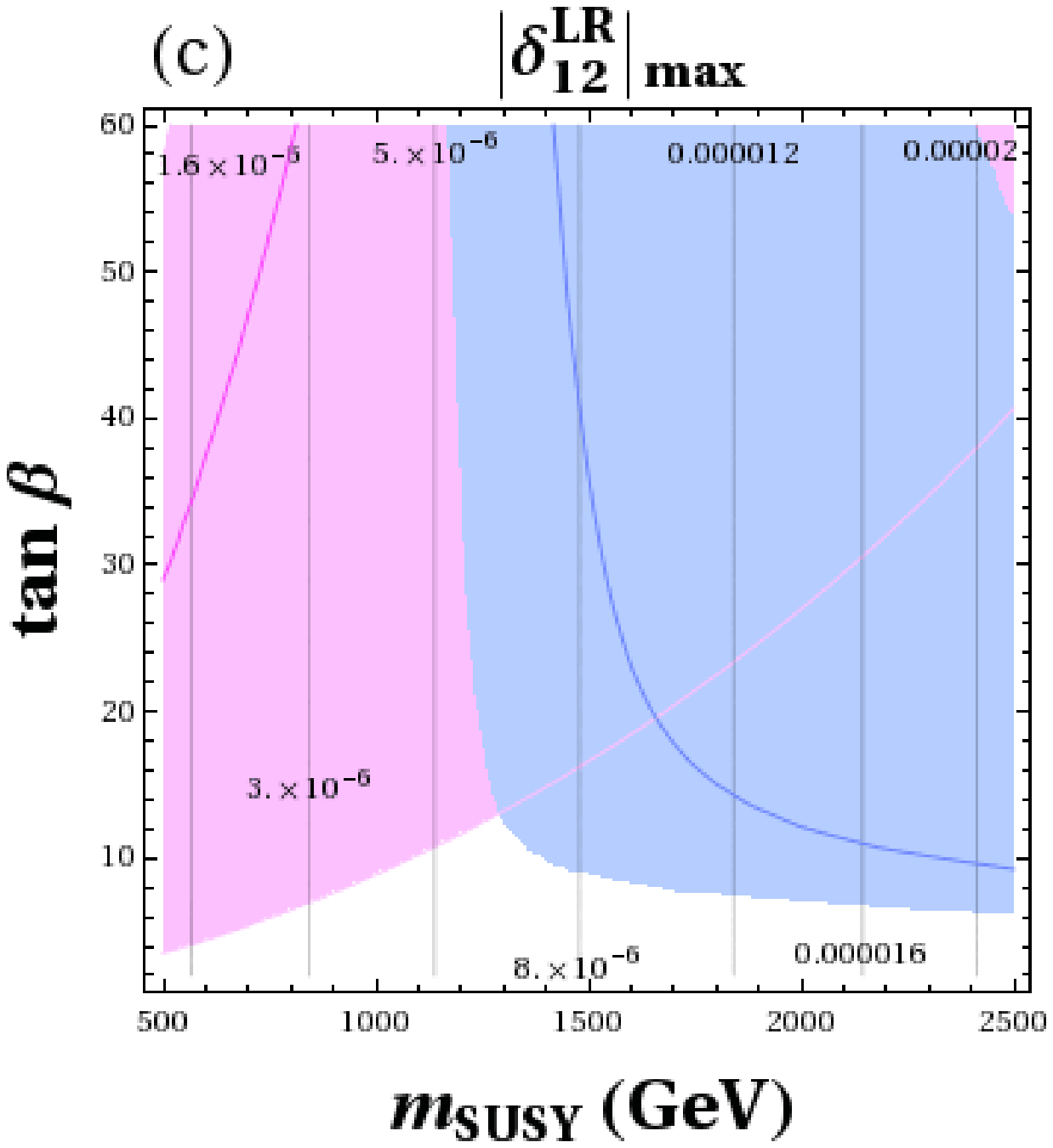,scale=0.60,clip=}
\psfig{file=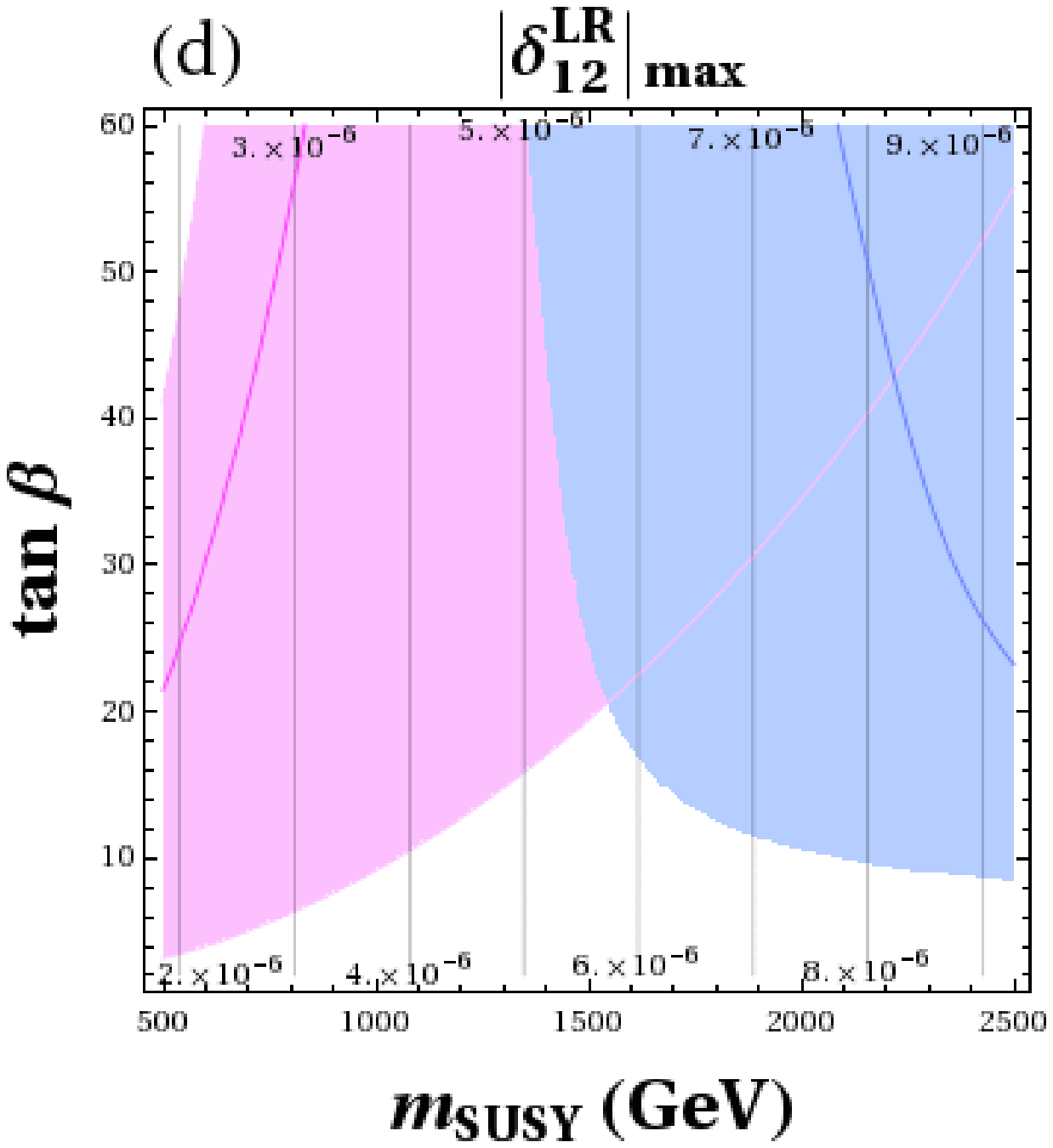,scale=0.60,clip=}
\end{center}
\caption{ Contourlines in the 
($m_{\rm SUSY}$, $\tb$) plane of maximum slepton mixing
  $|\delta_{12}^{LR}|_{\rm max}$ that are allowed by LFV searches in
  $\mu \to e \gamma$. All inputs and explanations are as in
  \reffi{msusytb-LL12}.} 
\label{msusytb-LR12}
\vspace{4em}
\end{figure}

\begin{figure}[ht!]
\vspace{4em}
\begin{center}
\psfig{file=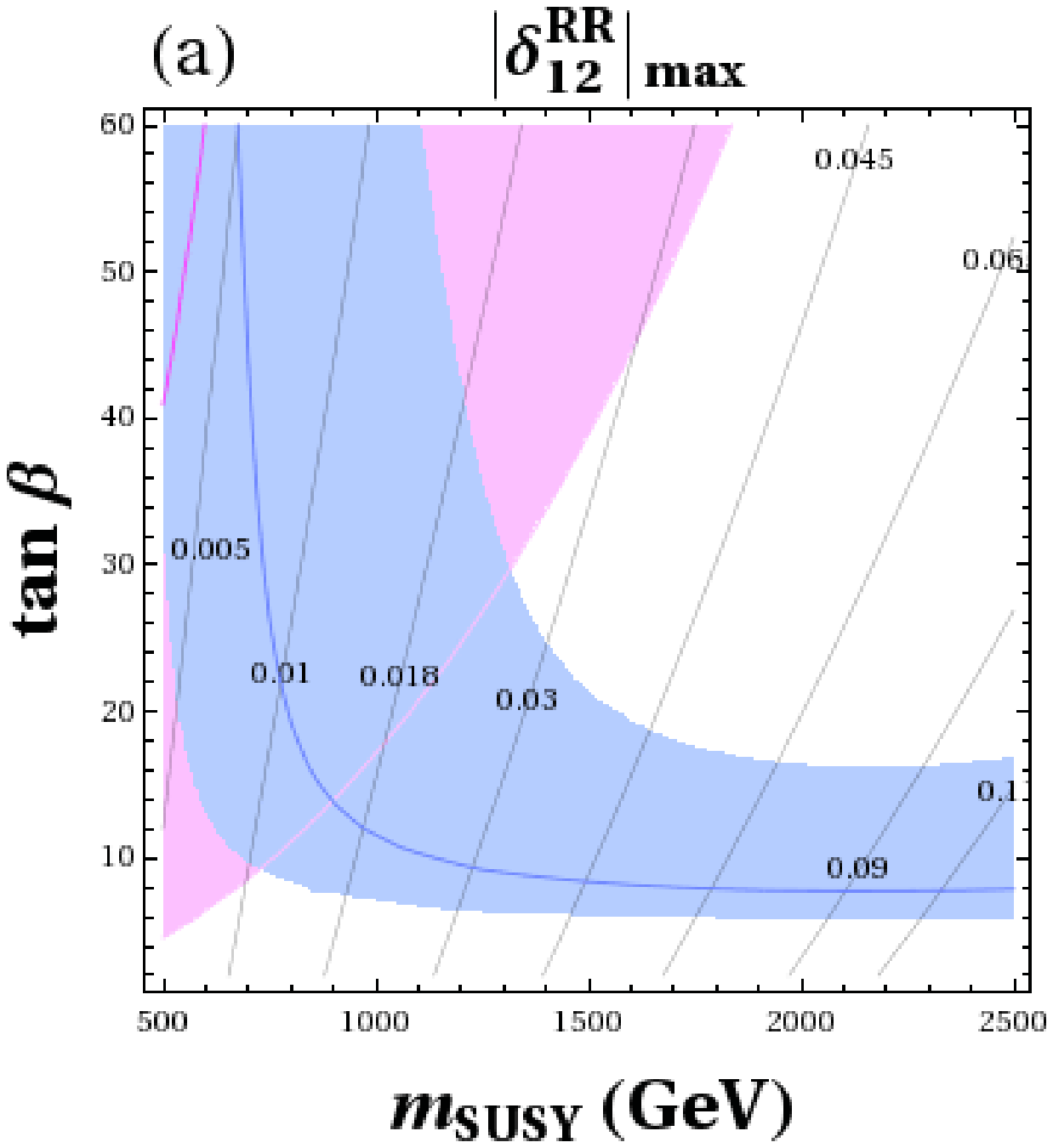,scale=0.60,clip=}
\psfig{file=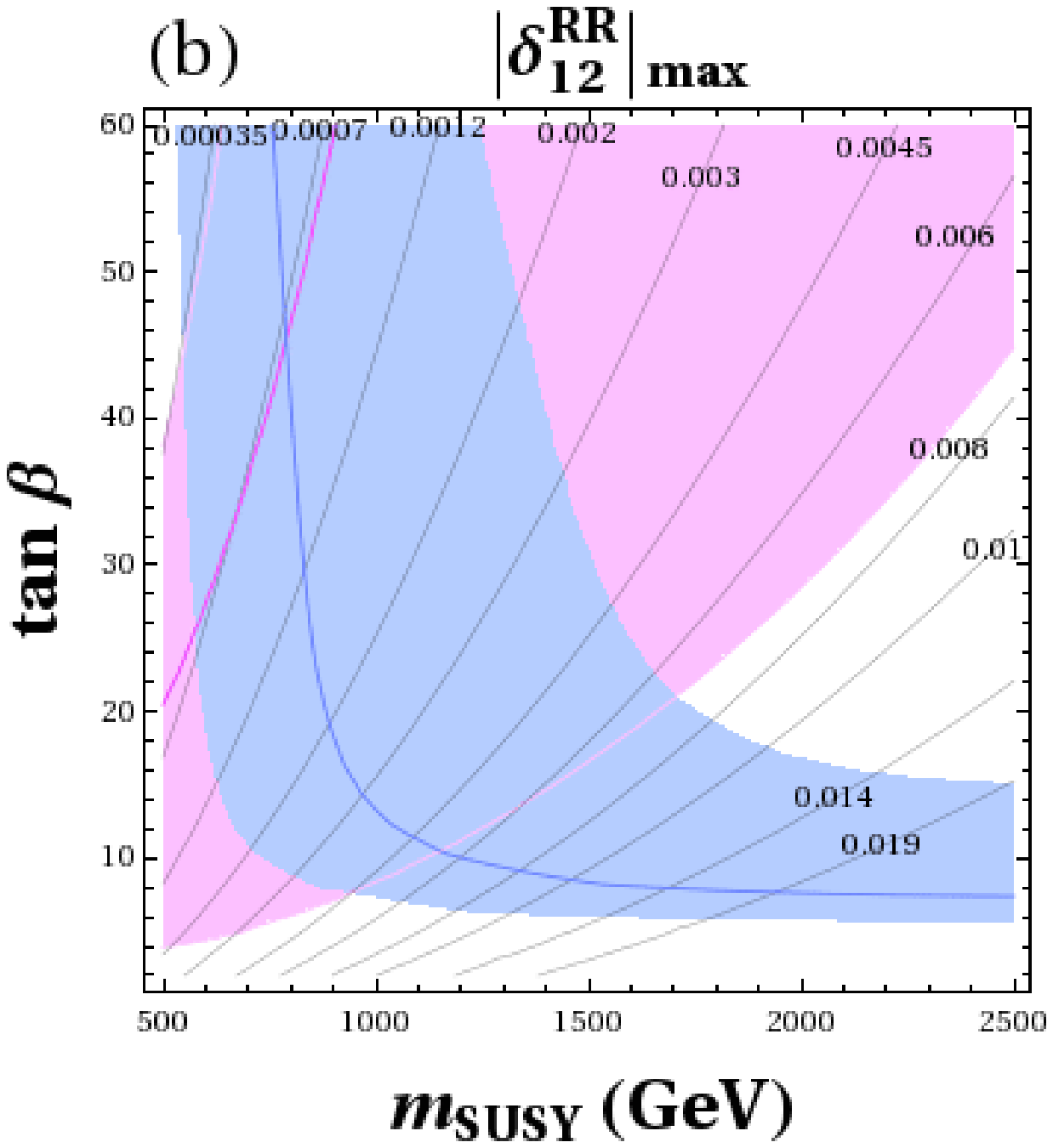,scale=0.60,clip=}\\
\psfig{file=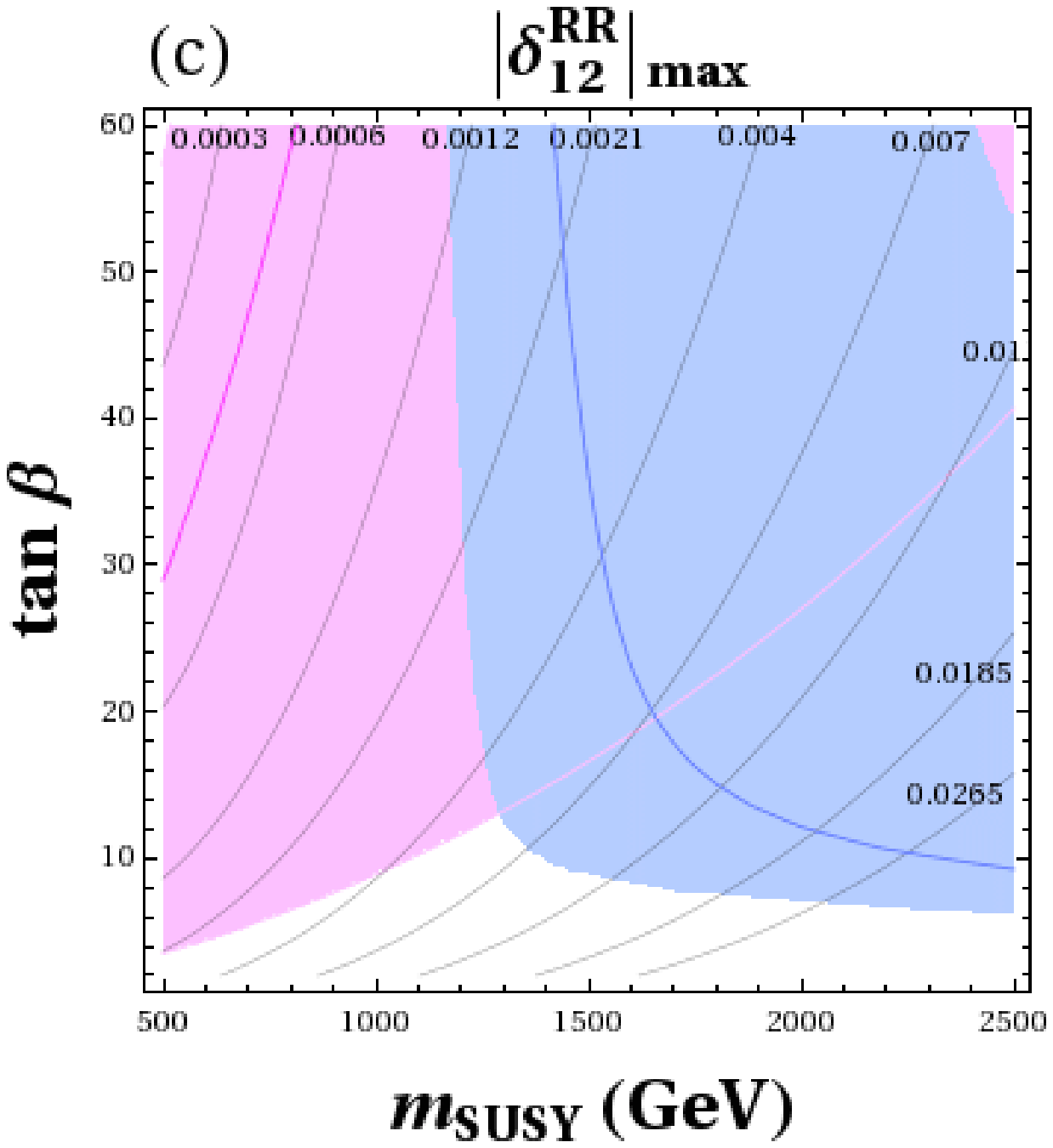,scale=0.60,clip=}
\psfig{file=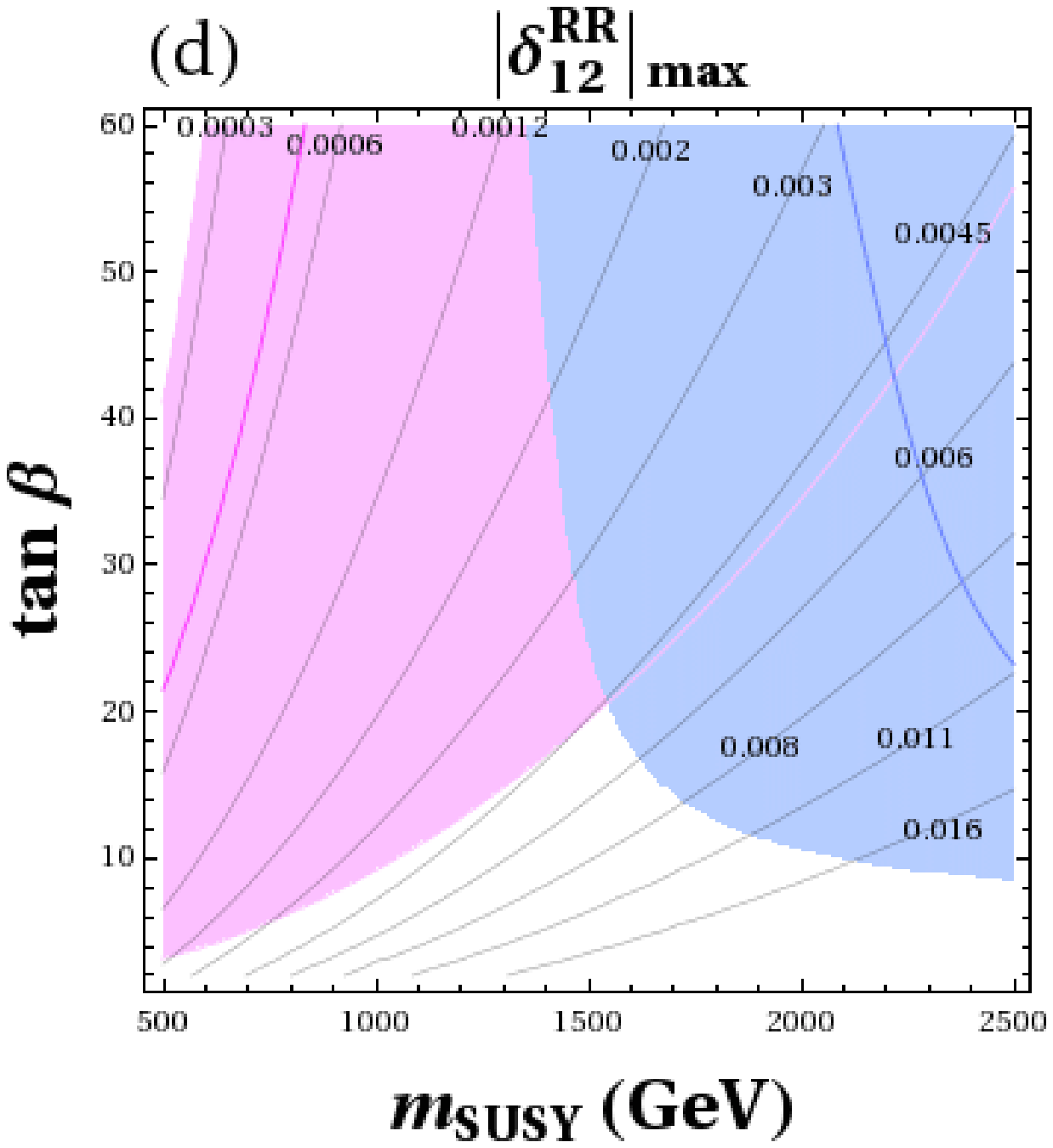 ,scale=0.60,clip=}
\end{center}
\caption{ Contourlines in the 
($m_{\rm SUSY}$, $\tb$) plane of maximum slepton mixing
  $|\delta_{12}^{RR}|_{\rm max}$ that are allowed by LFV searches in
  $\mu \to e \gamma$. All inputs and explanations are as in
  \reffi{msusytb-LL12}.} 
\label{msusytb-RR12}
\vspace{4em}
\end{figure}

\begin{figure}[ht!]
\vspace{3em}
\begin{center}
\psfig{file=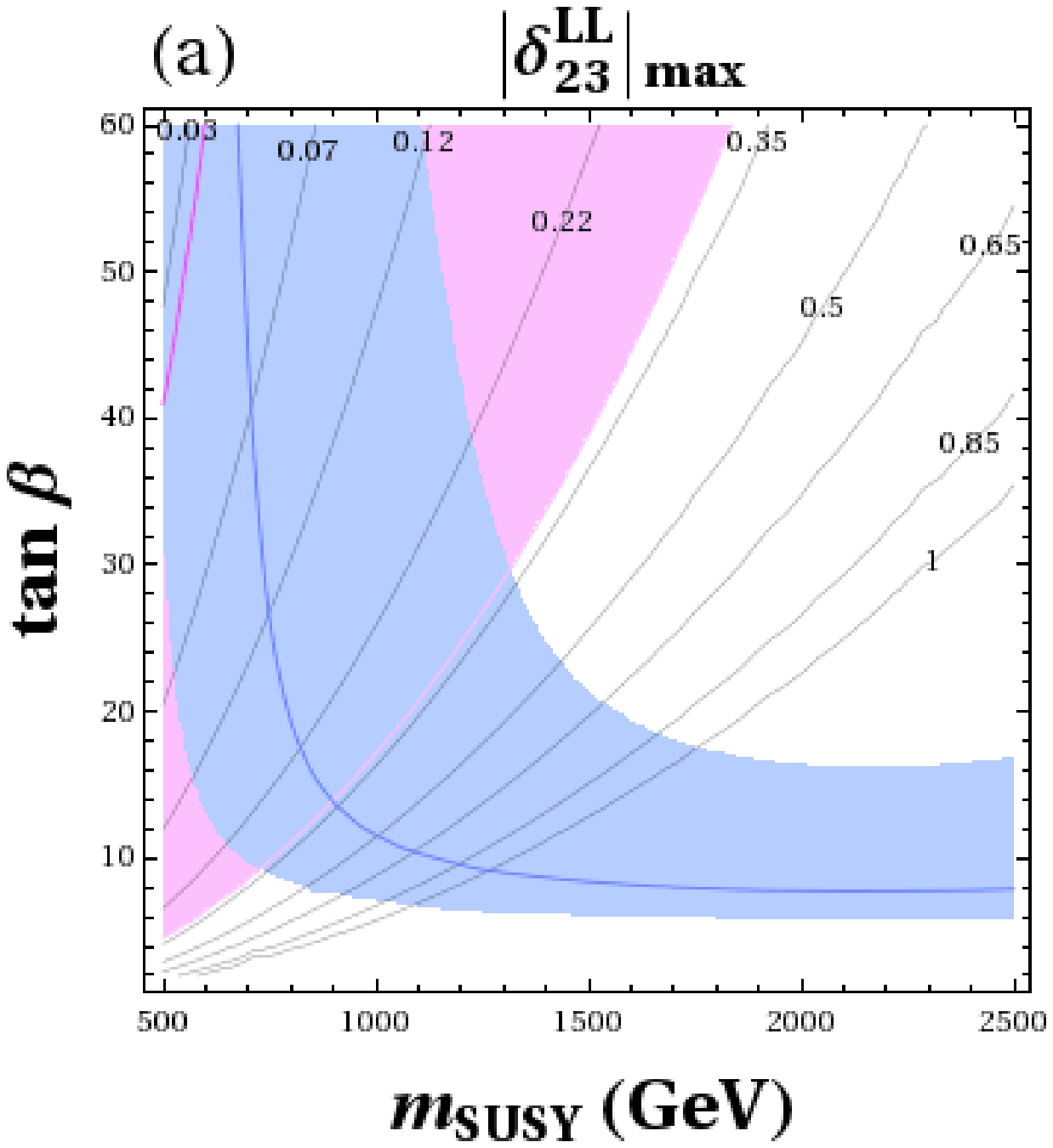,scale=0.60,clip=}
\psfig{file=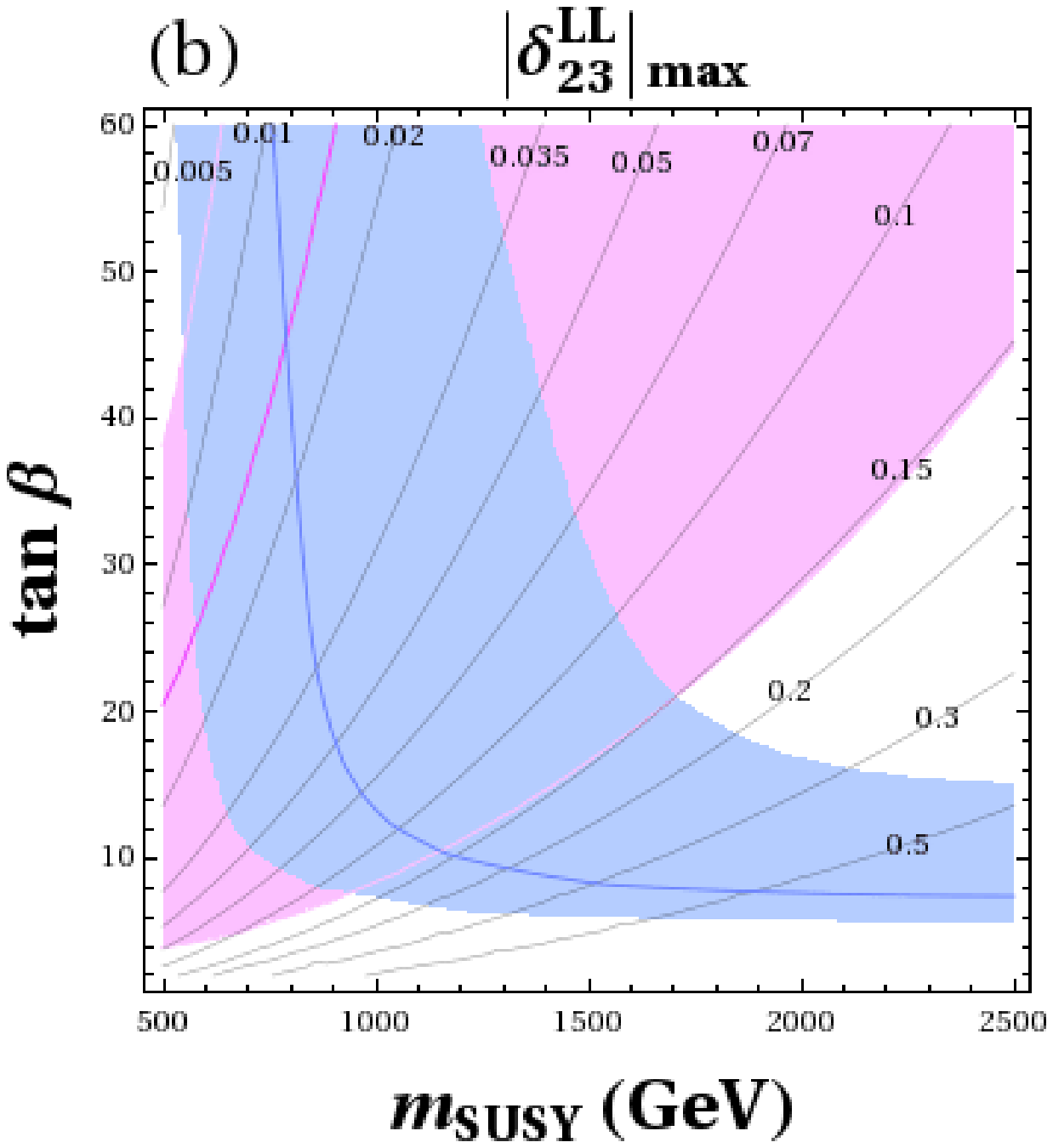,scale=0.60,clip=}\\
\psfig{file=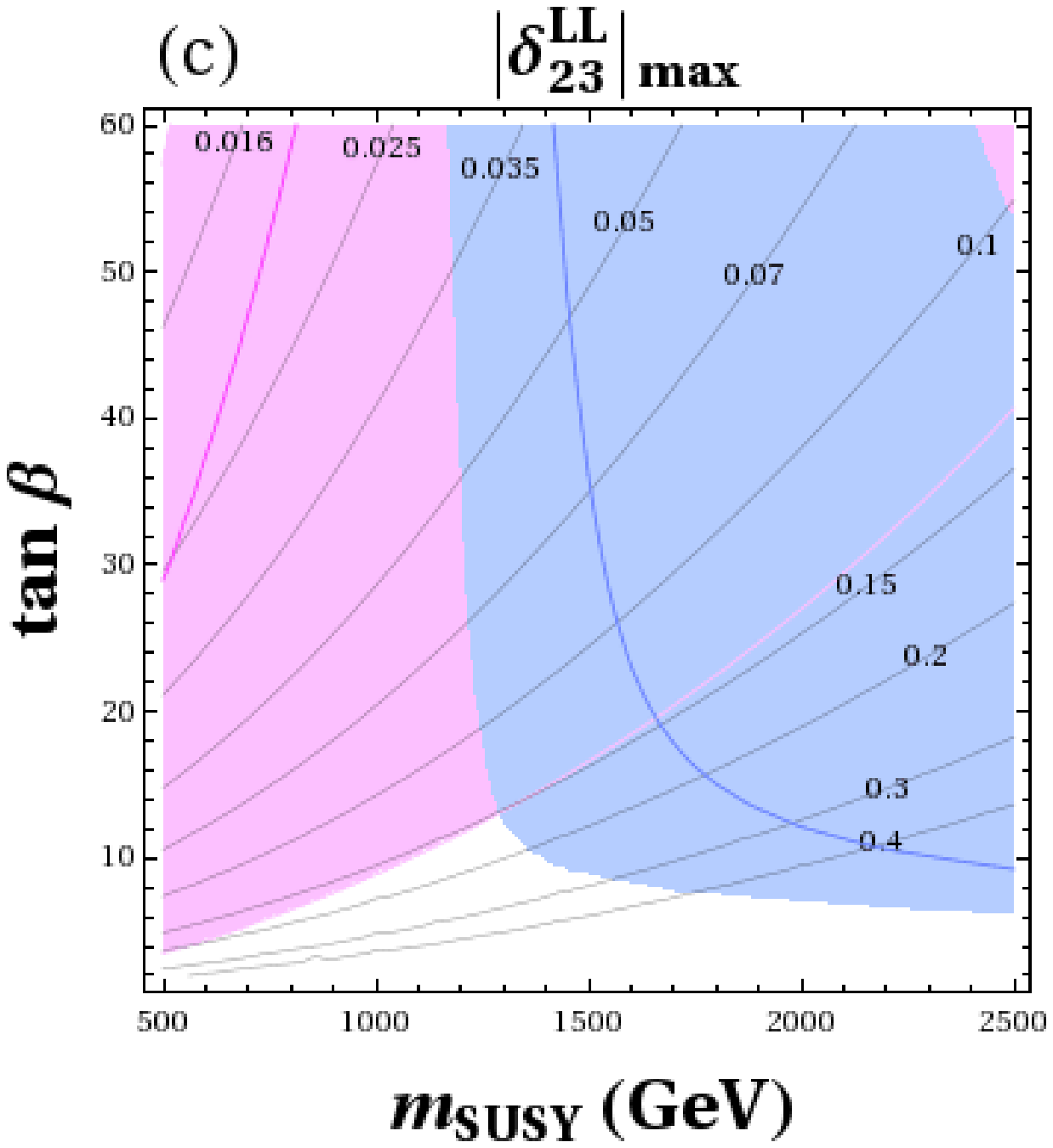,scale=0.60,clip=}
\psfig{file=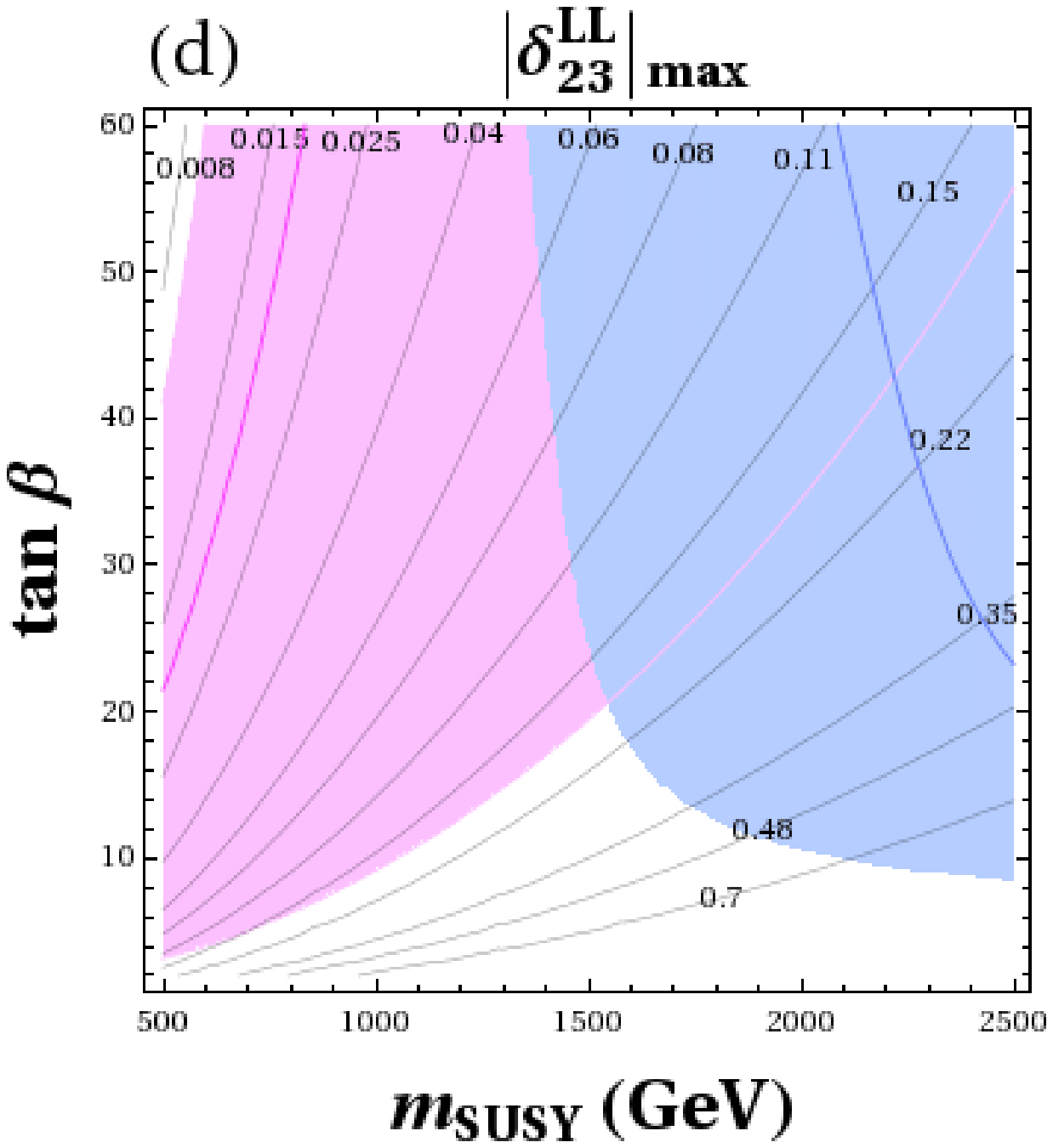 ,scale=0.60,clip=}
\end{center}
\caption{ Contourlines in the 
($m_{\rm SUSY}$, $\tb$) plane of maximum slepton mixing
  $|\delta_{23}^{LL}|_{\rm max}$ that are allowed by LFV searches in
  $\tau \to \mu \gamma$. All inputs and explanations are as in
  \reffi{msusytb-LL12}. Similar results/plots (not shown) are
    obtained for
  contourlines of maximum slepton mixing  $|\delta_{13}^{LL}|_{\rm max}$
  that are allowed by LFV searches in $\tau \to e \gamma$.} 
\label{msusytb-LL23}
\vspace{3em}
\end{figure}

\begin{figure}[ht!]
\vspace{3em}
\begin{center}
\psfig{file=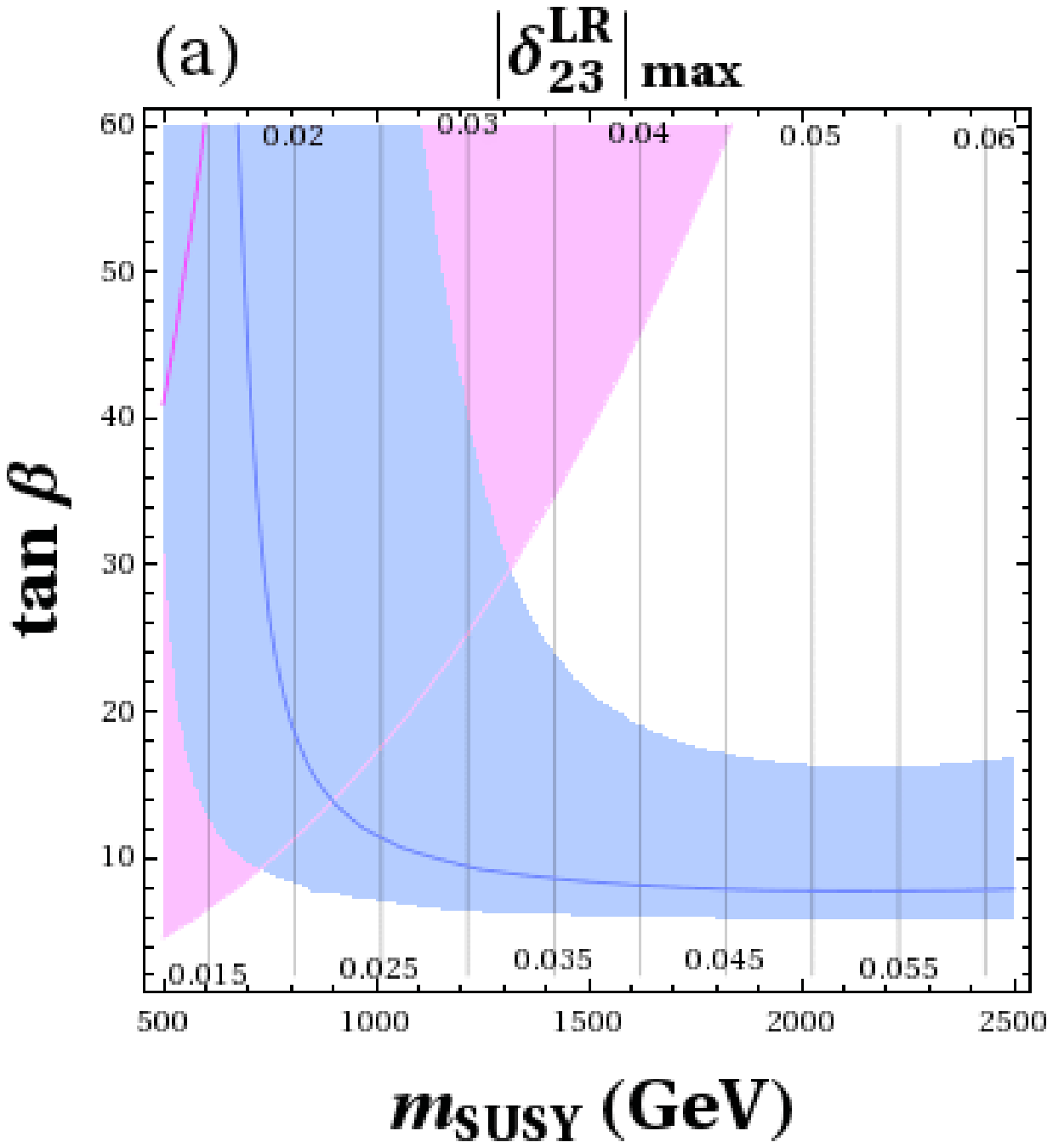,scale=0.60,clip=}
\psfig{file=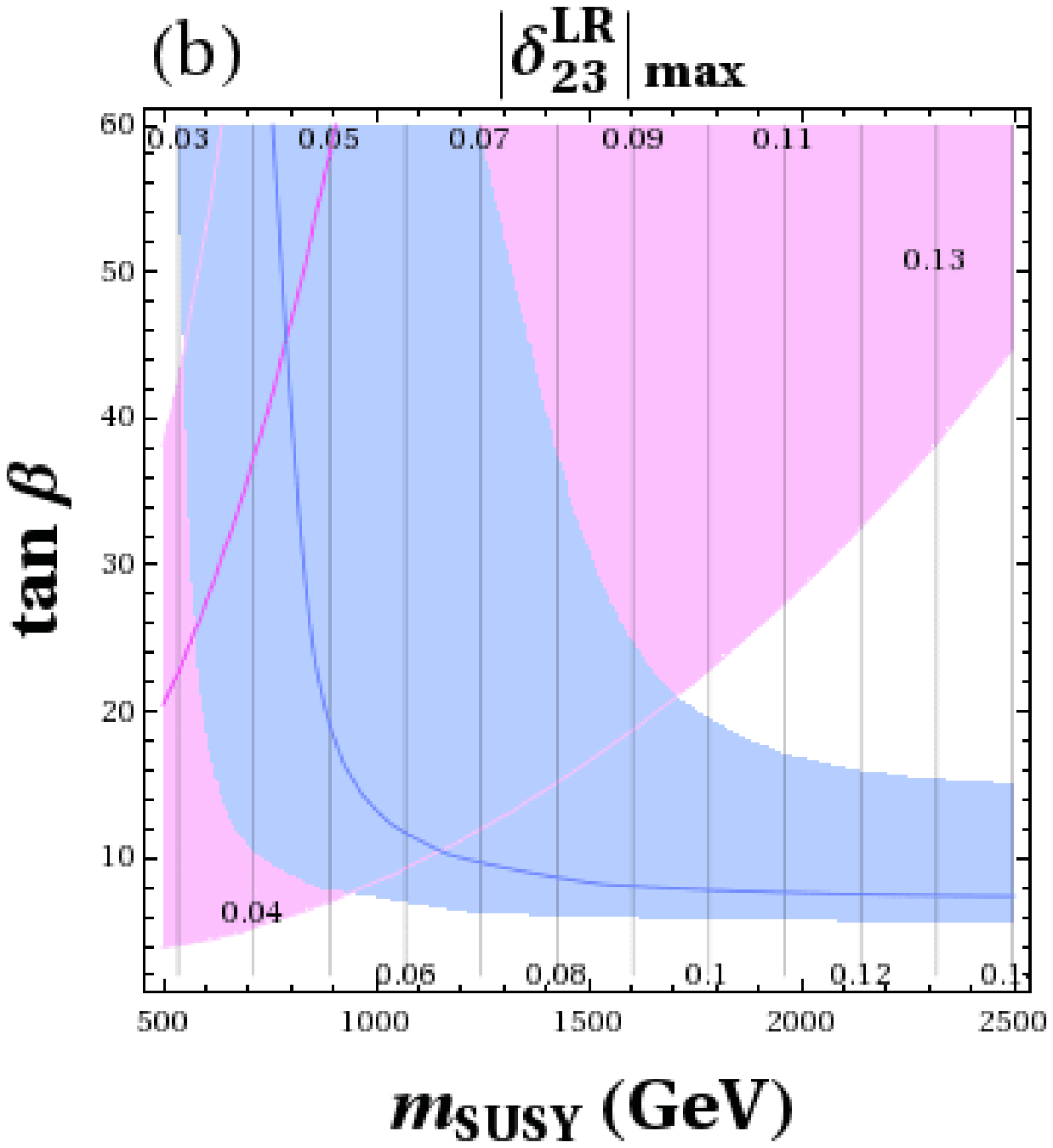,scale=0.60,clip=}\\
\psfig{file=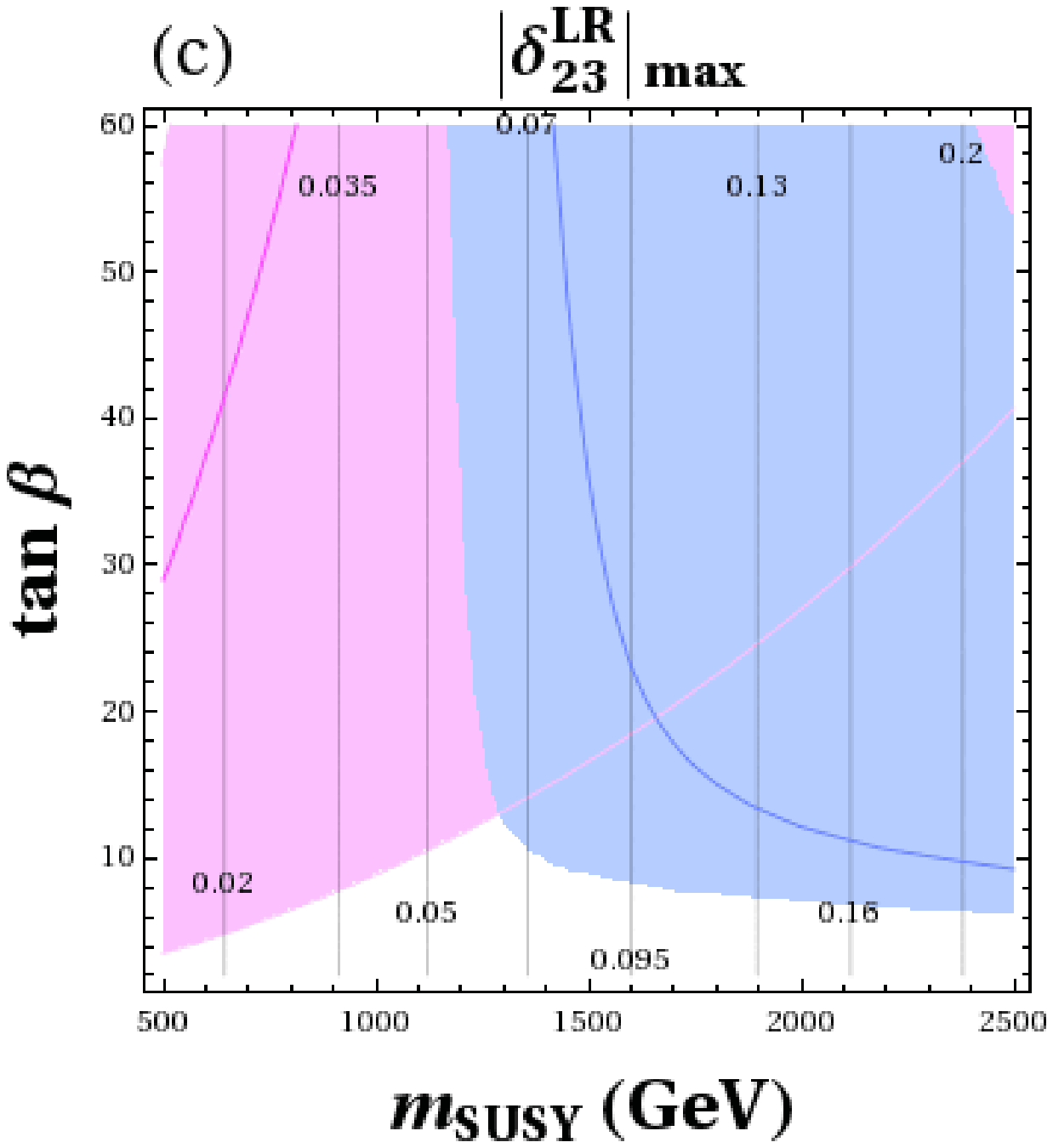,scale=0.60,clip=}
\psfig{file=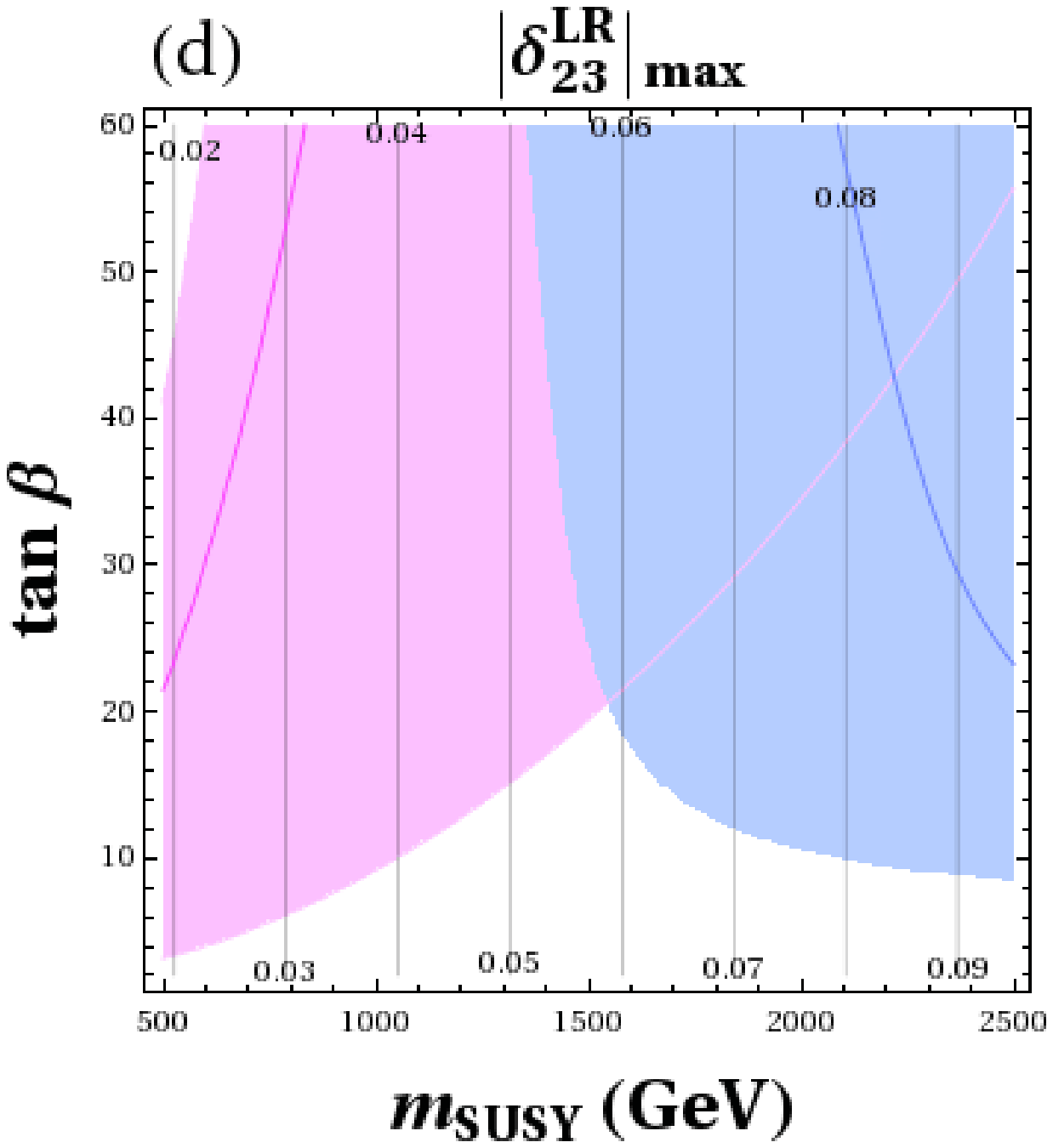,scale=0.60,clip=}
\end{center}
\caption{ Contourlines in the 
($m_{\rm SUSY}$, $\tb$) plane of maximum slepton mixing
  $|\delta_{23}^{LR}|_{\rm max}$ that are allowed by LFV searches in
  $\tau \to \mu \gamma$. All inputs and explanations are as in
  \reffi{msusytb-LL12}. Similar results/plots (not shown) are
    obtained for
  contourlines of maximum slepton mixing  $|\delta_{13}^{LR}|_{\rm max}$
  that are allowed by LFV searches in $\tau \to e \gamma$.} 
\label{msusytb-LR23}
\vspace{3em}
\end{figure}

\begin{figure}[ht!]
\vspace{3em}
\begin{center}
\psfig{file=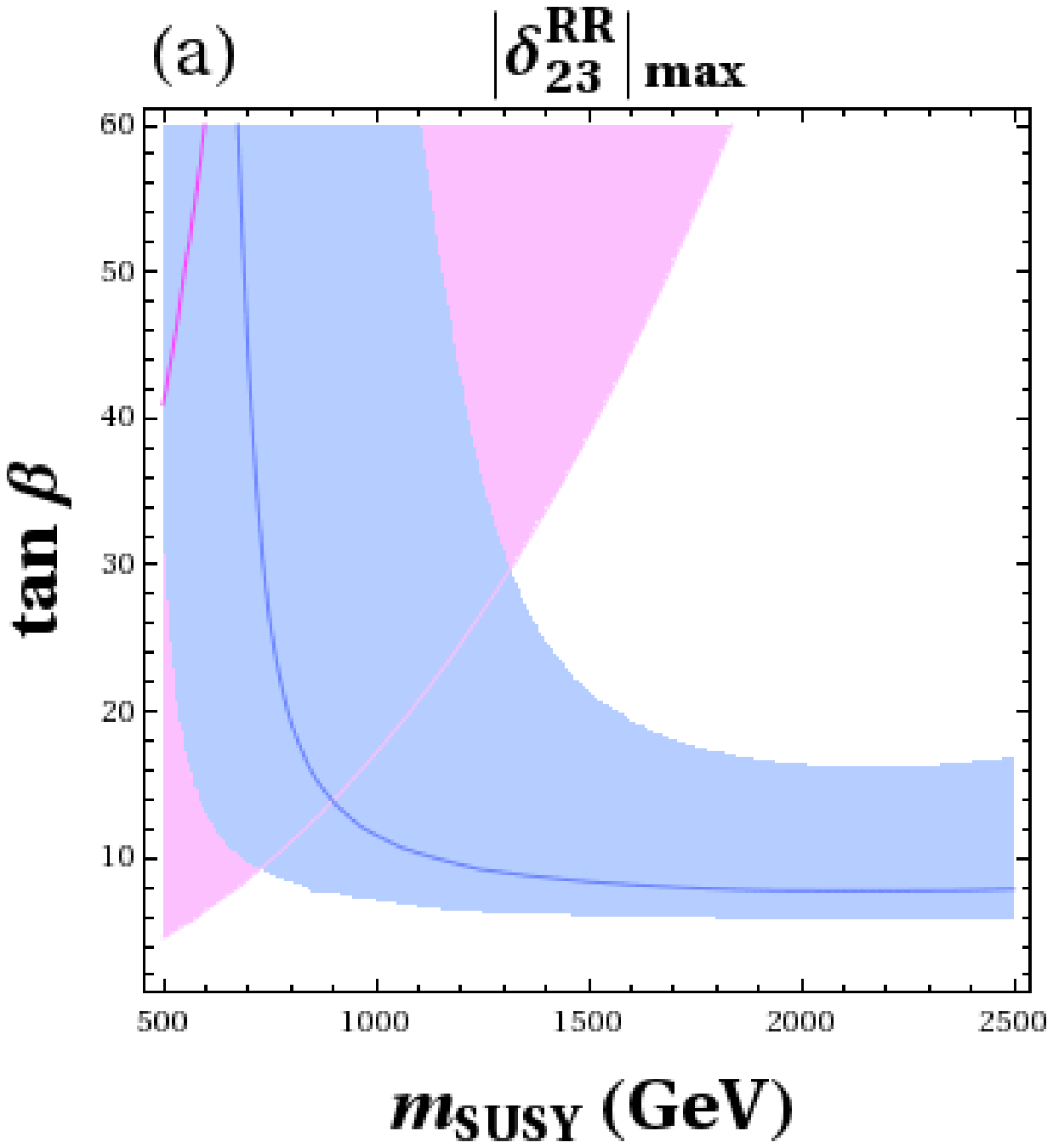,scale=0.60,clip=}
\psfig{file=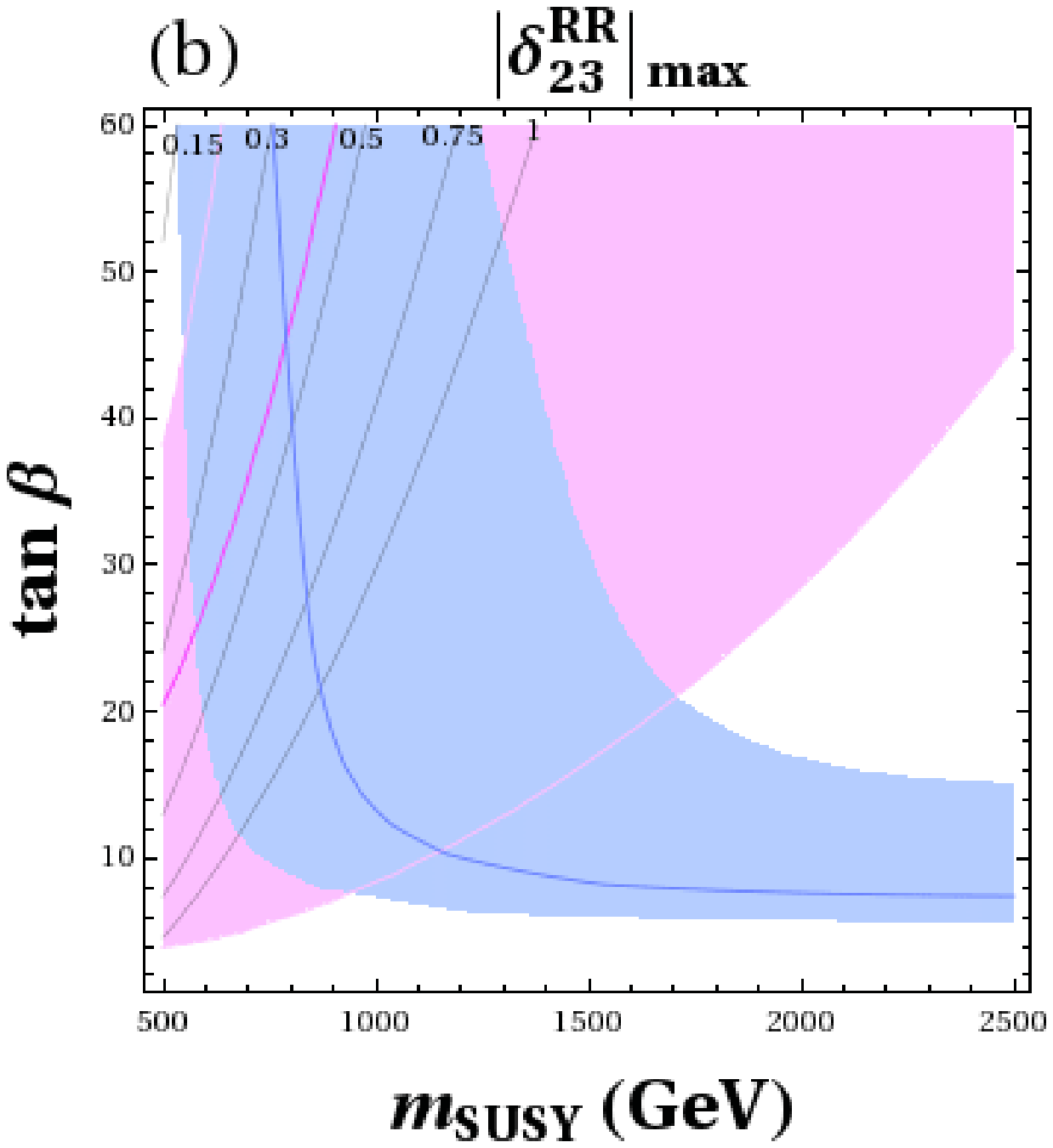,scale=0.60,clip=}\\
\psfig{file=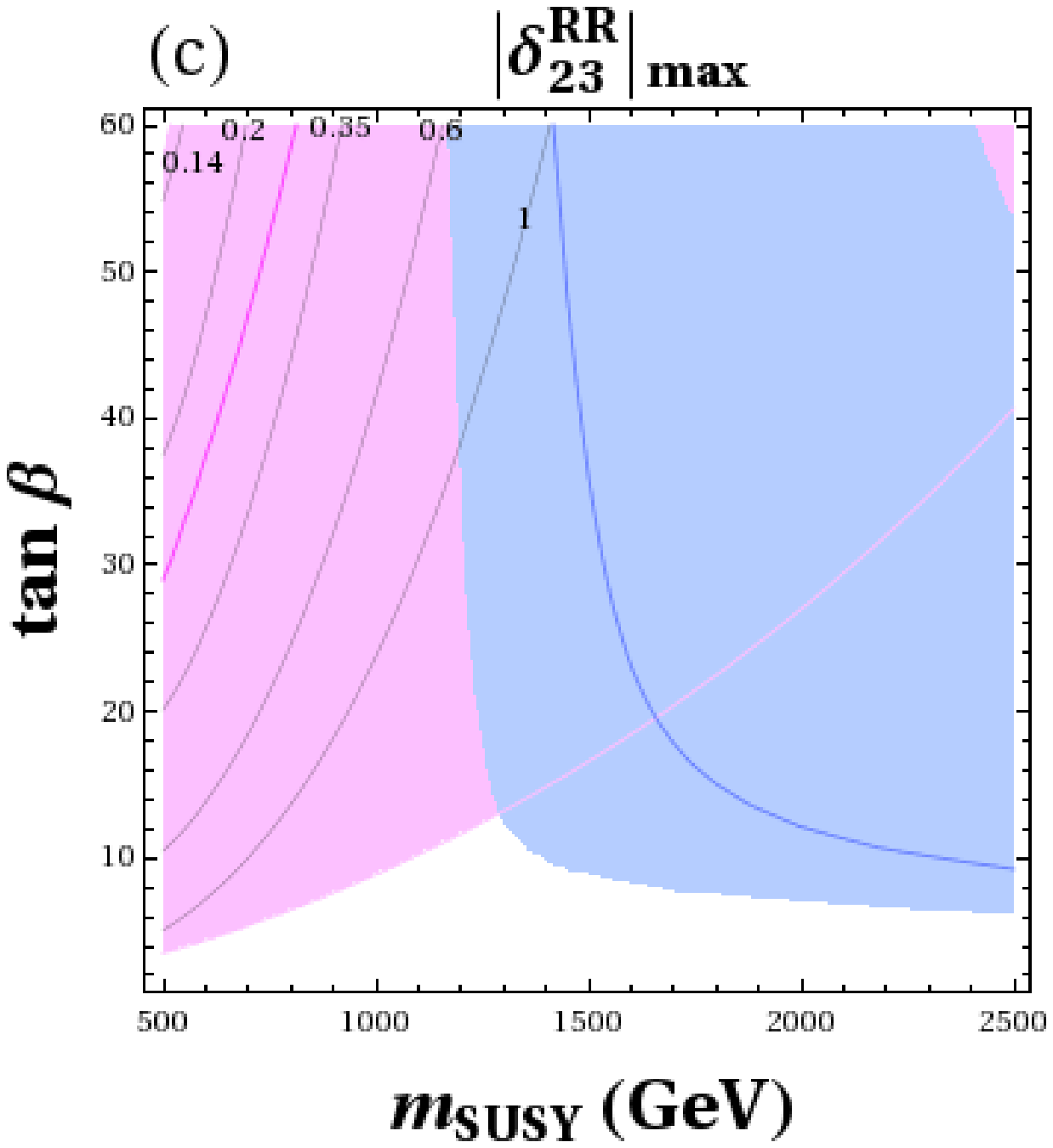,scale=0.60,clip=}
\psfig{file=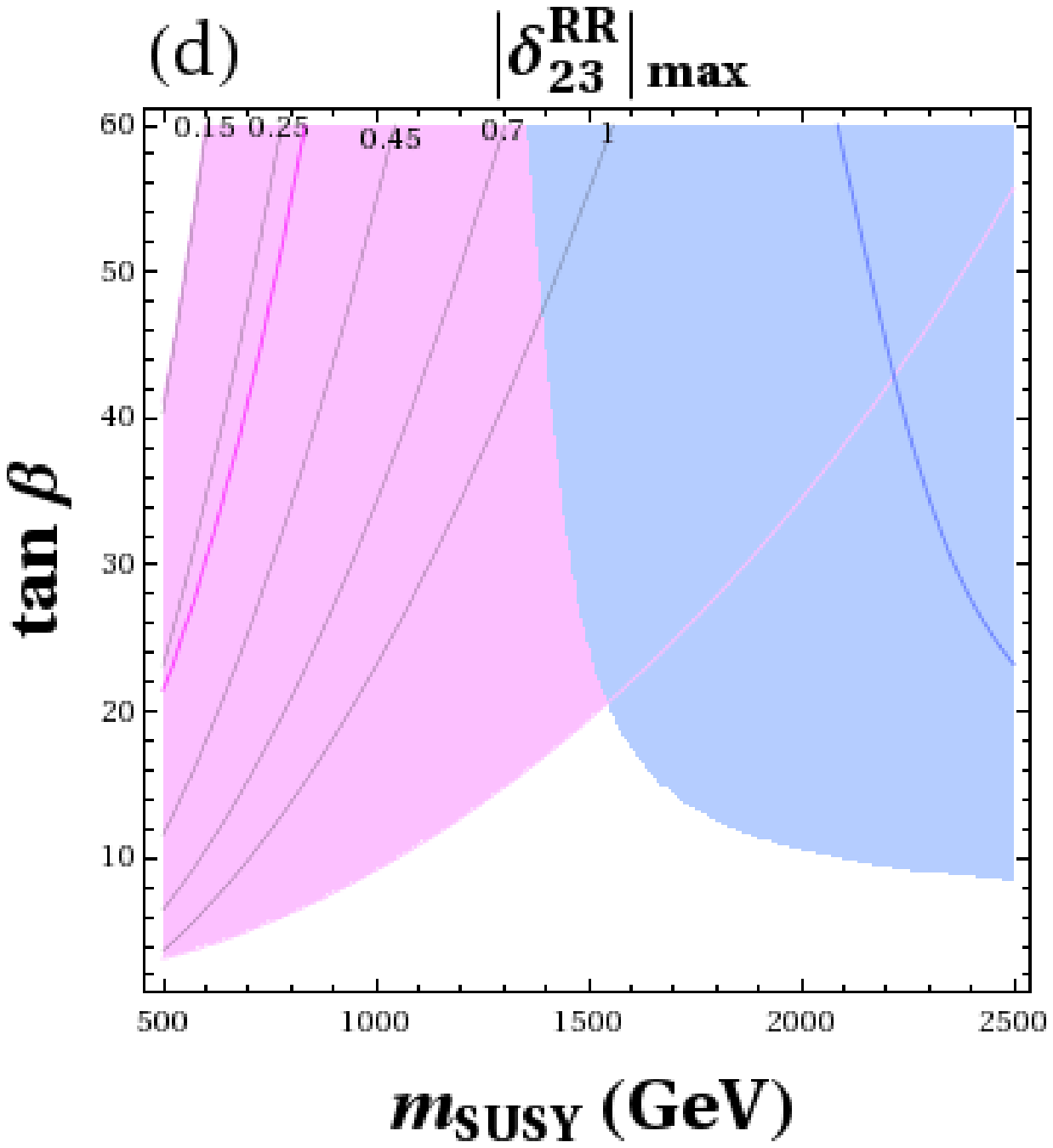,scale=0.60,clip=}
\end{center}
\caption{ Contourlines in the 
($m_{\rm SUSY}$, $\tb$) plane of maximum slepton mixing
  $|\delta_{23}^{RR}|_{\rm max}$ that are allowed by LFV searches in
  $\tau \to \mu \gamma$. All inputs and explanations are as in
  \reffi{msusytb-LL12}. Notice that only contourlines with $|\delta_{23}^{RR}|_{\rm max} \leq 1$ are included. In the scenario {\bf (a)} 
the countourlines with $|\delta_{23}^{RR}|_{\rm max} \leq 1$ are out of the region in the parameter space shown in this figure. Similar 
results/plots (not shown) are obtained for
  contourlines of maximum slepton mixing  $|\delta_{13}^{RR}|_{\rm max}$
  that are allowed by LFV searches in $\tau \to e \gamma$.} 
\label{msusytb-RR23}
\end{figure}

We present the numerical results of our analysis in framework 2 that are
shown in \reffis{msusytb-LL12} through 
\ref{msusytb-RR23}. Figures \ref{msusytb-LL12}, \ref{msusytb-LR12} and
\ref{msusytb-RR12} show the bounds for the slepton mixing of 12-type as
extracted from present $\mu \to e \gamma$ searches. Figures
\ref{msusytb-LL23}, \ref{msusytb-LR23} and \ref{msusytb-RR23} show the
bounds for the slepton mixing of 23-type as extracted from present $\tau
\to \mu \gamma$ searches. It should be noted that the bounds for
the slepton mixings of 13-type (not shown here) are equal (in the MIA)
to those of 23-type.  
In each plot we show the resulting contourlines in the 
($m_{\rm SUSY}$, $\tb$) plane of maximum allowed slepton mixing. 
In addition we also show in each plot the areas in the
pMSSM-4 parameter space for that particular scenario that lead 
to values of the lightest Higgs boson mass compatible with LHC data, and
at the same time to predictions of the muon anomalous magnetic
moment also compatible with 
data. As in the previous framework~1, we use here again
\fh~\cite{feynhiggs}  to evaluate 
$\Mh$ and {\it SPHENO}~\cite{Porod:2003um} to evaluate  
$(g-2)_\mu$ (where \fh\ gives very similar results). 
The shaded areas in pink are the regions leading to a
$(g-2)_\mu^{\rm SUSY}$ prediction, from the SUSY one-loop contributions,
in the allowed by data $(3.2,57.2) \times 10^{-10}$ interval. The
interior pink contourline corresponds to setting $(g-2)_\mu^{\rm SUSY}$
exactly at the central value of the discrepancy $(g-2)_\mu^{\rm
  exp}-(g-2)_\mu^{\rm SM}=30.2 \times 10^{-10}$. 
The shaded overimposed areas in blue are the regions leading to a $\Mh$
prediction within the  $(123,127)\gev$ interval. The interior blue
contourline corresponds to the particular $\Mh=125\gev$ value.  

From these plots in the ($m_{\rm SUSY}$, $\tb$) plane one can draw
the following conclusions:
\begin{itemize}
\item[1.-]  For each scenario (a), (b), (c) and (d) one can derive the
  corresponding upper bound for each $|\delta^{AB}_{ij}|$ at a given
  ($m_{\rm SUSY}$, $\tb$) point in this plane.  
\item[2.-] The maximal allowed values of the $\delta^{LL}_{ij}$'s and
  $\delta^{RR}_{ij}$'s scale with $m_{\rm SUSY}$ and $\tb$ approximately
  as expected, growing with increasing $m_{\rm SUSY}$ as $\sim m_{\rm
    SUSY}^2$ and decreasing with increasing (large) $\tb$ as 
    $\sim 1/\tb$. The maximal allowed values of the $\delta^{LR}_{ij}$'s
  (and similarly $\delta^{RL}_{ij}$'s) are independent on $\tb$ and grow
  approximately as $\sim m_{\rm SUSY}$  with increasing $m_{\rm
    SUSY}$. This is in agreement with the qualitative behavior 
found in the approximation formulas, 
  \refeqs{MIA-L} and (\ref{MIA-R}) of the MIA results in the simplest case
  of only one mass scale, $m_S$.  
\item[3.-] The intersections between the allowed areas by the required
  $(g-2)_\mu$ and $\Mh$ intervals move from the left side, $m_{\rm
  SUSY}\sim 500-1300 \gev$ to the right side of the plots,
  $m_{\rm SUSY}\sim 1300-2500 \gev$ from scenarios (a) through
  (d). This is clearly the consequence of the fact that
$(g-2)_\mu$ requires a rather light SUSY-EW sector, i.e.\
light charginos, neutralinos and sleptons, and a rather large $\tb$, and
that $\Mh$ requires a rather heavy SUSY squark sector.
Here we are using a common reference SUSY scale  $m_{\rm SUSY}$,
relating all 
the SUSY sparticle masses, both in the SUSY-EW and SUSY-QCD sectors, 
leading to this ``tension''. (A more lose connection between these
two sectors would yield a more relaxed combination of the $(g-2)_\mu$
and $\Mh$ experimental results.)
In fact, in our plots one can observe that the particular
contourlines 
for the ``prefered'' values of $(g-2)_\mu$ and $\Mh$ by data (i.e.\ the
interior blue and pink contourlines) only cross in scenario (b) at
$m_{\rm SUSY}$ around $800\gev$ and $\tb \sim 45$ and get close,
although not crossing, in scenario (a) at  $m_{\rm SUSY}\sim 650\gev$
and very large $\tb\sim 60$. 
However, taking the uncertainties into account the overlap regions
  are quite substantial.
\item[4.-] By assuming a favored region in the ($m_{\rm SUSY}$,$\tb$)
  parameter space given by the intersect of the two $(g-2)_\mu$ (in
  pink) and $\Mh$ (in blue) areas, one can extract improved bounds for
  the slepton mixing deltas valid in these intersects. 
Those bounds give a rough idea of which parameter regions in the
  pMSSM-4 are in better agreement with the experimental data on
  $(g-2)_\mu$ and $\Mh$.
The following
  intervals for the maximum allowed  $|\delta^{AB}_{ij}|$ values can be
  deduced from our plots in these intersecting regions:

Scenario (a): 

$|\delta^{LL}_{12}|_{\rm max} \sim (6,60) \times 10^{-5} $
 
$|\delta^{LR}_{12}|_{\rm max} \sim (1.2,3.2) \times 10^{-6} $

$|\delta^{RR}_{12}|_{\rm max} \sim  (3,25) \times 10^{-3} $

$|\delta^{LL}_{23}|_{\rm max} \sim  (3,35) \times 10^{-2}$

$|\delta^{LR}_{23}|_{\rm max} \sim (1,3.2) \times 10^{-2} $

$|\delta^{RR}_{23}|_{\rm max} \sim (10) \times 10^{-1} $

Scenario (b):

$|\delta^{LL}_{12}|_{\rm max} \sim (1.5,27) \times 10^{-5} $
 
$|\delta^{LR}_{12}|_{\rm max} \sim (3,9.2) \times 10^{-6} $

$|\delta^{RR}_{12}|_{\rm max} \sim  (0.35,7) \times 10^{-3} $

$|\delta^{LL}_{23}|_{\rm max} \sim  (0.7,15) \times 10^{-2}$

$|\delta^{LR}_{23}|_{\rm max} \sim (3,9.5) \times 10^{-2} $

$|\delta^{RR}_{23}|_{\rm max} \sim (2,10) \times 10^{-1} $

Scenario (c):

$|\delta^{LL}_{12}|_{\rm max} \sim (5,22) \times 10^{-5} $
 
$|\delta^{LR}_{12}|_{\rm max} \sim (5,22) \times 10^{-6} $

$|\delta^{RR}_{12}|_{\rm max} \sim  (1.2,10) \times 10^{-3} $

$|\delta^{LL}_{23}|_{\rm max} \sim  (3,15) \times 10^{-2}$

$|\delta^{LR}_{23}|_{\rm max} \sim (5,22) \times 10^{-2} $

$|\delta^{RR}_{23}|_{\rm max} \sim (6,10) \times 10^{-1} $

Scenario (d):

$|\delta^{LL}_{12}|_{\rm max} \sim (10,30) \times 10^{-5} $
 
$|\delta^{LR}_{12}|_{\rm max} \sim (5,9) \times 10^{-6} $

$|\delta^{RR}_{12}|_{\rm max} \sim  (1.2,4) \times 10^{-3} $

$|\delta^{LL}_{23}|_{\rm max} \sim  (5,20) \times 10^{-2}$

$|\delta^{LR}_{23}|_{\rm max} \sim (5,9.5) \times 10^{-2} $

$|\delta^{RR}_{23}|_{\rm max} \sim (7,10) \times 10^{-1} $

\end{itemize}

It should be noted that in the previous upper bounds, the
particular $10 \times 10^{-1}$ value appearing in
$|\delta^{RR}_{23}|_{\rm max}$ really means 1 or larger that 1, since we
have not explored out of the $-1 \leq \delta^{AB}_{ij} \leq 1$
intervals. Particularly, in scenario (a) which has the heaviest
gauginos, we find that all values in the $-1 \leq \delta^{RR}_{23} \leq
1$ interval are allowed by LFV data.  

Finally, one can shortly summarize the previous $|\delta^{AB}_{ij}|_{\rm
  max}$ intervals found from LFV searches, by just  signaling the
typical intervals for each delta, in the favored by LHC and
$(g-2)_\mu$ data MSSM parameter space region, where the predictions 
in all scenarios lay at:   
$|\delta^{LL}_{12}|_{\rm max} \sim {\cal O} (10^{-5},10^{-4}) $, 
$|\delta^{LR}_{12}|_{\rm max} \sim {\cal O} (10^{-6},10^{-5}) $,
$|\delta^{RR}_{12}|_{\rm max} \sim {\cal O} (10^{-3},10^{-2}) $,
$|\delta^{LL}_{23}|_{\rm max} \sim {\cal O} (10^{-2},10^{-1}) $,
$|\delta^{LR}_{23}|_{\rm max} \sim {\cal O} (10^{-2},10^{-1}) $,
$|\delta^{RR}_{23}|_{\rm max} \sim {\cal O} (10^{-1},10^{0}) $. 
Very similar general bounds as for the 23 mixing are found for the
13 mixing.

 \newpage

\section{Conclusions}
\label{sec:conclusions}

We presented an up-to-date comparison of the most recent experimental
limits on LFV observables and their
predictions within the MSSM. 
The LFV observables include 
$\br(\mu \to e \ga)$ (in particular including the latest MEG results), 
$\br(\tau \to \mu \ga)$,
$\br(\tau \to e \ga)$, 
$\br(\mu \to 3 e)$,
$\br(\tau \to 3 \mu)$,
$\br(\tau \to 3 e)$,
$\br(\tau \to \mu \eta)$,
$\br(\tau \to e \eta)$ and
$\CR(\mu-e, {\rm nuclei})$.
Within the MSSM the calculations were performed at the full one-loop
level with the full (s)lepton flavor structure, i.e.\ not relying on the
mass insertion or other approximations. The results have been combined
into a Fortran code allowing for a fast joint evaluation. For
convenience we also summarized the relevant approximation formulas which
have been shown to be valid for not too large values of the LFV
parameters, which are given as $\de_{ij}^{AB}$ with $A,B = L,R$ and 
$i,j = 1, 2, 3$.

In the first part we analyzed six representative scenarios which are in
agreement with current bounds on the SUSY and Higgs searches at the
LHC. We derived the most up-to-date bounds on $\de_{ij}^{AB}$ within
these six scenarios, thus giving an idea of the overall size of these
parameters taking the latest experimental bounds into account. 
As shown in previous analyses, the observables $\br(l_i \to l_j \ga)$
continue to give the most stringent constraints on $\de_{ij}^{AB}$ for
all $A,B = L,R$. Apart from bounds on single $\de_{ij}^{AB}$'s we also
derived bounds on two parameters simultaneously, and studied where either a 
{\em positive} or a {\em negative} interference of the two $\de$'s can
be observed. As a prime example, in the case of mixings of 23 type, we
found that due to a negative interference, values of
$|\de_{23}^{LL,LR}|$ as large as  
$|\de_{23}^{LL,LR}| \approx 0.5$ are allowed in our scenarios S1 to S6
from $\br(\tau \to \mu \ga)$ 
when the two $\de$'s are allowed to vary simultaneously. On the other
hand, we also found that a relevant positive 
interference can be observed when $\de$'s of different
generation combinations are combined. In particular, we have found
important restrictions from $\mu \to e \gamma$ and $\mu-e$ conversion to
several delta pairings of the (23,13) type which are more stringent than
the ones from the single delta analysis.  

In the second part we analyzed four different two-dimensional scenarios,
which are characterized by universal scales for the SUSY electroweak
scale, $\mEW$, that determines the masses of the scalar leptons, and for
the SUSY QCD scale, $\mQCD$, that determines the masses of the scalar
quarks. As additional free parameters we kept $\mu$ and $\tb$, thus we
are investigating a special version of the pMSSM-4.
Within this simplified model it is possible to analyze the
behavior of the LFV observables with respect to the latest experimental
results of the measurement of the lightest MSSM Higgs boson mass, $\Mh$,
and the anomalous magnetic moment of the muon, \gmt. Fixing the relation
between the masses in the gaugino/higgsino sector and $\mEW$, $\mQCD$,
we obtained results for the overall behavior of the general size of
limits on the $\de_{ij}^{AB}$, which are in agreement with the
experimental results for $\Mh$ and \gmt.
In this way a general idea of the upper bounds on the deltas in these
more general scenarios can be obtained. We find
$|\delta^{LL}_{12}|_{\rm max} \sim {\cal O} (10^{-5},10^{-4}) $, 
$|\delta^{LR}_{12}|_{\rm max} \sim {\cal O} (10^{-6},10^{-5}) $,
$|\delta^{RR}_{12}|_{\rm max} \sim {\cal O} (10^{-3},10^{-2}) $,
$|\delta^{LL}_{23}|_{\rm max} \sim {\cal O} (10^{-2},10^{-1}) $,
$|\delta^{LR}_{23}|_{\rm max} \sim {\cal O} (10^{-2},10^{-1}) $,
$|\delta^{RR}_{23}|_{\rm max} \sim {\cal O} (10^{-1},10^{0}) $,
with very similar general bounds for the 13 mixing as for the 23 mixing.


\vspace{-0.5em}
\subsection*{Acknowledgments}

M.J.H.\ wishes to thank Ernesto Arganda for interesting discussions on
recent LFV and MSSM phenomenology.  
We thank M.~Rehman for helpful discussions.
The work of S.H.\ was supported in part by CICYT (grant FPA
2010--22163-C02-01) and by the Spanish MICINN's Consolider-Ingenio 2010
Program under grant MultiDark CSD2009-00064. 
M.J.H. and M.A.-C. acknowledge partial support from the European Union FP7 ITN
INVISIBLES (Marie Curie Actions, PITN- GA-2011- 289442), from the CICYT through the
project FPA2009-09017 and from CM (Comunidad Autonoma de Madrid)  through the project HEPHACOS S2009/ESP-1473. The work is also supported in part by the Spanish Consolider-Ingenio 2010 Programme CPAN (CSD2007-00042).



\end{document}